\documentclass[a4paper,fleqn,usenatbib]{mnras}

\usepackage{newtxtext,newtxmath}

\usepackage[T1]{fontenc}
\usepackage{ae,aecompl}

\usepackage{graphicx}	
\usepackage{amsmath}	
\usepackage{amssymb}	
\usepackage{CJKutf8}	

\newcommand{\chandra}{{\it Chandra\/}}

\newcommand{\xmm}{\hbox{\it XMM-Newton\/}}

\newcommand{\galex}{{\it GALEX\/}}

\newcommand{\xray}{\hbox{X-ray}}  
\newcommand{\cdfs}{\hbox{CDF-S}}

\newcommand{\lx}{L_{\rm X}}
\newcommand{\nh}{N_{\rm H}}

\newcommand{\kbol}{k_{\rm bol}}

\newcommand{\mbh}{M_{\rm BH}}

\newcommand{\lxbar}{\overline{L_{\rm X}}}
\newcommand{\bharbar}{\overline{\rm BHAR}}

\newcommand{\mstar}{M_{\star}}

\newcommand{\mhalo}{M_{\rm halo}}

\newcommand{\ks}{K_{\rm S}}
\newcommand{\photoz}{\hbox{photo-$z$}}
\newcommand{\sigmaz}{\sigma_{|\Delta z|/(1+z)}}
\newcommand{\nmad}{\sigma_{\rm NMAD}}

\title[Black-hole growth dependence on environment]{Does black-hole growth depend on the cosmic environment?}

\author[G. Yang et al.]{
G. Yang \begin{CJK*}{UTF8}{gbsn}(杨光)\end{CJK*},$^{1,2}$\thanks{E-mail: gxy909@psu.edu (GY)}
W. N. Brandt,$^{1,2,3}$
B. Darvish,$^{4}$
C.-T. J. Chen \begin{CJK*}{UTF8}{bsmi}(陳建廷)\end{CJK*},$^{1,2}$
F. Vito,$^{1,2}$ \newauthor
D. M. Alexander,$^5$
F.~E. Bauer,$^{6,7,8}$
and
J. R. Trump$^{9}$
\\
$^{1}$Department of Astronomy and Astrophysics, 525 Davey Lab, The Pennsylvania State University, University Park, PA 16802, USA\\
$^{2}$Institute for Gravitation and the Cosmos, The Pennsylvania State University, University Park, PA 16802, USA\\
$^{3}$Department of Physics, 104 Davey Laboratory, The Pennsylvania State University, University Park, PA 16802, USA\\
$^{4}$Cahill Center for Astrophysics, California Institute of Technology, 1216 East California Boulevard, Pasadena, CA 91125, USA \\
$^{5}$Centre for Extragalactic Astronomy, Department of Physics, Durham University, South Road, Durham DH1 3LE, UK \\
$^{6}$Instituto de Astrof{\'{\i}}sica and Centro de Astroingenier{\'{\i}}a, Facultad de F{\'{i}}sica, Pontificia Universidad Cat{\'{o}}lica de Chile, Casilla 306, Santiago 22, Chile \\
$^{7}$Millennium Institute of Astrophysics (MAS), Nuncio Monse{\~{n}}or S{\'{o}}tero Sanz 100, Providencia, Santiago, Chile \\
$^{8}$Space Science Institute, 4750 Walnut Street, Suite 205, Boulder, Colorado 80301 \\
$^{9}$Department of Physics, 2152 Hillside Road, U-3046, University of Connecticut, Storrs, CT 06269, USA
}

\date{Accepted XXX. Received YYY; in original form ZZZ}

\pubyear{2018}

\begin{document}
\label{firstpage}
\pagerange{\pageref{firstpage}--\pageref{lastpage}}
\maketitle

\begin{abstract}
It is well known that environment affects galaxy evolution, 
which is broadly related to supermassive black hole (SMBH) growth. 
We investigate whether SMBH evolution also depends on host-galaxy 
local (sub-Mpc) and global ($\approx 1\text{--}10$~Mpc) environment.  
We construct the surface-density field (local environment) and 
cosmic web (global environment) in the COSMOS field at 
$z=0.3\text{--}3.0$. 
The environments in COSMOS range from the field to clusters 
($\mhalo \lesssim 10^{14}\ M_\odot$), {covering the 
environments where ${\approx 99\%}$ of galaxies in the 
Universe reside}.
We measure sample-averaged SMBH accretion rate ($\bharbar$) from 
\xray\ observations, and study its dependence on overdensity and 
cosmic-web environment at different redshifts while controlling for 
galaxy stellar mass ($\mstar$).
Our results show that $\bharbar$ does not significantly depend
on overdensity or cosmic-web environment once $\mstar$ is controlled, 
indicating that environment-related physical mechanisms (e.g.  
tidal interaction and ram-pressure stripping) might not 
significantly affect SMBH growth.
We find that $\bharbar$ is strongly related to host-galaxy 
$\mstar$, regardless of environment. 
\end{abstract}

\begin{keywords}
galaxies: evolution -- large-scale structure of Universe -- 
galaxies: active -- galaxies: nuclei -- 
quasars: supermassive black holes -- X-rays: galaxies 
\end{keywords}



\section{Introduction}\label{sec:intro}
The environments of galaxies play a crucial role in their
evolution \citep[e.g.][]{de_lucia06, conselice14, somerville15}. 
In the local universe, denser regions are preferentially 
populated by early-type quiescent galaxies, while less-dense
regions are more likely to host late-type star-forming 
galaxies \citep[e.g.][]{dressler80, balogh04, kauffmann04}. 
This environmental dependence of star-forming/quiescent 
types exists at $z\lesssim 1$, although it is less clear 
at higher redshifts \citep[e.g.][]{cooper06, elbaz07, peng10, 
scoville13, darvish16}. 

Several possible environment-related mechanisms could affect 
galaxy evolution. 
Cold gas, the fuel of star formation, could flow into 
galaxies through cosmic filaments \citep[e.g.][]{keres05, dekel09};
frequent tidal interactions in denser regions could 
effectively deplete cold gas (e.g. \hbox{\citealt{farouki81}}; 
\hbox{\citealt{moore98}}); 
the strong ram pressure in clusters can strip cold 
gas from galaxies and suppress subsequent star formation 
(e.g. \citealt{gunn72, ebeling14, poggianti16});
and mergers, which can fundamentally change galaxy 
properties, happen more frequently in high-density regions 
\citep[e.g.][]{hopkins06, lin10}.
These physical processes might also affect active galactic
nucleus (AGN) activity, as the growth of supermassive black 
holes (SMBHs) also relies on the supply of cold gas 
\citep[e.g.][]{alexander12, vito14, poggianti17}.

Optical observations of low-redshift ($z\lesssim 1$) quasars 
disagree on whether they tend to reside in 
high-density or low-density regions compared to normal galaxies 
(e.g. \citealt{serber06, strand08, lietzen09}). 
This disagreement might be caused if these works did not
carefully control for host-galaxy properties.
\citet{karhunen14} found quasars do not show a significant 
dependence on environment compared to normal galaxies with 
matched redshift and host-galaxy luminosities. 
At high redshift ($z \gtrsim 3$), optical observations are 
limited to rare luminous quasars, and deep spectroscopic 
observations are often needed to measure their environment.
Therefore, these studies are often limited to small sample 
sizes and statistically significant conclusions cannot be 
obtained \citep[e.g.][]{banados13, overzier16, balmaverde17}.

Optical selection is often biased to luminous broad-line 
(BL) quasars, especially at high redshift. 
These BL quasars are rare and not well representative of 
the whole AGN population.
\xray\ emission can trace AGN activity down to a modest 
level and is widely used to investigate SMBH growth 
over the majority of cosmic history 
\citep[e.g.][]{brandt15, xue17}. 
Studies of AGN activity vs.\ environment
found that, at low redshift ($z\lesssim 1$), 
the \xray\ AGN fraction in rare rich clusters is generally 
lower than that in the field
(e.g. \hbox{\citealt{martini09}}; \hbox{\citealt{ehlert14}};
{but also see, e.g. \hbox{\citealt{haggard10}}}).
At higher redshifts, relevant studies are often constrained 
to rare protoclusters with limited AGN/galaxy sample sizes.
Their results suggest that AGN activity tends to be enhanced 
in these protoclusters (e.g. \hbox{\citealt{lehmer09}}; 
\hbox{\citealt{digby_north10}}; \hbox{\citealt{lehmer13}}; 
\hbox{\citealt{martini13}}; \hbox{\citealt{umehata15}}; 
\hbox{\citealt{alexander16}}; {but also see \citealt{macuga18}}).

However, this apparent environmental dependence might only 
be a secondary effect, and SMBH growth might be more 
fundamentally related to host-galaxy properties which
are themselves related to environment. 
{For example, \xray\ AGN activity 
is strongly related to host-galaxy stellar mass (${\mstar}$) 
rather than color (e.g. \hbox{\citealt{xue10}}) or star-formation rate 
(SFR; e.g. \hbox{\citealt{yang17}}),
and thus ${\mstar}$ must be carefully controlled when 
assessing AGN dependence on other host-galaxy properties. 
On the other hand, massive galaxies tend to reside in 
high-density regions \citep[e.g.][]{coil06, coil17}.
Therefore, to avoid such ${\mstar}$-related biases,
a large sample of AGNs and galaxies is 
needed to investigate the accretion-environment relation 
while controlling for host-galaxy ${\mstar}$.
}


In this paper, we study the dependence of 
sample-averaged SMBH accretion rate ($\bharbar$) on 
galaxy overdensity and cosmic-web environment 
while controlling for $\mstar$. 
The sample-averaged SMBH accretion is employed to 
approximate long-term average SMBH accretion for a galaxy sample 
\citep[e.g.][]{chen13, hickox14, yang17, yang18}, because 
AGNs plausibly have strong variability on timescales 
of $\sim 10^2\text{--}10^7$~yr \citep[e.g.][]{martini03, 
novak11, sartori18}.
Here, we define the overdensity as the galaxy surface number 
density relative to the median value at a given redshift and
cosmic-web environment as a galaxy's 
association to the field, a filament, or a cluster.
The overdensity and cosmic-web environment are assessed 
on physical scales of \hbox{sub-Mpc} and 
$\approx 1\text{--}10$~Mpc, respectively. 
Hereafter, we refer to overdensity and cosmic-web 
environment as ``local'' and ``global'' environments, 
respectively.

Our aim is to probe a wide redshift range of 
$z=0.3\text{--}3.0$ with large samples of \xray\ AGNs 
($\approx 2,000$) and galaxies ($\approx 170,000$).
In particular, this range covers $z\approx 1.5\text{--}2.5$, 
the peak of cosmic AGN and star-formation activity, when various 
physical processes such as galaxy mergers and AGN feedback
likely play an important role in shaping SMBH and galaxy 
coevolution 
\citep[e.g.][]{conselice14, madau14, brandt15, king15a}. 

Our analyses are based on the Cosmic Evolution Survey
\citep[COSMOS; e.g.][]{scoville07, mccracken12}. 
COSMOS has been intensively covered by spectroscopic and 
multiwavelength imaging observations 
\citep[e.g.][]{lilly09, laigle16}.
Over 20,000 sources have secure spectroscopic redshifts 
(spec-$z$) while other sources have reliable photometric 
redshifts (photo-$z$) derived from high-quality 
UV-to-IR data (up to 32 bands; e.g. \hbox{\citealt{laigle16}}).
The UV-to-IR data also make it possible to assess host-galaxy
properties such as $\mstar$ and star-forming/quiescent type
(e.g. \hbox{\citealt{ilbert13}}; \hbox{\citealt{davidzon17}}).
Deep \chandra\ \xray\ observations ($\approx$~160~ks exposure), 
which can be used to measure SMBH growth, are also available 
from the \hbox{COSMOS-Legacy} survey \citep{civano16}.
The excellent \xray\ positions from \chandra\ 
($\approx 0.5 \arcsec$) enables reliable
matching between \xray\ and optical sources 
\citep[][]{marchesi16}.
 
Thanks to its relatively large area ($\approx 2$~deg$^2$) and
deep panchromatic coverage, 
COSMOS is one of the major fields for environment studies. 
State-of-the-art techniques have been applied to COSMOS
to derive reliable measurements of the surface-density field 
up to $z\approx 3$ \citep[e.g.][]{scoville13, darvish15}. 
The statistical properties of the resulting density field 
such as mean densities and density ranges agree 
with the predictions from cosmological simulations 
\citep[e.g.][]{scoville13}.
Based on the density field, \citet{darvish17} 
utilized a new technique to construct a measurement of the cosmic 
web \citep{aragon_calvo07}. 
This method allows the mapping of sources to clusters, filaments, 
and the field. 

This paper is structured as follows. 
In \S\ref{sec:analyses}, we describe our data analyses.
In \S\ref{sec:res}, we present our results.
We discuss our results in \S\ref{sec:discuss} and 
summarize our study in \S\ref{sec:summary}.

Throughout this paper, we assume a cosmology 
with $H_0=70$~km~s$^{-1}$~Mpc$^{-1}$, $\Omega_M=0.3$, 
and $\Omega_{\Lambda}=0.7$, and a Chabrier
initial mass function \citep[][]{chabrier03}.
Quoted uncertainties are at the $1\sigma$\ (68\%)
confidence level, unless otherwise stated. 
We express $\mstar$, $\mbh$, and $\mhalo$ (halo mass) 
in units of $M_\odot$ and $\bharbar$ in units of 
$M_\odot$~yr$^{-1}$.
$\lx$\ indicates AGN \xray\ luminosity 
at rest-frame \hbox{2--10 keV} and is in units of 
erg~s$^{-1}$. 
All lengths/distances are in physical (proper) 
scale, unless otherwise stated.

\section{Data Analyses}\label{sec:analyses}

\subsection{Galaxy Sample Selection}\label{sec:sample}
Our data are based on the COSMOS2015 survey 
\citep[][]{laigle16}.
We only utilize sources within both the COSMOS and 
UltraVISTA regions, and remove objects in masked regions
(e.g. bad pixels in detectors).
These sources cover an area of $\approx1.4$~deg$^2$
(see Fig.~1 and Tab.~7 in \citealt{laigle16}).
The UltraVISTA region has deep NIR imaging data 
that are essential in estimating \photoz\ and 
$\mstar$ (\S\ref{sec:Mstar}).
We restrict our study to the $\approx 170,000$ sources 
brighter than $\ks=24$ (the $3\sigma$ limiting magnitude
of the COSMOS2015 catalog) to avoid large uncertainties 
of \photoz\ for faint sources.
The basic properties of our sample are listed in 
Tab.~\ref{tab:sample}.
Our analyses (\S\ref{sec:res}) are performed for 
the three redshift bins ($z=0.3\text{--}1.2$, 1.2--2, and 
2--3) listed in Tab.~\ref{tab:sample}.
These redshift bins cover comoving volumes of 
$7 \times 10^6$~Mpc$^3$, $1.2 \times 10^7$~Mpc$^3$, and 
$1.7 \times 10^7$~Mpc$^3$, respectively.
Tab.~\ref{tab:catalog} shows a portion of our source catalogs, 
and the full version is available as supplementary material.

We obtain spec-$z$ for $\approx 20,000$ sources in our sample
(see Tab.~\ref{tab:sample}; \citealt{marchesi16, delvecchio17}; 
Salvato et al.\ in prep.).\footnote{{In the late stages 
of this work, a new spec-$z$ data set, the DEIMOS 10k catalog, 
was released \citep{hasinger18}. 
This catalog could increase our spec-${z}$ sample by 
${\approx 10\%}$,
unlikely to affect our qualitative results.
}}
For sources without spec-$z$, we adopt the \hbox{photo-$z$}
measurements from the COSMOS2015 catalog.
These measurements are derived from high-quality 
UV-to-NIR photometric data including 18 broad bands,
12 medium bands, and 2 narrow bands (see Tab.~1 
in \citealt{laigle16}).
The medium bands can effectively improve the \hbox{photo-$z$} 
quality, enabling reliable environment studies in COSMOS 
(e.g. \hbox{\citealt{darvish15}}; \hbox{\citealt{darvish17}}).
{When compared to different \hbox{spec-${z}$} catalogs, the  
\hbox{photo-${z}$} have ${\sigma_{\rm NMAD} 
\approx 0.007\text{--}0.06}$ and outlier 
(${|\Delta z|/(1+z_{\rm spec})>0.15}$) fraction 
${\eta \approx 0.5\%\text{--}10\%}$ 
(see Tab.~5 in \citealt{laigle16})},
where $\sigma_{\rm NMAD}$ is defined as 
$1.48\times \mathrm{median}(\frac{|\Delta z- \mathrm{median}(\Delta z)|}
{1+z_\mathrm{spec}}$) (e.g. \hbox{\citealt{yang14}}).
{When compared with the recently released DEIMOS 10k 
\hbox{spec-${z}$} catalog \citep{hasinger18}, the COSMOS2015 
\hbox{photo-${z}$} have ${\sigma_{\rm NMAD}=0.015}$
and ${\eta=8\%}$, further demonstrating the high
\hbox{photo-${z}$} quality of the COSMOS2015 catalog}.
We consider all galaxies (including \xray\ detected and 
undetected) when deriving $\bharbar$ 
(see \S\ref{sec:bhar}).

\begin{table*}
\begin{center}
\caption{Summary of sample properties}
\label{tab:sample}
\begin{tabular}{cccccccccccc}\hline\hline
Redshift & $N_{\rm gal}$/$N_{\rm spec}$ & $\nmad$ & $\eta$  &
$\log M_{\star, \rm med}$ & Frac$_{\rm Q}$ & $N_{\rm slice}$ &
$\log(1+\delta)$ & $N_{\rm field}/N_{\rm fila}/N_{\rm clu}$ & 
$N_{\rm X}$ & $E_{\rm rest}$ (keV) & $\log\lx$ \\ 
(1) & (2) & (3) & (4) & (5) & (6) & (7) & (8) & (9) & (10) & (11) & (12) \\  
\hline
0.3--1.2 & 94,152/18,099 & 0.011 & 2\% & 9.3 & 13\% & 180 & ($-0.16$,0.18) & 38,840/ 48,960/ 6,352 & 889 & (0.9,12.5) & (42.6,43.3) \\
1.2--2.0 & 48,981/1,322 & 0.020 & 3\% & 9.8 & 7\% & 80 & ($-0.13$,0.13) & 20,365/ 28,616/ -- & 701 & (1.3,17.7) & (43.3,43.9) \\
2.0--3.0 & 22,828/412 & 0.059 & 10\% & 9.9 & 2\% & 50 & ($-0.14$,0.13) & 14,265/ 8,563/ -- & 429 & (1.7,23.6) & (43.7,44.2) \\
\hline 
\end{tabular}
\end{center}
\begin{flushleft}
{\sc Note.} ---
(1) Redshift bins.
(2) Numbers of galaxies and \hbox{spec-$z$} sources in our sample ($\ks < 24$). 
(3) \hbox{Photo-$z$} uncertainty (compared to \hbox{spec-$z$}).
(4) \hbox{Photo-$z$} outlier fraction.
(5) Median stellar mass. 
(6) Fraction of quiescent galaxies. 
(7) Number of $z$-slices.
(8) The overdensity (25\%,75\%) percentile range.
(9) Number sources in the field/filament/cluster environments.
We do not assign cluster environment at $z>1.2$ due to its generally 
weak signals (see \S\ref{sec:cosmic_web}). 
(10) Number of \xray\ detected sources.
(11) Rest-frame \xray\ energy sampled at median redshift. 
(12) The (25\%,75\%) percentile range of $\log\lx$ for \xray\ detected sources.
\end{flushleft}
\end{table*}

\begin{table*}
\begin{center}
\caption{Source catalog}
\label{tab:catalog}
\begin{tabular}{ccccccccccc}\hline\hline
RA & DEC & $\ks$ & $z$ & $z_{\rm lo}$ & $z_{\rm up}$ & $\log M_{\star}$ & Type$_{\rm gal}$ &
$\log(1+\delta)$ & Web & $\log\lx$ \\
(1) & (2) & (3) & (4) & (5) & (6) & (7) & (8) & (9) & (10) & (11) \\ \hline 
149.411496 & 2.712315 & 22.8 & 0.706 & 0.655 & 0.774 &  9.18 & 1 & $-0.343$ & 2 & $-99.00$ \\
149.411504 & 2.765237 & 23.5 & 0.975 & 0.951 & 0.990 &  8.06 & 1 & $-0.256$ & 2 & $-99.00$ \\
149.411576 & 2.336084 & 21.0 & 1.783 & 1.729 & 1.818 & 11.22 & 1 & $ 0.292$ & 1 & $-99.00$ \\
149.411578 & 2.306681 & 21.3 & 1.359 & 1.241 & 1.433 & 10.71 & 0 & $-0.019$ & 1 & $-99.00$ \\
149.411581 & 2.411649 & 21.5 & 0.389 & 0.381 & 0.398 &  9.28 & 1 & $ 0.149$ & 1 & $-99.00$ \\
149.411603 & 2.243533 & 21.9 & 1.502 & 1.469 & 1.543 & 10.32 & 1 & $ 0.285$ & 1 & $-99.00$ \\
149.411643 & 2.290855 & 23.6 & 1.185 & 1.171 & 1.198 &  9.07 & 1 & $ 0.128$ & 1 & $-99.00$ \\
149.411643 & 2.592744 & 23.6 & 0.880 & 0.825 & 0.942 &  9.24 & 1 & $ 0.017$ & 1 & $-99.00$ \\
149.411659 & 2.319370 & 23.6 & 1.022 & 0.812 & 1.148 &  9.23 & 1 & $-0.621$ & 2 & $-99.00$ \\
149.411661 & 2.410365 & 22.5 & 1.063 & 1.063 & 1.063 &  9.11 & 1 & $-0.192$ & 1 & $ 43.63$ \\ \hline 
\end{tabular}
\end{center}
\begin{flushleft}
{\sc Note.} ---
Only a portion of this table is shown here, and the full version is available 
as supplementary materials. 
The table is sorted in ascending order of RA. 
(1) \& (2) Source J2000 coordinates. 
(3) $\ks$ AB magnitude from the COSMOS2015 catalog \citep{laigle16}. 
(4), (5), \& (6) Redshift and redshift 1$\sigma$ lower and upper limits
	(\S\ref{sec:sample}).
	For \hbox{spec-$z$} sources, the lower and upper limits are set
	the same as the redshift value. 
(7) Stellar mass (\S\ref{sec:Mstar}).
(8) Galaxy type (0: quiescent; 1: star-forming; \S\ref{sec:Mstar}). 
(9) Overdensity (\S\ref{sec:density_field}).
(10) Cosmic-web environment (0: cluster; 1: filament; 2: field; 
\S\ref{sec:cosmic_web}).
We do not assign cluster environment at $z>1.2$ due to its generally 
weak signals.
(11) \xray\ luminosity (rest-frame 2\text{--}10~keV; \S\ref{sec:xray_det}).
For \xray\ undetected sources, the values are set to ``$-99.00$''.
\end{flushleft}
\end{table*}

\subsection{Stellar Mass}\label{sec:Mstar}
\begin{figure}
\includegraphics[width=\linewidth]{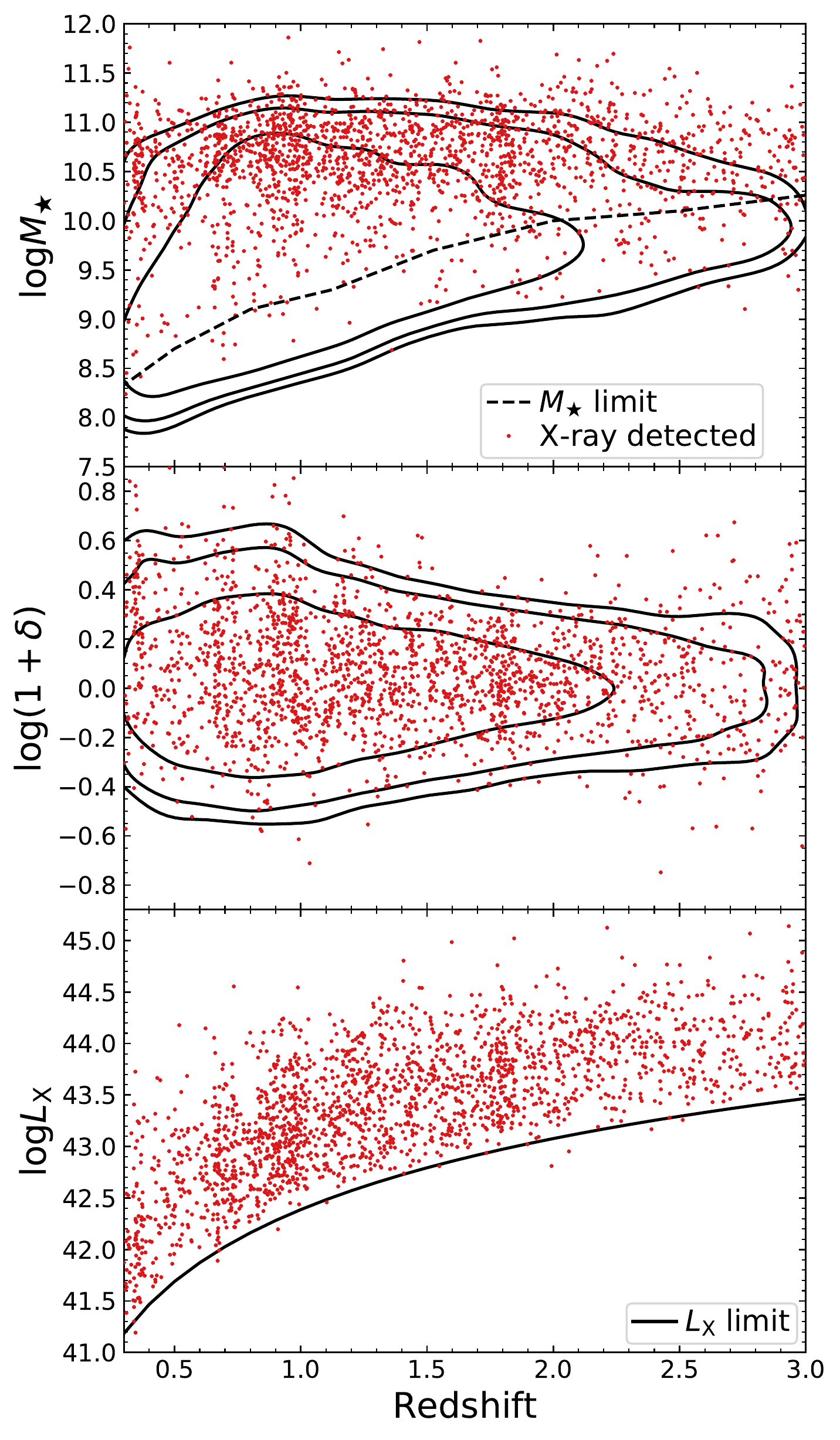}
\caption{$\mstar$, overdensity, and $\lx$ as a 
function of redshift.
The contours encircle 68\%, 90\%, and 95\% of all 
sources (\S\ref{sec:sample}), respectively. 
The red points represent \xray\ detected sources.
In the top panel, the dashed curve indicates the 
$\mstar$ completeness limit from \citet{laigle16}.
In the bottom panel, the solid curve represents 
the limiting $\lx$.
}
\label{fig:all_vs_z}
\end{figure}

To estimate $\mstar$, we perform spectral energy 
distribution (SED) fitting with {\sc cigale} 
\citep{noll09, serra11} at \hbox{$z_{\rm spec}$} 
or \hbox{$z_{\rm photo}$} (\S\ref{sec:sample}).
The input photometry is from the COSMOS2015 catalog 
(\S\ref{sec:sample}). 
We do not adopt the $\mstar$ measurements from the
COSMOS2015 catalog directly, mainly because the 
our redshifts are not exactly the same as those 
in the COSMOS2015 catalog (\S\ref{sec:sample}).
We employ nebular and dust emission in 
{\sc cigale} \citep{noll09, draine07}.
We apply the extinction law from \cite{calzetti00} 
with $E(B-V)$ ranging from $0-1$. 
Following \cite{yang18}, we use a $\tau$ model
of the star formation history (SFH) with 
$\log(\tau/{\rm yr})$ ranging from 8 to 10.5. 
We allow stellar metallicity values of $Z=0.0001, 0.0004, 
0.004, 0.008, 0.02, 0.05$, where $Z$ is the 
mass fraction of metals. 
Our $\mstar$ measurements have a systematic of 0.002~dex 
and a scatter of 0.11~dex compared to those in the
COSMOS2015 catalog. 
For the 239 BL AGNs (identified by 
\citealt{marchesi16}), we also adopt an additional 
BL AGN component following the settings in Tab.~1 
of \cite{ciesla15}.
The resulting $\mstar$ values are typically 
$\approx 0.3$~dex different from those obtained 
with only galaxy templates (see \S~2.1.3 of \citealt{yang18}).
Fig.~\ref{fig:all_vs_z} displays $\mstar$ vs.\ 
redshift for our sample. 
We also show the $\mstar$ completeness limit 
corresponding to $\ks=24$ from \citet{laigle16} in
Fig.~\ref{fig:all_vs_z}. 
The completeness limit is estimated based on an empirical
method which does not assume a specific galaxy template.
The limiting $\log\mstar$ at $z=1.2$, 2, and 3 are 
9.3, 10.0, and 10.3, respectively.
In \S\ref{sec:res}, we perform analyses for 
$\mstar$ above these limits in three redshift bins 
of $z=0.3\text{--}1.2$, 1.2--2, and 2--3, respectively. 

We classify a source as a quiescent galaxy if its 
rest-frame colors satisfy $\mathrm{NUV}-r > 3(r-J)+1$ and 
$\mathrm{NUV}-r > 3.1$, otherwise we classify 
it as a star-forming galaxy 
\citep[][]{williams09, ilbert10}.\footnote{Here, the NUV 
specifically refers to the \galex\ band centered at 2300~\AA.} 
Here, the rest-frame colors are obtained from 
our SED fitting. 
This color-based selection helps to avoid misclassifying 
dust-reddened star-forming galaxies as quiescent 
galaxies \citep[e.g.][]{ilbert10, ilbert13}. 
The fractions of quiescent galaxies in different redshift
ranges are listed in Tab.~\ref{tab:sample}.
This classification is used to estimate \xray\
emission from \xray\ binaries (XRBs; see \S\ref{sec:bharbar}). 
Our color-color scheme is not appropriate for galaxies hosting 
BL AGNs due to the strong AGN UV-to-NIR emission. 
Following \citet{yang17}, we set the hosts of BL AGNs 
as star-forming galaxies. 
Setting them as either star-forming or quiescent galaxies has 
negligible effects to our results, as BL~AGNs are only a small 
population compared to the entire galaxy sample 
($\approx 0.1\%$).


\subsection{Environment}\label{sec:env}
We build the surface-density field and cosmic-web estimates 
in this section. 
The technical details are presented in 
\citet{darvish15, darvish17}, and we briefly describe 
the procedures in \S\ref{sec:density_field} and 
\S\ref{sec:cosmic_web}. 
In Appendix~\ref{app:ill}, we explain our environment 
measurements in a straightforward way, especially for 
readers who are not familiar with environmental 
studies. 
{As demonstrated in \S\ref{sec:phy}, the physical 
environment-SFR relation clearly exists in our sample, 
supporting the robustness of our environment measurements.
}

{Some studies suggest that there might be different
environmental effects for ``central'' vs.\ ``satellite'' 
galaxies in a dark-matter halo 
(e.g. \hbox{\citealt{li06}}; \hbox{\citealt{hickox09}}).
We do not label our sources as central or 
satellite galaxies, because most galaxies 
(${\approx 80\%\text{--}90\%}$) 
at low redshift (${z \lesssim 0.6}$) are observed to be 
isolated or reside in small groups (galaxy members 
${\lesssim 5}$) and their central/satellite classification 
is challenging due to factors like photo-${z}$ 
uncertainties and survey sensitivity 
(e.g. \hbox{\citealt{knobel09, knobel12}}).  
At higher redshift, the fraction of 
isolated or small-group galaxies is even higher as the 
large-scale structure is still in development 
(e.g. \hbox{\citealt{springel05}}; \hbox{\citealt{overzier16}}). 
Considering that our sources cover a wide redshift range 
of ${z=0.3\text{--}3}$, a detailed unbiased 
central/satellite classification is beyond the scope of 
this work.}
  
\subsubsection{Density Field (Local Environment)}\label{sec:density_field}
We adopt the ``weighted adaptive kernel smoothing'' 
method to construct the surface-density field 
that probes \hbox{sub-Mpc} physical scales.
As demonstrated by intensive simulations, the 
performance of this method is excellent (see \S5 and \S6 
of \citealt{darvish15}).
The density field is calculated for all sources, 
including normal galaxies and \xray\ detected sources.

We first calculate $\sigmaz$ as a function of redshift.
$\sigmaz$ is derived within $z\pm0.2$ at each redshift. 
$\sigmaz$ is $\approx 0.01$ at low
redshift ($z\lesssim 1$) and rises to $\approx 0.04$ toward 
high redshift ($z\gtrsim 2$). 
{This level of photo-${z}$ accuracy is sufficient 
for reliable cosmic-environment characterization (e.g. 
\hbox{\citealt{scoville13}}; \hbox{\citealt{darvish15}}).}
We then define a series of redshift slices ($z$-slices) 
with widths of $\pm 1.5(1+z)\sigmaz$. 
This width is suggested by \citet{malavasi16}.
The $z$-slices are designed in a way that  
$\gtrsim 90\%$ of each $z$-slice is overlapping with its 
next $z$-slice.
Such dense design is to appropriately consider the 
\hbox{photo-$z$} distribution of galaxies close 
to the boundaries of each $z$-slice (see \S3.1 of 
\hbox{\citealt{darvish15}}).
The numbers of $z$-slices in different redshift ranges 
are listed in Tab.~\ref{tab:sample}.
For each $z$-slice, we calculate the weight 
for each source, defined as the percentage of the 
redshift probability distribution function within 
this $z$-slice.
We assign a weight of 100\% to sources with available
spectroscopic redshifts.
To reduce computational time, at each redshift, we only 
include sources with weight at least 10\%. 
To derive the surface-density field for each $z$-slice, 
we utilize a 2D Gaussian kernel whose 
width adaptively decreases in denser regions,
ranging from $\approx 0.2$~Mpc (1\% percentile) to 
$\approx 0.9$~Mpc (99\% percentile).
The algorithm requires an input ``global smoothing 
width''. 
We adopt the value of 0.5~Mpc which 
is the typical virial radius of \xray\ clusters in COSMOS 
($\log \mhalo \approx13\text{--}14$; 
e.g. \hbox{\citealt{finoguenov07, 
george11}}).\footnote{\citet{darvish15} tested global
smoothing widths from 0.1 to 2.0~Mpc and did not find 
a significant change in the resulting density field.}
Following \citet{darvish17}, we filter out sources 
near ($<1$~Mpc) the edge of the field and/or large 
masked regions in the COSMOS2015 survey \citep{laigle16}, 
because density measurements for these sources are 
unreliable.
The procedures above yield measurements of surface number 
density ($\Sigma$, in units of Mpc$^{-2}$) for each source. 

\begin{figure}
\includegraphics[width=\linewidth]{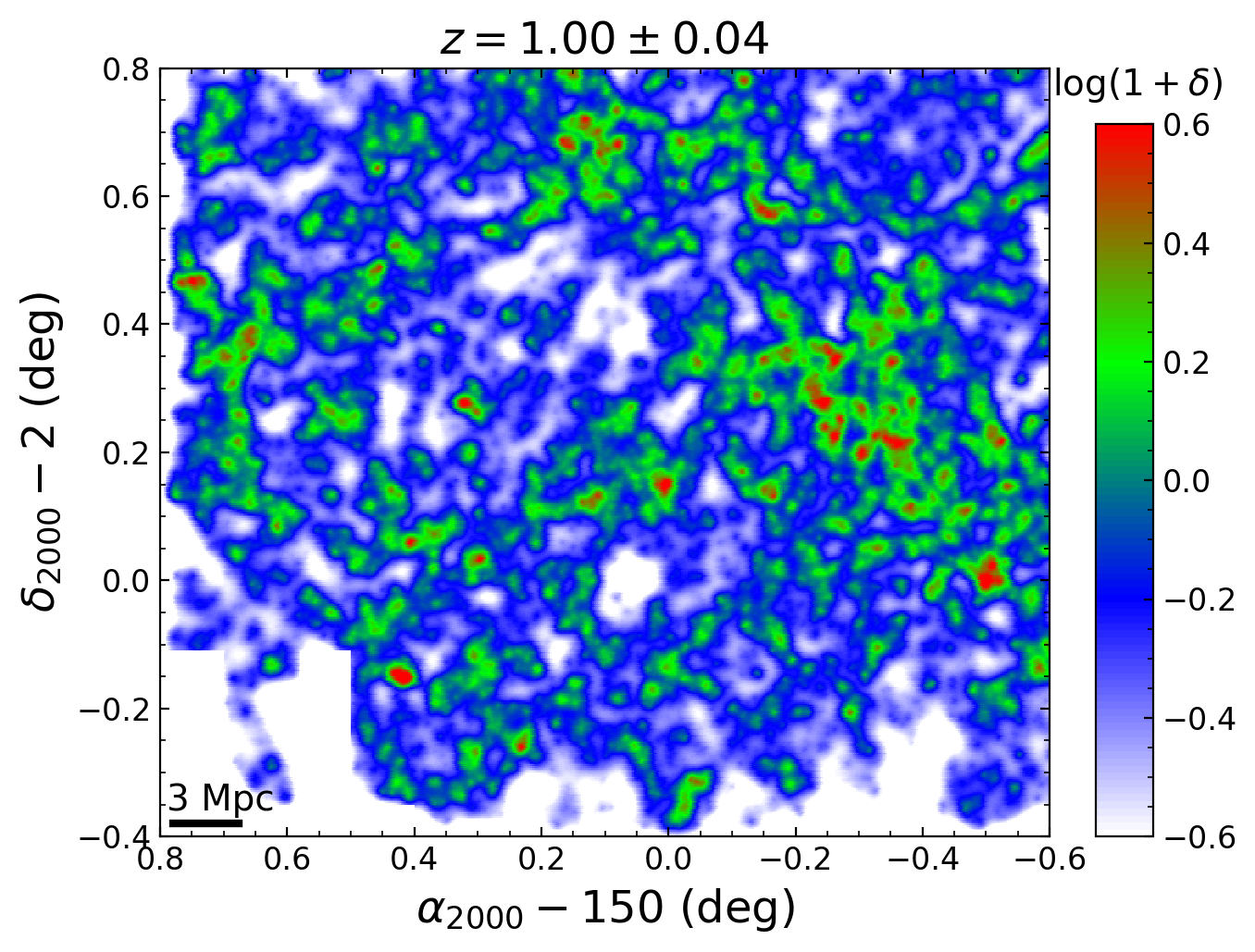}
\includegraphics[width=\linewidth]{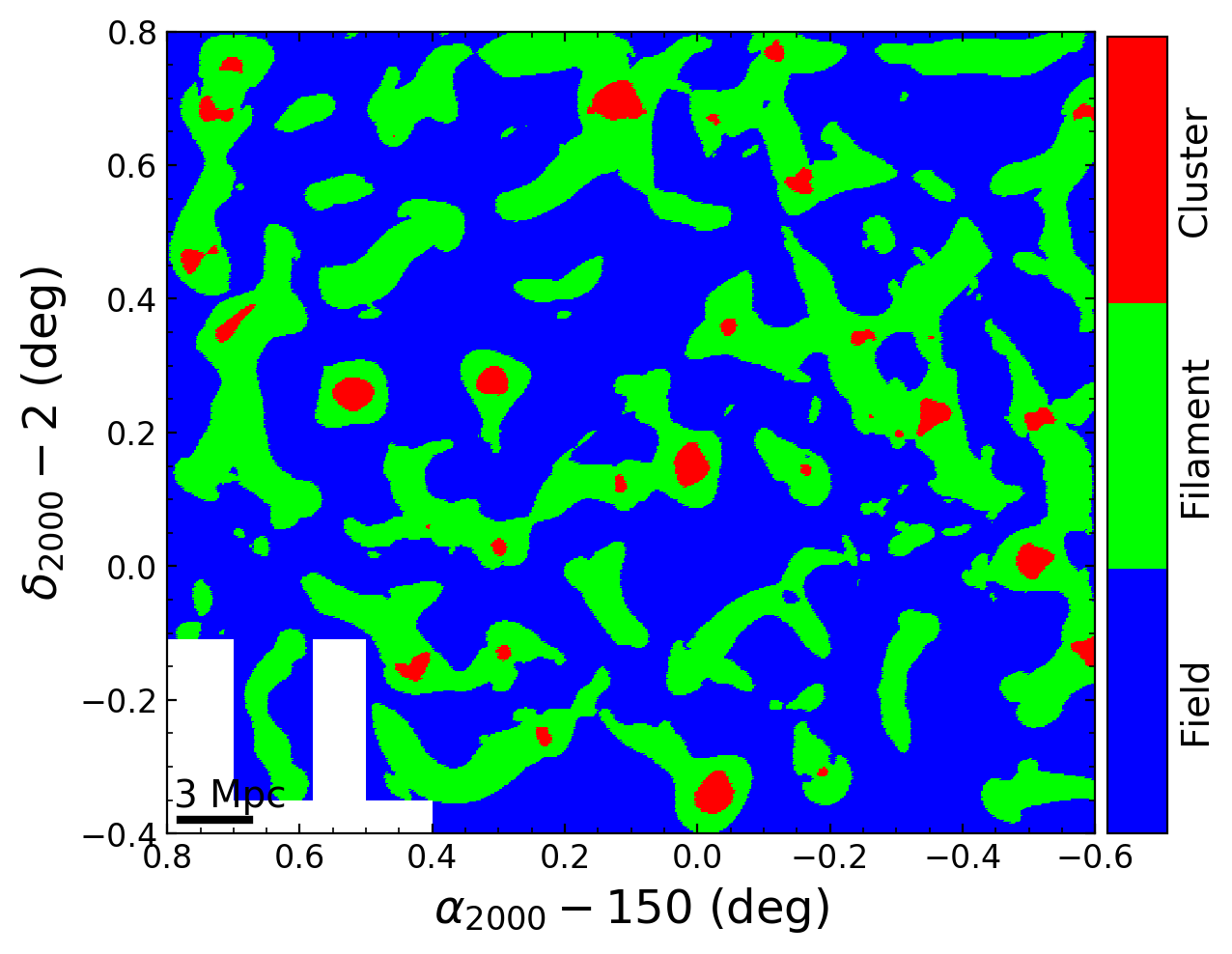}
\caption{The maps of overdensity (top) and cosmic web 
(bottom) for the $z$-slice at $z=1.00\pm0.04$ derived 
from our galaxy sample (see \S\ref{sec:env}). 
A physical scale of 3~Mpc is marked at the lower left
corner in each panel. 
From the field to cluster environment, the overdensity 
tends to be higher. 
{However, this is only a statistical trend. 
For example, high overdensity (top) does not 
necessarily correspond to cluster (bottom), and vice 
versa (see \S\ref{sec:cosmic_web}).}
The clusters identified in COSMOS are relatively low-mass
systems ($\log\mhalo \lesssim 14$; see 
\S\ref{sec:density_field}).
The white patches at the lower left corner are masked 
regions where NIR imaging data are not available 
\citep{mccracken12}. 
}
\label{fig:det_web_z1}
\end{figure}

\begin{figure}
\includegraphics[width=\linewidth]{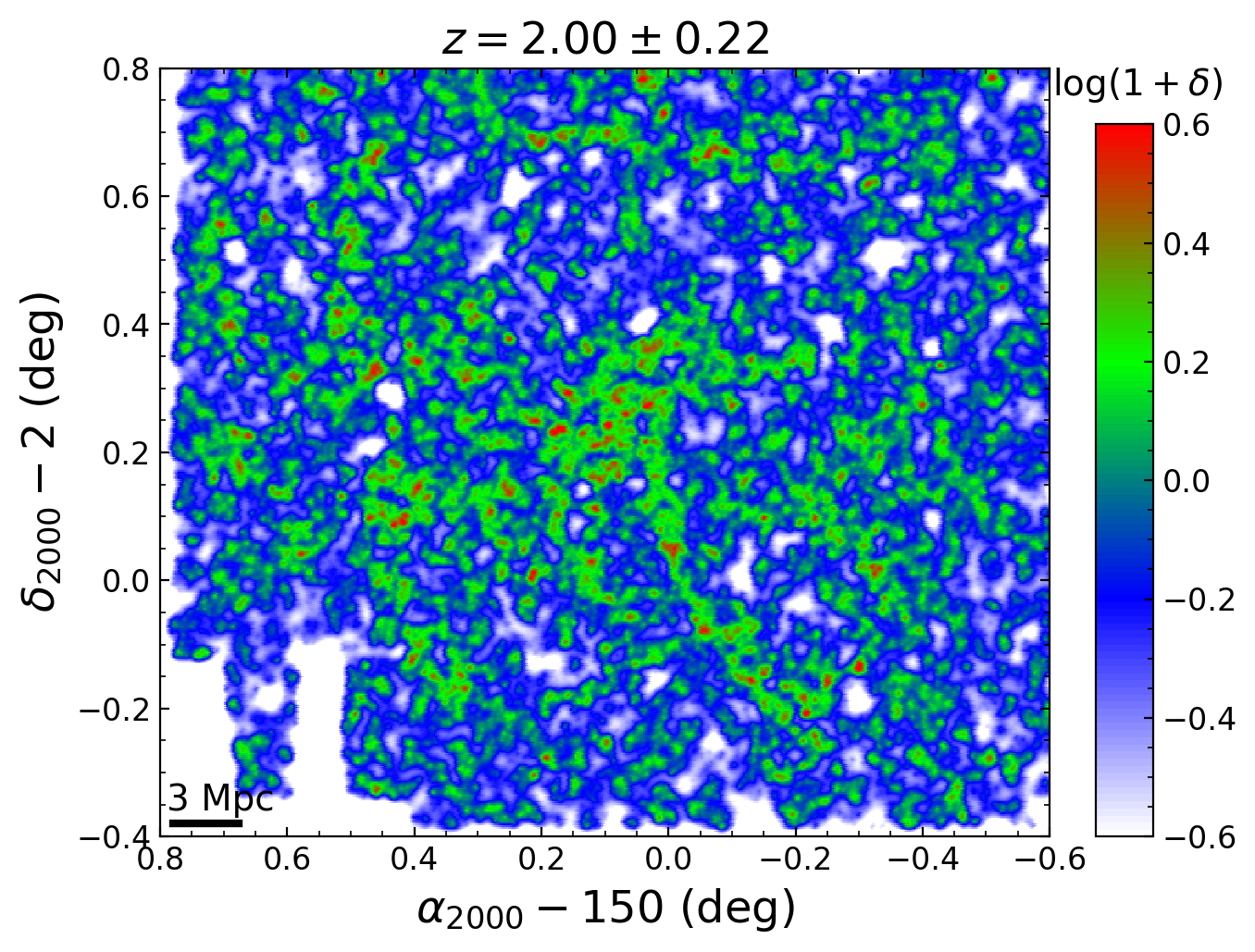}
\includegraphics[width=\linewidth]{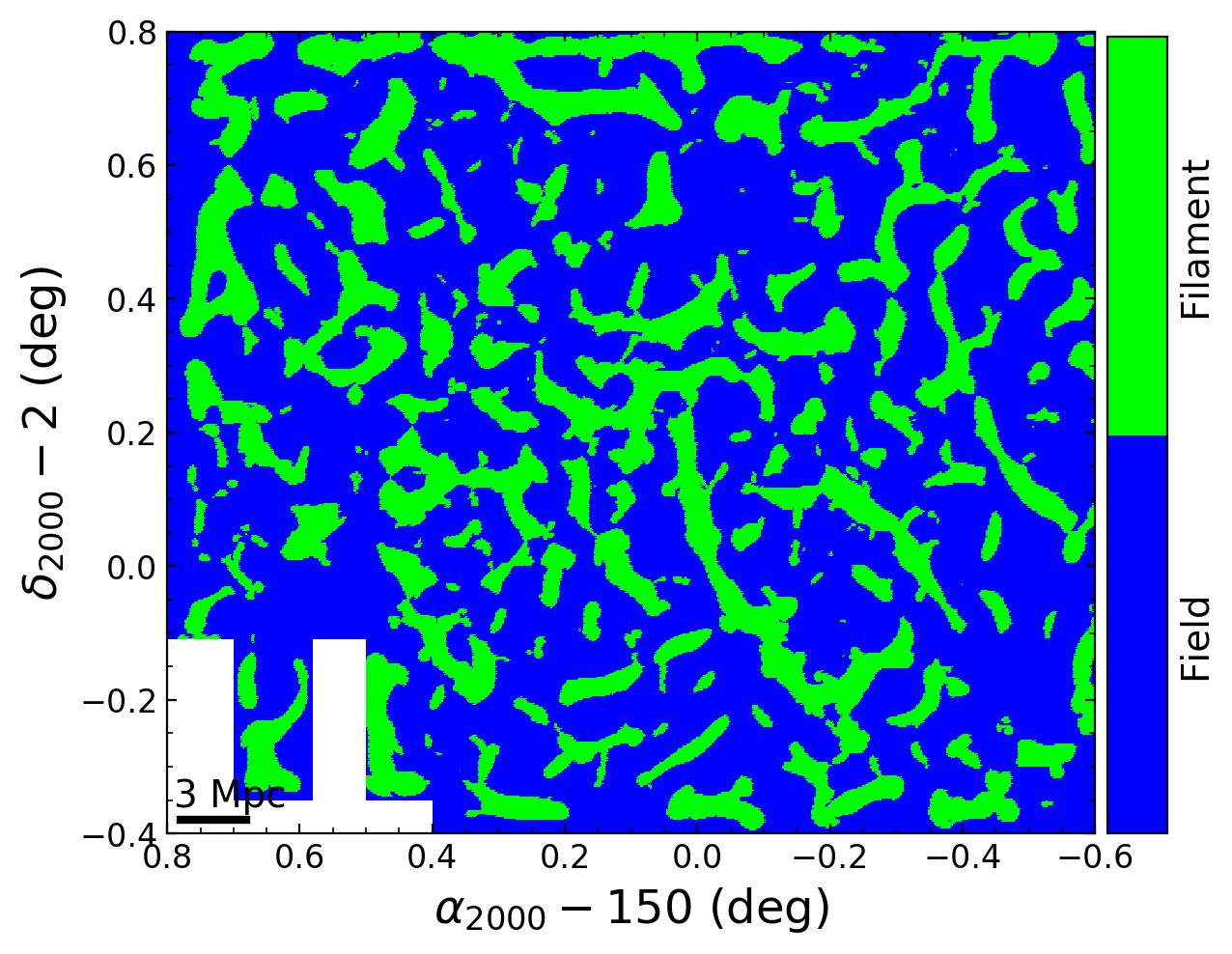}
\caption{Same format as Fig.~\ref{fig:det_web_z1} but
for the $z$-slice centered at $z=2.00\pm0.22$. 
Unlike in Fig.~\ref{fig:det_web_z1}, we do not assign 
cluster environment due to its generally weak signals 
(see \S\ref{sec:cosmic_web}).
}
\label{fig:det_web_z2}
\end{figure}

We quantify the local environment for each source via the 
dimensionless overdensity parameter, defined as 
\begin{equation}
\label{eq:over_density}
1+\delta = \frac{\Sigma}{\Sigma_{\rm median}},
\end{equation}
where $\Sigma_{\rm median}$ is the median $\Sigma$ at 
each redshift. 
To minimize the effects of cosmic variance, $\Sigma_{\rm median}$ 
is calculated within $z\pm0.2$ at redshift $z$.
Figs.~\ref{fig:det_web_z1} and \ref{fig:det_web_z2} show 
the overdensity maps for two $z$-slices centered at $z=1.0$ 
and 2.0, respectively. 
The above overdensity measurements are based on sources
with $K_{\rm s}<24$ (see \S\ref{sec:sample}). 

\begin{figure*}
\includegraphics[width=\linewidth]{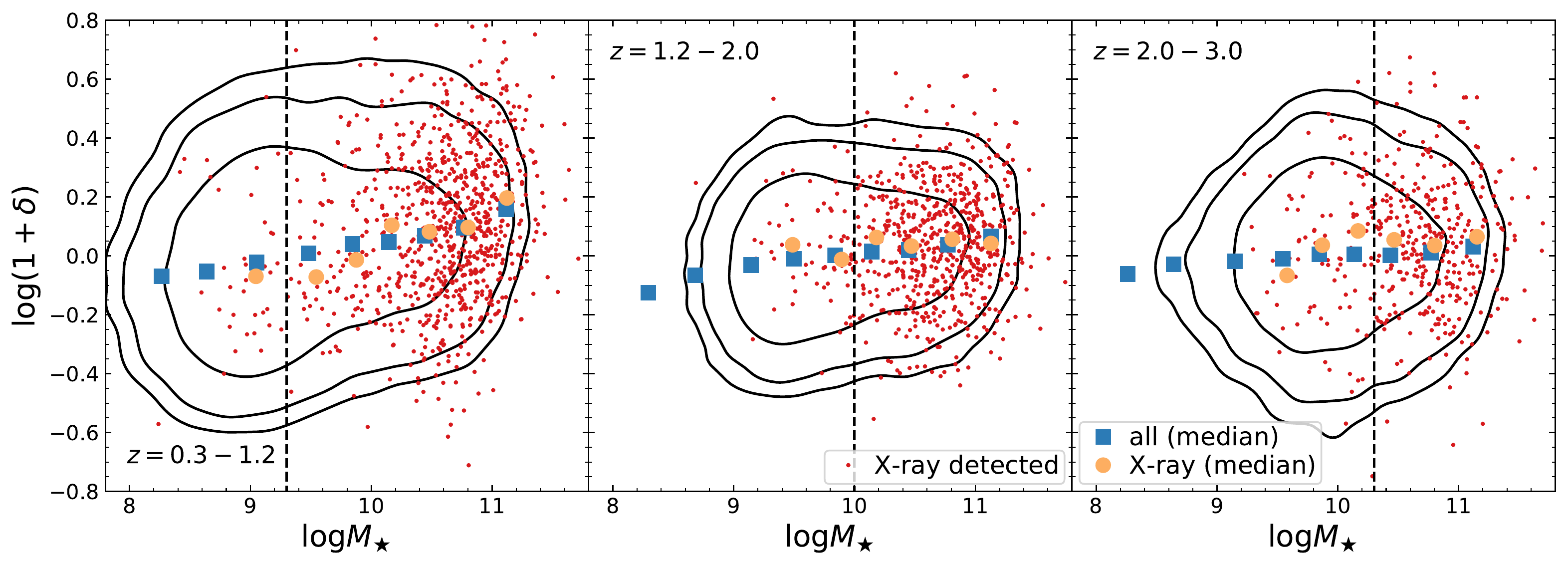}
\caption{Overdensity vs.\ $\mstar$ for different redshift 
ranges. 
The contours include 68\%, 90\%, and 95\% of all 
galaxies, respectively. 
The red points represent \xray\ detected sources.
The blue squares and orange circles indicate the median overdensity 
in each $\mstar$ bin (\S\ref{sec:split_det}) for all
galaxies and \xray\ detected sources, respectively.
We only calculate the medians for bins with more than 
20 sources to avoid large uncertainties. 
The vertical dashed lines represent the $\mstar$ 
completeness limits at $z=1.2$, 2, and 3, respectively
(\S\ref{sec:Mstar}).
The \xray\ detected sources tend to have high $\mstar$
but do not show an obvious dependence on overdensity.
}
\label{fig:det_vs_M}
\end{figure*}

Figs.~\ref{fig:all_vs_z} and \ref{fig:det_vs_M} show
overdensity as a function of redshift and $\mstar$,
respectively (see Tab.\ref{tab:sample} for typical 
overdensity ranges of our sample). 
There are positive trends between overdensity and $\mstar$
at $z \lesssim 2$, consistent with previous work 
\citep[e.g.][]{darvish17}.

In our overdensity estimation above, we have appropriately weighted sources 
according to their \hbox{photo-$z$} uncertainties (see \S\ref{sec:env}). 
However, this weighting technique does not account for catastrophic 
\hbox{photo-$z$} outliers. 
The outlier fractions are $\approx 0.5\%\text{--}10\%$, when compared 
with different \hbox{spec-$z$} catalogs (see \S\ref{sec:sample} and 
Tab.~\ref{tab:sample}).
The \hbox{photo-$z$} outliers increase the noise level in the derivation 
of density field. 
\citet{darvish15} assessed the effects of outliers via simulations 
(see their \S5.1).
They concluded that an outlier fraction of 10\% has little effect on 
the derived density field. 
Therefore, our results should not be qualitatively affected by the 
\hbox{photo-$z$} outliers.

The typical stellar mass of our sample is relatively small 
($\log\mstar \lesssim 10$; see Tab.~\ref{tab:sample}). 
We have also tested using only a subsample of $\log\mstar>10$ galaxies 
when estimating the density field. 
Our results (\S\ref{sec:res}) do not change qualitatively. 
The total stellar mass included in this subsample is $\approx 80\%$ 
of that included in the entire sample. 
However, the subsample consists of only $\approx 30\%$ of our sources, 
inevitably leading to stronger Poisson noise in the density-field
estimation. 

\subsubsection{Cosmic Web (Global Environment)}
\label{sec:cosmic_web}
Based on the density field derived in \S\ref{sec:density_field}, 
we extract a cosmic-web estimate with the multi-scale morphology filter 
(MMF) algorithm \citep[e.g.][]{aragon_calvo07, darvish14, darvish17}. 
The basic idea is to measure the geometry of the density field 
around each point in the $z$-slice. 
If the geometry is similar to that of a typical cluster/filament, 
then the point's environment is classified as cluster/filament;
otherwise it is classified as the field. 

Specifically, we first derive the Hessian matrix (second-order
partial derivatives) of the density field for each point in 
a $z$-slice (see \S3.4.1 of \citealt{darvish17} for details).
We then calculate two eigenvalues for each Hessian matrix.
The eigenvalues describe the density-field geometry around 
the point. 
Based on the signs, ratios, and normalizations of the two 
eigenvalues, a cluster signal ($S_c$) and a filament signal 
($S_f$) are obtained for the point. 
Here, the signals are two numbers within $0\text{--}1$,
where larger values indicate higher chances of lying in  
a cluster/filament. 
To account for the multi-scale nature of clusters and 
filaments, the above procedures are repeated but each
time the density field is smoothed with a Gaussian 
kernel of different physical scales (0.25, 0.50, 0.75, 
1.00, 1.50, 2.00~Mpc). 
The final cluster/filament signal for each point is 
assigned as the largest value among those obtained 
with different smoothing scales. 

After obtaining the signal maps for each $z$-slice, we 
need to apply appropriate signal thresholds to identify 
clusters and filaments. 
A higher signal threshold generally increases reliability 
but decreases completeness in structure selection. 
We adopt the redshift-dependent thresholds from \S3.4.2 of 
\citet{darvish17}, which are designed to balance 
between reliability and completeness.
These thresholds are $T_c = 0.0639z+0.1142$ (cluster) and 
$T_f = 0.0253z+0.0035$ (filament). 
Following \citet{darvish17}, we assign a cluster environment
to a point if it has $S_c \geq S_f$  and $S_c \geq T_c$.
If a point is not assigned as a cluster environment but 
has $S_f \geq T_f$, we assign it as a filament environment. 
If an object is not assigned as cluster or filament 
environment, we assign it as the field environment. 

The above criterion has been successfully applied to 
$z\lesssim 1.2$ sources \citep[e.g.][]{darvish17}. 
Fig.~\ref{fig:det_web_z1} shows the cosmic-web map for
the $z$-slice centered at $z=1.0$. 
The clusters are roughly round with typical physical sizes of
$\sim 1$~Mpc.
The filaments are elongated, with a typical length 
of $\sim 10$~Mpc.
However, we find, at $z\gtrsim 1.2$, the clusters are 
often dominated by noise. 
This is understandable as clusters are still forming at
high redshifts and they exist in the form of protoclusters
\citep[e.g.][]{kravtsov12, overzier16}.
The protoclusters have weaker signals and are generally
beyond our detection sensitivity. 
Therefore, we do not assign cluster environments for 
redshift ranges of $z=1.2\text{--}2.0$ and $z=2.0\text{--}3.0$.
Specifically, if an object at $z=1.2\text{--}3.0$ has 
$S_f \geq T_f$, we assign it as a filament environment;
otherwise, we assign it as a field environment. 
Fig.~\ref{fig:det_web_z2} displays the cosmic-web map 
for the $z$-slice centered at $z=2.0$.
The numbers of sources associated with different cosmic-web
environments are summarized in Tab.~\ref{tab:sample}.

\begin{figure}
\includegraphics[width=\linewidth]{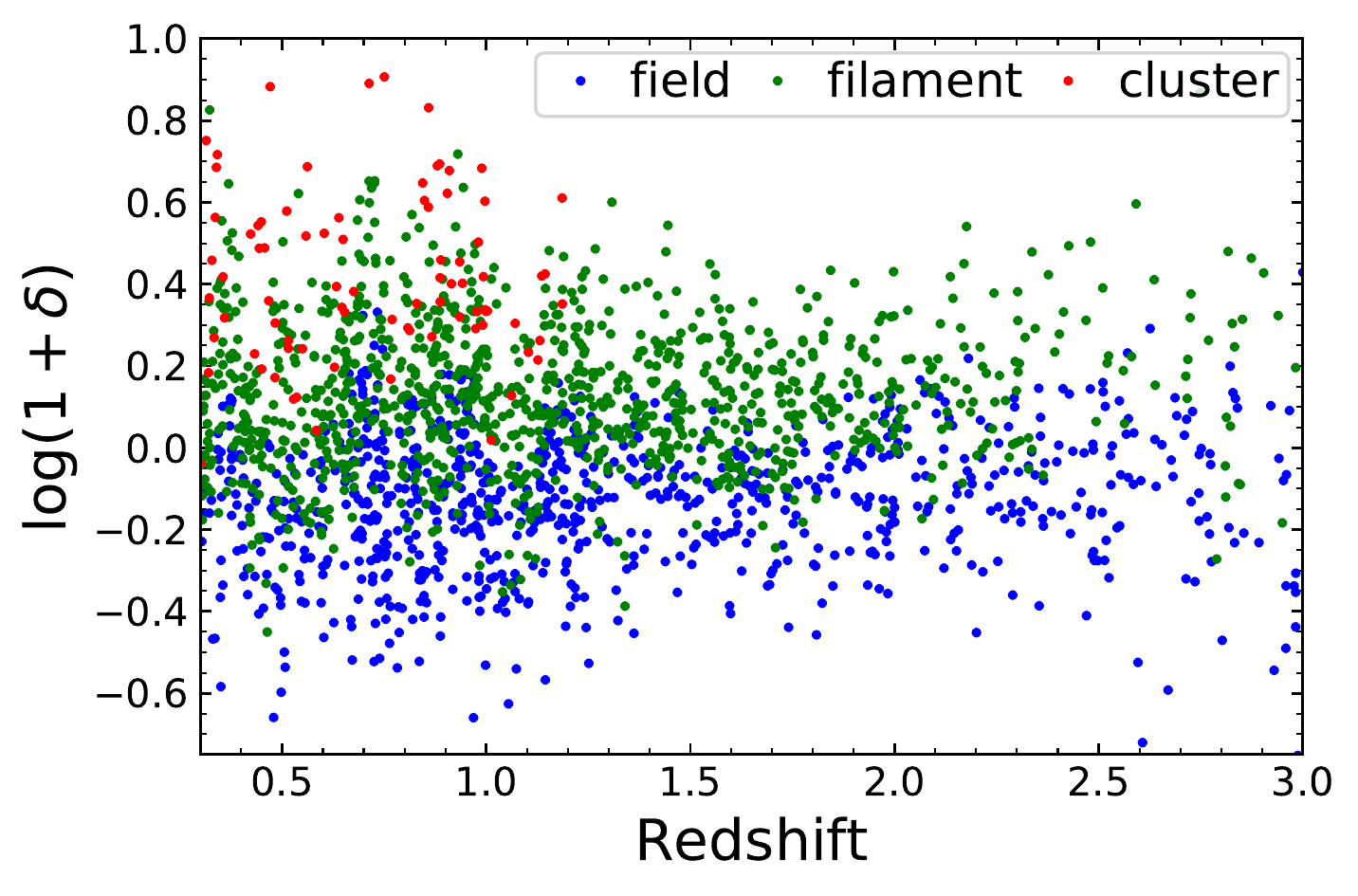}
\caption{Overdensity vs.\ redshift for 2,000 randomly 
selected sources in our sample.
Different colors indicate sources with different 
cosmic-web environments. 
The clusters identified in COSMOS are relatively low-mass
systems ($\log\mhalo \lesssim 14$; see 
\S\ref{sec:density_field}).
We do not assign cluster environment at redshifts 
above $z=1.2$ due to its generally weak signals 
(\S\ref{sec:cosmic_web}).
Although the overdensity tends increase moving from 
the field to clusters, there are significant overlaps
between different cosmic-web environments. 
}
\label{fig:det_vs_z}
\end{figure}

Fig.~\ref{fig:det_vs_z} shows the overdensity as a 
function of redshift for different cosmic-web environments.
{Although the overdensity generally rises from the field
to clusters, there are substantial overlapping areas 
in the overdensity-redshift parameter space. 
This overlap is understandable as the overdensity and cosmic-web
measurements describe cosmic environment on different 
scales (\hbox{sub-Mpc vs.\ ${\approx 1\text{--}10}$~Mpc}).
The overlap highlights the importance of our 
MMF algorithm in the construction of the cosmic web, 
and a simple overdensity-based algorithm
would not be feasible for cosmic-web association.
Readers might worry that the overlap might smear 
out potentially weak trends between ${\bharbar}$ 
and environment.
However, this is not an issue in our analyses, because 
we assess ${\bharbar}$ dependence on both overdensity and 
cosmic-web environment individually and reach consistent 
results (see \S\ref{sec:res}). 
}

{\citet{george11} have associated galaxies with 
\xray\ selected clusters at ${z<1}$ using a probabilistic method. 
We match their cluster-member candidates (member 
probability above 70\%) to our sample with a ${0.5\arcsec}$
matching radius. 
As expected, the 1,851 matched galaxies generally have high 
overdensity values of 
${\log(1+\delta)=0.34\text{--}0.66}$ (25\%--75\% percentile)
compared to our overall sample (see Tab.~\ref{tab:sample}).
For these 1,851 galaxies, 44\% and 50\% are assigned as cluster 
and filament, respectively, in our catalog. 
The 44\% filament objects tend to lie in the 
boundary between clusters and filaments in our ${z}$-slices,
where the cluster/filament classification is sensitive 
to the methodology. 
The other 6\% of galaxies are assigned as the field environment 
in our catalog. 
This disagreement is likely caused by the differences in 
adopted redshift measurements, as these galaxies' redshift values 
in the two catalogs differ by ${\approx 3\%}$.
In comparison, the other 94\% of galaxies (\citealt{george11}
cluster candidates assigned as cluster/filament galaxies) 
have redshift differences of only ${\approx 0.7\%}$.
Our adopted photo-${z}$ values should have improved 
quality compared to those adopted in \citet{george11}, 
who used photo-${z}$ from an earlier COSMOS catalog 
\hbox{\citep{ilbert09}}.
Note that it is natural that most of our cluster galaxies are 
not identified by \citet{george11}, because their \xray\ selected 
clusters are not complete. 
}

%

\subsection{Black Hole Accretion Rate}\label{sec:bhar}
We derive $\bharbar$ for samples of sources broadly following 
the procedures in \S2.3 of \cite{yang17}. 
Briefly, we first calculate the total sample-averaged $\lx$ ($\lxbar$)
considering both \xray\ detected sources (\S\ref{sec:xray_det}) 
and undetected sources (\S\ref{sec:xray_undet}). 
{The undetected sources are considered with an \xray\
stacking technique. 
The stacking procedure is necessary to avoid biases due to 
the limited \xray\ survey sensitivity, as faint AGNs become 
undetected toward high redshift.}
We obtain the AGN $\lxbar$ by subtracting the $\lxbar$ 
component contributed by XRBs (\S\ref{sec:bharbar}). 
Finally, we convert AGN $\lxbar$ to $\bharbar$ in units of 
$M_\odot$~yr$^{-1}$.

\subsubsection{X-ray Detected Sources}\label{sec:xray_det}
We select all \xray\ detected sources using the \hbox{COSMOS-Legacy} 
\xray\ survey \citep{civano16}. 
The \hbox{COSMOS-Legacy} survey, conducted by \chandra, is the 
deepest \xray\ survey available for COSMOS, and can sample 
most of cosmic accretion power. 
For instance, at $z\approx 1.5\text{--}2.5$, the peak of cosmic AGN 
activity, it covers $\lx$ ranging from $\approx 10$ times 
below the knee luminosity of the \xray\ luminosity function 
to $\approx 3$ times above the 
knee luminosity, corresponding to $\approx 80\%$ of 
the total $\lx$ (integrated from the \xray\ luminosity 
function). 

There are $\approx$~2,020 \xray\ sources matched to the 
optical/NIR COSMOS2015 catalog based on a likelihood-ratio 
technique by \citet[][see Tab.~\ref{tab:sample}]{marchesi16}.
Most ($\gtrsim 90\%$) of these \xray\ sources should be AGNs 
considering their relatively high \xray\ luminosity 
($\log\lx>42.5$; e.g. \hbox{\citealt{xue16}}; 
\hbox{\citealt{luo17}}).
Besides the $\approx 240$ BL AGNs, there are $\approx 730$
\xray\ sources with \hbox{spec-$z$} measurements 
(see \S\ref{sec:sample}).
We use these $\approx 730$ non-BL \xray\ sources to 
assess the \hbox{photo-$z$} quality for the $\approx 1,050$
\xray\ sources with only \hbox{photo-$z$} measurements, 
because \citet{yang18} estimated most ($\approx 80\%$) of the 
\hbox{photo-$z$} sources should be non-BL AGNs.
The \hbox{photo-$z$} have $\nmad=0.018$ and $\eta = 7.5\%$ 
(see \ref{sec:sample}), comparable 
to the AGN \hbox{photo-$z$} quality in the literature
\citep[e.g.][]{luo10, hsu14, yang14}.

A source might be detected in multiple \xray\ bands. 
In this case, we choose, in order of priority, hard-band 
(\hbox{2--7 keV}), full-band (\hbox{0.5--7 keV}), and soft-band 
(\hbox{0.5--2 keV}) fluxes, for the $\lx$ calculation below.
This order of detection bands is to minimize the effects 
of \xray\ obscuration.
The fractions of \xray\ sources with fluxes from the 
hard, full, and soft bands are 63\%, 34\%, and 3\%, 
respectively.

The \xray\ fluxes in the \hbox{COSMOS-Legacy} catalog are only 
corrected for Galactic absorption.
\citet{marchesi16} have estimated intrinsic absorption 
column densities ($\nh$) based on hardness ratios.
We do not apply these absorption corrections, because 
the majority ($\approx 70\%$) of sources have poorly constrained 
$\nh$ values (consistent with zero at a 90\% confidence
level) mainly due to limited numbers of counts. 
Instead, we evaluate the level of absorption corrections for 
COSMOS-like sources using the ultradeep 7~Ms catalog of the
\textit{Chandra} Deep Field-South (\cdfs; \citealt{luo17}).
Such sources have $\approx$~44 times more counts in \cdfs\ 
than COSMOS and absorption corrections have been estimated 
individually (e.g. \hbox{\citealt{yang16}}; 
\hbox{\citealt{liu17}}; \hbox{\citealt{luo17}}).
These COSMOS-like sources are selected via applying the 
COSMOS flux limits \citep{civano16} to \cdfs.
We find that these COSMOS-like sources have a median absorption 
correction factor (intrinsic flux divided by observed flux) 
of $\approx 1.2$.
The corresponding uncertainty caused by absorption correction 
is generally smaller than the statistical uncertainties of $\bharbar$
(\S\ref{sec:bharbar}), and thus absorption should not bias our 
conclusions.
The relatively low level of absorption corrections is mainly 
due to our choice of bands.
Another reason is that we can sample ultra-hard 
($\approx 10\text{--}20$~keV, rest-frame) penetrating \hbox{X-rays} for 
most sources (see Tab.~\ref{tab:sample}).

We convert the \xray\ fluxes to $\lx$ assuming a power-law model 
with a photon index of $\Gamma=1.7$, which is the typical intrinsic 
slope of distant AGNs (e.g. \hbox{\citealt{marchesi16b}}; 
\hbox{\citealt{yang16}}; \hbox{\citealt{liu17}}).
Our conclusions do not change if we adopt a slightly different
$\Gamma$ value (e.g. $\Gamma=1.4$).
The resulting $\lx$ values as a function of redshift are displayed
in Fig.~\ref{fig:all_vs_z} (also see Tab.~\ref{tab:sample} for
typical $\lx$ ranges). 
Fig.~\ref{fig:all_vs_z} also shows the estimated $\lx$ limit, 
assuming a \hbox{0.5--10~keV} flux threshold of 
$8.9\times 10^{-16}$~erg~cm$^{-2}$~s$^{-1}$ \citep{civano16}.

\subsubsection{X-ray Undetected Sources}\label{sec:xray_undet}
We perform \xray\ stacking to calculate $\lxbar$ for \xray\ 
undetected sources in our samples. 
We use the full-band \xray\ data. 
{The full band is the most sensitive in the sense that 
it detects the largest number of \xray\ sources \citep{civano16}, 
while the hard band is the least sensitive.
Also, compared to the soft band, the full band is less affected 
by \xray\ obscuration. 
Tab.~\ref{tab:sample} shows the rest-frame energy ranges 
corresponding to the full band.
Using the soft band or hard band for stacking does not change our 
conclusions qualitatively.}

Based on the full-band \xray\ image and exposure map, 
we broadly follow the procedures in \S2 of \citet{vito16}.
First, we mask the \xray\ image for both detected extended 
and point-like sources. 
Since the extended-source catalog for the \hbox{COSMOS-Legacy} 
survey is not available, we use the \xray\ cluster catalog 
from \xmm\ observations of COSMOS \citep{finoguenov07}.
For point-like sources, we use the catalog from 
\hbox{\citet{civano16}}.
Since the \xray\ clusters are masked, we cannot perform 
stacking analyses for some of the densest environments, and 
this could potentially bias our results for $\bharbar$ 
dependence on environment. 
However, most of the $\bharbar$ in our sample is contributed 
by the \xray\ detected sources (see \S\ref{sec:split_det}). 
{Indeed, our qualitative results do not change even if 
we only consider ${\bharbar}$ contributed from 
\xray\ detected sources.} 
Therefore, we argue that the masking of \xray\ clusters, and other 
technical details of the stacking, should not be critical to 
our analyses (see \S\ref{sec:res}). 

For each detected source (including extended and point-like 
sources),
we mask its surrounding area with a radius of $R_{\rm msk}$. 
We adopt $R_{\rm msk}=r_{500}$ for extended sources, 
where $r_{500}$, in the range of 
$\approx 0.5\arcmin\text{--}3\arcmin$, 
is the estimated cluster radius provided by 
\hbox{\citet{finoguenov07}}.
We adopt $R_{\rm msk}=20\arcsec$ for point-like sources, 
as \citet{vito16} show such a radius is large enough to include 
nearly all \xray\ flux even for the brightest sources 
(thousands of counts) at the largest off-axis angles.
We do not adopt a masking radius that depends on off-axis
angle, because one source is often observed by multiple 
pointings and has different off-axis angles in different 
pointings.
The masked regions (including those for extended and 
point-like sources) cover a total of $\approx$~20\% of the 
survey area. 
We fill each masked region with the background randomly sampled 
from the corresponding background region, defined as the 
annulus with inner and outer radii of $R_{\rm msk}$ and 
$2R_{\rm msk}$, respectively. 
This is performed with the ``dmfilth'' command in the \chandra\
data-analysis package {\sc ciao}.
In the analyses below, we treat the masked regions as 
background. 

We then derive net count rates for \xray\ undetected 
galaxies (\S\ref{sec:sample}) utilizing the masked \xray\ 
image and exposure map. 
We only calculate the net count rates for each source that 
is not within or close to the masked regions, i.e., its distance 
($d$) to every masked source should satisfy 
$d>R_{\rm msk}+R_{\rm phot}$, where $R_{\rm phot}$ is the 
radius used to perform \xray\ photometry.
About $30\%$ of sources are discarded in this step, and we 
account for \xray\ emission from these sources in 
\S\ref{sec:bharbar}.

For each object, we obtain its total \xray\ counts enclosed 
within a radius of $R_{\rm phot}=4\arcsec$ and the average 
exposure time for this area. 
The value $R_{\rm phot}=4\arcsec$ is chosen because the 
signal-to-noise ratio (S/N) of stacking becomes substantially
lower for larger $R_{\rm phot}$ values. 
We calculate the total count rate by dividing the total 
counts by the average exposure time in the $R_{\rm phot}$ 
circle.
The total count rate includes not only the \xray\ emission 
from the source but also the background.
Therefore, we need to subtract the background properly. 
We choose the background region as an annulus with inner and
outer radii of $10\arcsec$ and $20\arcsec$, 
respectively, and calculate the background count rate. 
We obtain the net count rates by subtracting the background count 
rates from the total count rates. 

Since we are performing aperture photometry with a limited 
aperture size ($R_{\rm phot}=4\arcsec$), there is 
a fraction of \xray\ emission falling outside of our photometric 
aperture. 
To estimate this effect, we re-calculate net counts but
with a very large photometric radius of $R_{\rm phot}=10\arcsec$.
We find that the final stacked count rates for 
$R_{\rm phot}=10\arcsec$ are systematically higher than 
those for $R_{\rm phot}=4\arcsec$ by a factor of $\approx 1.4$. 
Thus, $R_{\rm phot}=4\arcsec$ corresponds to a radius for 
$1/1.4\approx 70\%$ encircled-energy fraction (EEF) on average.
The 70\% EEF is reasonable considering that the typical 50\% EEF 
radius is $3\text{--}4 \arcsec$ for the detected sources 
(see Fig.~2 of \citealt{civano16}).
We correct the net count rate ($R_{\rm phot}=4\arcsec$) for 
each source by multiplying by 1.4 to obtain the final net count 
rate. 

Following \cite{yang17}, we obtain the average count rate for 
samples of sources and convert it to full-band \xray\ flux 
with a constant factor 
{(${9.5\times10^{-12}}$~erg~cm$^{{-2}}$~counts$^{{-1}}$).
The factor is calculated with {\sc pimms} assuming 
${\Gamma=1.7}$ (\S\ref{sec:xray_det}).}
We derive the average \xray\ luminosity 
($\overline{L_{\rm X,stack}}$) 
from the average flux and the average redshift of the stacked 
sample.
{Our conclusions do not change if we adopt ${\Gamma=1.4}$ 
for \xray\ undetected sources (resulting in a ${\approx 10\%}$ 
change of 
${\overline{L_{\rm X,stack}}}$ at ${z=1}$), as 
expected from the fact that total \xray\ emission is dominated by 
\xray\ detected sources in general.}

\subsubsection{Calculation of\/ $\bharbar$}\label{sec:bharbar}
We calculate $\bharbar$ for samples of sources following the 
recipe in \S2.3 of \citet{yang17}.
We first calculate the average AGN \xray\ luminosity for 
the sample as
\begin{equation}\label{eq:lxbar}
\begin{split}
\lxbar = \frac{ (\Sigma_{\rm det}\lx) + 
		N_{\rm non}\overline{L_{\rm X,stack}} -
		\Sigma_{\rm all}L_{\rm X,XRB}
}{N_{\rm det} + N_{\rm non}},
\end{split}
\end{equation} 
where $N_{\rm det}$ and $N_{\rm non}$ are numbers of 
\xray\ detected and undetected sources, respectively;
$L_{\rm X,XRB}$ is the luminosity from XRBs. 

In the numerator of Eq.~\ref{eq:lxbar}, the first 
term ($\Sigma_{\rm det}\lx$) is the total luminosity
of \xray\ detected sources (\S\ref{sec:xray_det}). 
The second term ($N_{\rm non}\overline{L_{\rm X,stack}}$)
accounts for the total luminosity of \xray\ undetected
sources. 
Stacked sources are only a subsample of the undetected
sources as some undetected sources (within or close to the
masked regions) are discarded in the 
stacking procedure (see \S\ref{sec:xray_undet}). 
The formula ($N_{\rm non}\overline{L_{\rm X,stack}}$)
assumes these discarded sources have the same average 
luminosity as the stacked sources. 

The third term ($\Sigma_{\rm all}L_{\rm X,XRB}$ 
in Eq.~\ref{eq:lxbar}) is to subtract the XRB component 
from the total luminosity. 
For star-forming galaxies (see \S\ref{sec:Mstar} for the
classification), we use 
$L_{\rm X,XRB}=\alpha \mstar + \beta \rm{SFR}$.
The coefficients ($\alpha$ and $\beta$) are functions 
of redshift from model 269 of \citet{fragos13};
model 269 is a theoretical XRB model which is preferred 
by observations of galaxies at $z\approx 0\text{--}2$ 
(\citealt{lehmer16}; typical uncertainties $\lesssim 0.3$~dex).
The $\mstar$ value is from our SED fitting 
(\S\ref{sec:Mstar}).
We approximate SFR by using the value from the star-forming 
main sequence in Eq.~6 of \citet{aird17b}.
For quiescent galaxies, we neglect the SFR term 
and estimate their XRB emission as $L_{\rm X,XRB}=\alpha \mstar$,
since this term dominates. 
The sample-averaged XRB emission is $\approx 0.6 \text{--} 1.3$~dex 
lower than the average AGN emission. 
It is even $\approx 0.3$~dex lower than the stacked \xray\
emission. 
Therefore, the details of subtracting the XRB component are
not critical to our analyses. 

Following \citet{yang17}, we convert $\lxbar$ to $\bharbar$ 
as 
\begin{equation}\label{equ:bhar}
\begin{split}
\bharbar 
= \frac{3.53 \lxbar}{10^{45}\ \rm{erg~s^{-1}}}
	    M_{\sun}\ \mathrm{yr}^{-1}.
\end{split}
\end{equation}   
This conversion assumes a constant bolometric correction
factor ($\kbol=22.4$; \citealt{vasudevan07}) and 
a constant radiation efficiency ($\epsilon=0.1$).
We calculate the uncertainties of $\bharbar$ with 
a bootstrapping technique (see \S2.3 of \citealt{yang17}).
The bootstrapping $\bharbar$\ errors are statistical 
uncertainties resulting from finite sampling.
 
{A radiation efficiency of ${\epsilon=0.1}$ 
is a typical value for the overall AGN population and is 
supported by observations (e.g. \hbox{\citealt{marconi04}};
\hbox{\citealt{davis11}}; \hbox{\citealt{brandt15}}).
Studies have found ${\kbol}$ depends on AGN 
luminosity (e.g. \hbox{\citealt{steffen06}}; 
\hbox{\citealt{hopkins07}}; \hbox{\citealt{lusso12}}). 
We do not adopt a ${\lx}$-dependent ${\kbol}$, 
because it cannot be applied to our stacking procedure.
Also, a ${\lx}$-dependent ${\kbol}$
requires careful subtraction of non-negligible XRB 
contributions for individual low-luminosity AGNs, and 
this task is challenging and beyond our work.
Therefore, we adopt the constant ${\kbol}$ 
for simplicity and consistency.
However, we have tested applying a ${\lx}$-dependent
${\kbol}$ \citep{hopkins07} for AGN-dominated 
\xray\ sources with ${\log\lx>43}$ and our 
results do not change qualitatively. 
This is as expected, because our main conclusions only 
depend on the relative values of ${\bharbar}$ in  
different environments which are not significantly affected
by different ${\kbol}$ schemes.
Admittedly, there might be systematics up to a factor of 
a few in our absolute values of ${\bharbar}$
due to the uncertainties of ${\kbol}$ and 
${\epsilon}$. 
We have also marked ${\lxbar}$ values in the 
relevant figures below, allowing readers to 
consider either ${\bharbar}$ or ${\lxbar}$
when viewing these.}

\section{Results}\label{sec:res}
\subsection{$\bharbar$ vs.\ Overdensity}
\label{sec:bhar_vs_det}

\subsubsection{Qualitative Tests}
\label{sec:split_det}
To probe the $\bharbar$ dependence on overdensity at 
different redshifts, we first bin sources into redshift 
ranges of $0.3\text{--}1.2$, $1.2\text{--}2.0$, and $2.0\text{--}3.0$,
respectively.
We restrict our analyses to sources above the $\mstar$
completeness limits, $\log\mstar=9.3$, 10, and 10.3, for 
redshift ranges of $0.3\text{--}1.2$, $1.2\text{--}2.0$, 
and $2.0\text{--}3.0$,
respectively (see Figs.~\ref{fig:all_vs_z} and \ref{fig:det_vs_M}). 
{Applying these ${\mstar}$ cuts is crucial, 
because incomplete ${\mstar}$ samples could lead to 
biased ${\bharbar}$ values.
For example, the ${\bharbar}$ at 
${\log\mstar\approx 8}$ in the bin of 
${z=0.3\text{--}1.2}$ would be strongly biased to 
${z\lesssim 0.5}$, above which
${\log\mstar\approx 8}$ galaxies remain
largely undetected (see Fig.~\ref{fig:all_vs_z} top). 
Therefore, such a ${\bharbar}$ value would not be representative 
of the entire redshift range of ${z=0.3\text{--}1.2}$.
}
Tab.~\ref{tab:pcor} lists the sizes of these refined samples
in different redshift ranges.
Hereafter, this rule applies to all the analyses, unless 
otherwise stated. 
The fractions of our \xray\ detected sources lying above the 
$\mstar$ cuts are 96\%, 90\%, and 77\% for redshift ranges of 
$0.3\text{--}1.2$, $1.2\text{--}2.0$, and $2.0\text{--}3.0$, 
respectively (see Fig.~\ref{fig:det_vs_M}).
Therefore, we are still capturing most accretion power after
applying the $\mstar$ cuts.
For each redshift bin, we further divide the sources 
into overdensity bins of $\log(1+\delta)<-0.3$, 
$\log(1+\delta)=-(0.3\text{--}0.1)$, $\log(1+\delta)=-0.1\text{--}0$, 
$\log(1+\delta)=0\text{--}0.1$, $\log(1+\delta)=0.1\text{--}0.3$,
and $\log(1+\delta)>0.3$.
These bin boundaries ($-0.3$, $-0.1$, 0, 0.1, and 0.3) roughly 
correspond to $\approx 10\%$, 30\%, 50\%, 70\%, and 90\% 
percentiles of the $\log(1+\delta)$ distribution of all objects. 

\begin{figure*}
\includegraphics[width=0.49\linewidth]{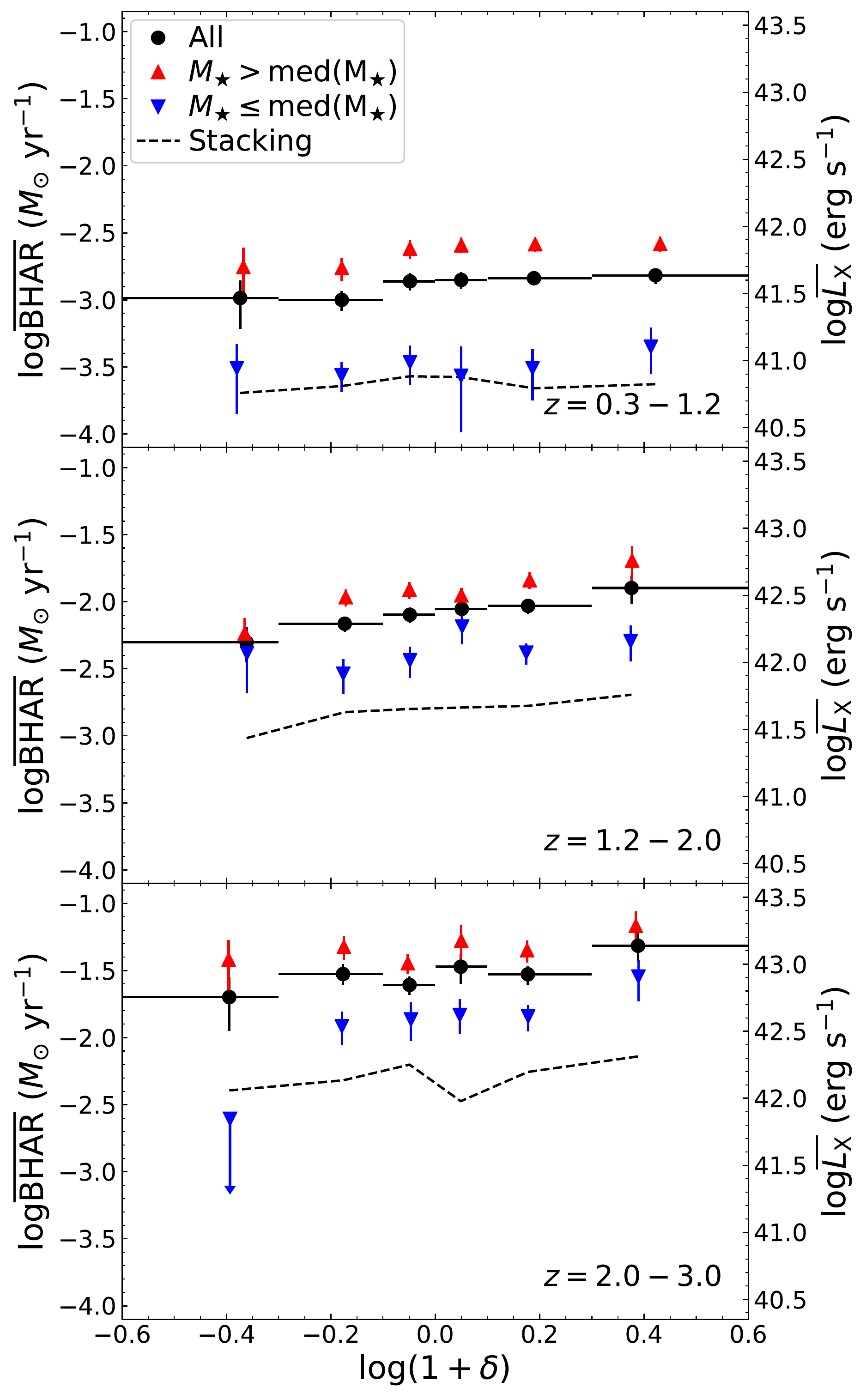}
\includegraphics[width=0.49\linewidth]{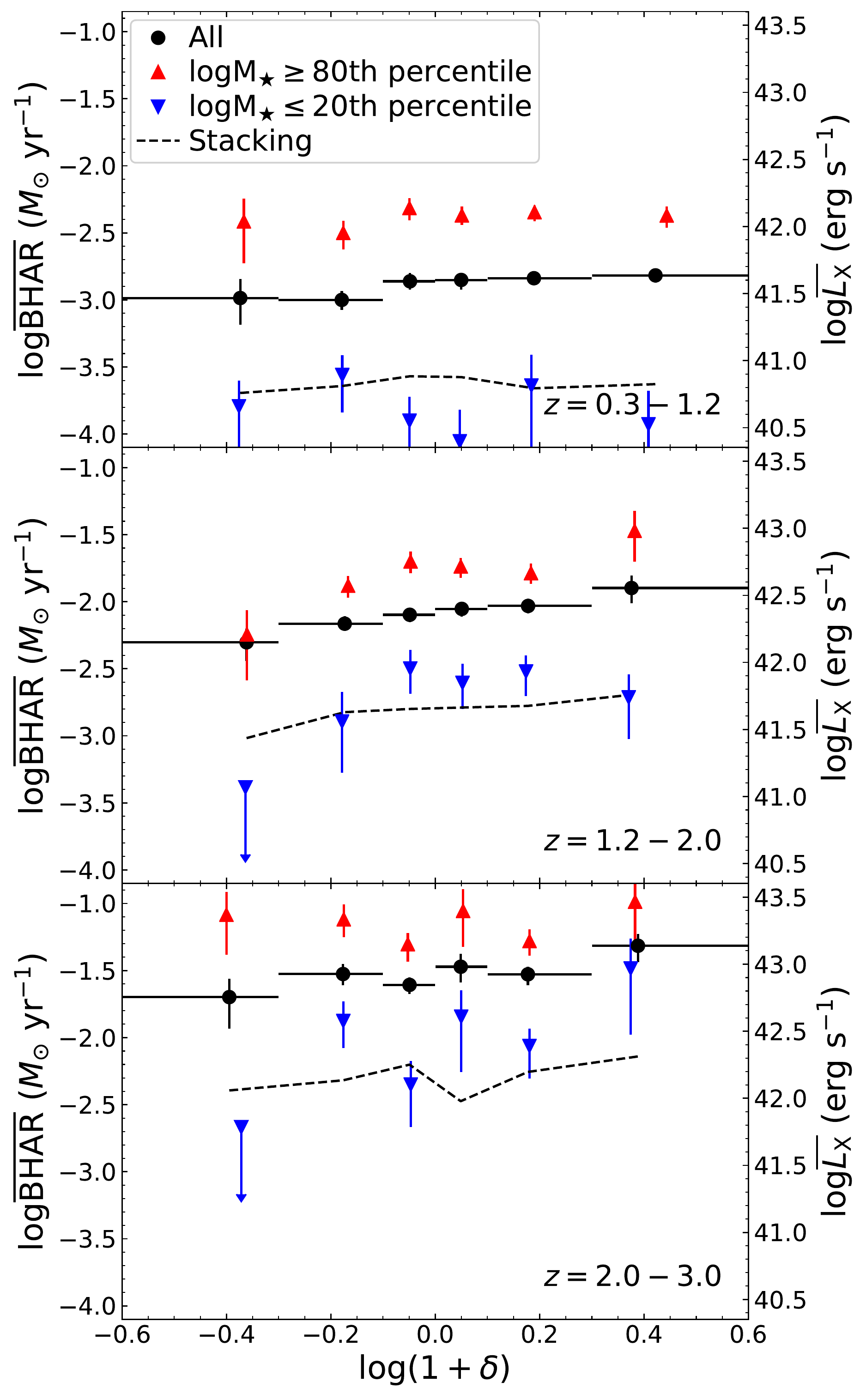}
\caption{$\bharbar$ as a function of overdensity for 
different redshift ranges. 
{In the left panels, each overdensity bin is split into 
two subsamples with ${\mstar}$ above and below the median 
value, respectively.
In the right panels, the subsamples include sources with 
the highest 20\% of ${\mstar}$ and the lowest 20\% of
${\mstar}$, respectively. 
${\lxbar}$ is marked on the right-hand 
side of each panel.}
The red upward and blue downward triangles indicate 
the subsamples with $\mstar$ above and below the
median value, respectively. 
The error bars are derived from a bootstrapping 
technique (see \S\ref{sec:bharbar}).
The dashed curves indicate $\bharbar$ contributed 
from \xray\ stacking (see \S\ref{sec:xray_undet}).
The high-$\mstar$ subsamples have significantly higher 
$\bharbar$ than the corresponding low-$\mstar$ subsamples.}
\label{fig:bhar_vs_delta}
\end{figure*}

We calculate $\bharbar$ with the methods in 
\S\ref{sec:bharbar} for all the bins and show the 
results in Fig.~\ref{fig:bhar_vs_delta}.
$\bharbar$ tends to be slightly higher toward high overdensity 
(black points), likely due to the positive dependence
between $\mstar$ and overdensity and the intrinsic 
$\bharbar$-$\mstar$ correlation 
(see Fig.~\ref{fig:det_vs_M}; e.g. \hbox{\citealt{xue10}};
\hbox{\citealt{georgakakis17}}; \hbox{\citealt{yang17, yang18}}; 
\hbox{\citealt{aird18}}). 
To show the $\bharbar$ dependence on $\mstar$, we divide
each overdensity sample into high-$\mstar$ and low-$\mstar$
subsamples. 
{In Fig.~\ref{fig:bhar_vs_delta}-left, the 
high-${\mstar}$ and low-${\mstar}$ 
subsamples have ${\mstar}$ above and below the 
median ${\mstar}$ of the overdensity sample, respectively;
in Fig.~\ref{fig:bhar_vs_delta}-right, the high-${\mstar}$ 
and low-${\mstar}$ subsamples include sources with the highest
20\% ${\mstar}$ and the lowest 20\% 
${\mstar}$ of the overdensity sample, respectively.}
{In both the left and right panels,}
the high-$\mstar$ subsamples have significantly higher
$\bharbar$ than their corresponding low-$\mstar$ subsamples,
indicating the previously known strong $\bharbar$-$\mstar$ correlation. 
{The ${\bharbar}$ differences between the 
high-${\mstar}$ and low-${\mstar}$ subsamples 
are generally larger in the right panels than in the corresponding 
left panels, as expected from the positive 
${\bharbar}$-${\mstar}$ correlation.}

To investigate the potential $\bharbar$-overdensity correlation 
for the $\mstar$ controlled sample, we divide the sources into 
bins of $\log\mstar=9.3\text{--}9.7$, \hbox{9.7--10}, \hbox{10--10.3}, 
\hbox{10.3--10.6}, \hbox{10.6--11}, and \hbox{11--11.5} for each 
redshift range. 
The bin widths are $\approx 0.4$~dex and the $\mstar$
limits at $z=1.2$, 2.0, and 3.0 (\S\ref{sec:Mstar}) are 
chosen as the boundaries of the $\mstar$ bins. 
We then split each $\mstar$ sample into high-overdensity
and low-overdensity subsamples in a similar way as 
in Fig.~\ref{fig:bhar_vs_delta}. 

\begin{figure*}
\includegraphics[width=0.49\linewidth]{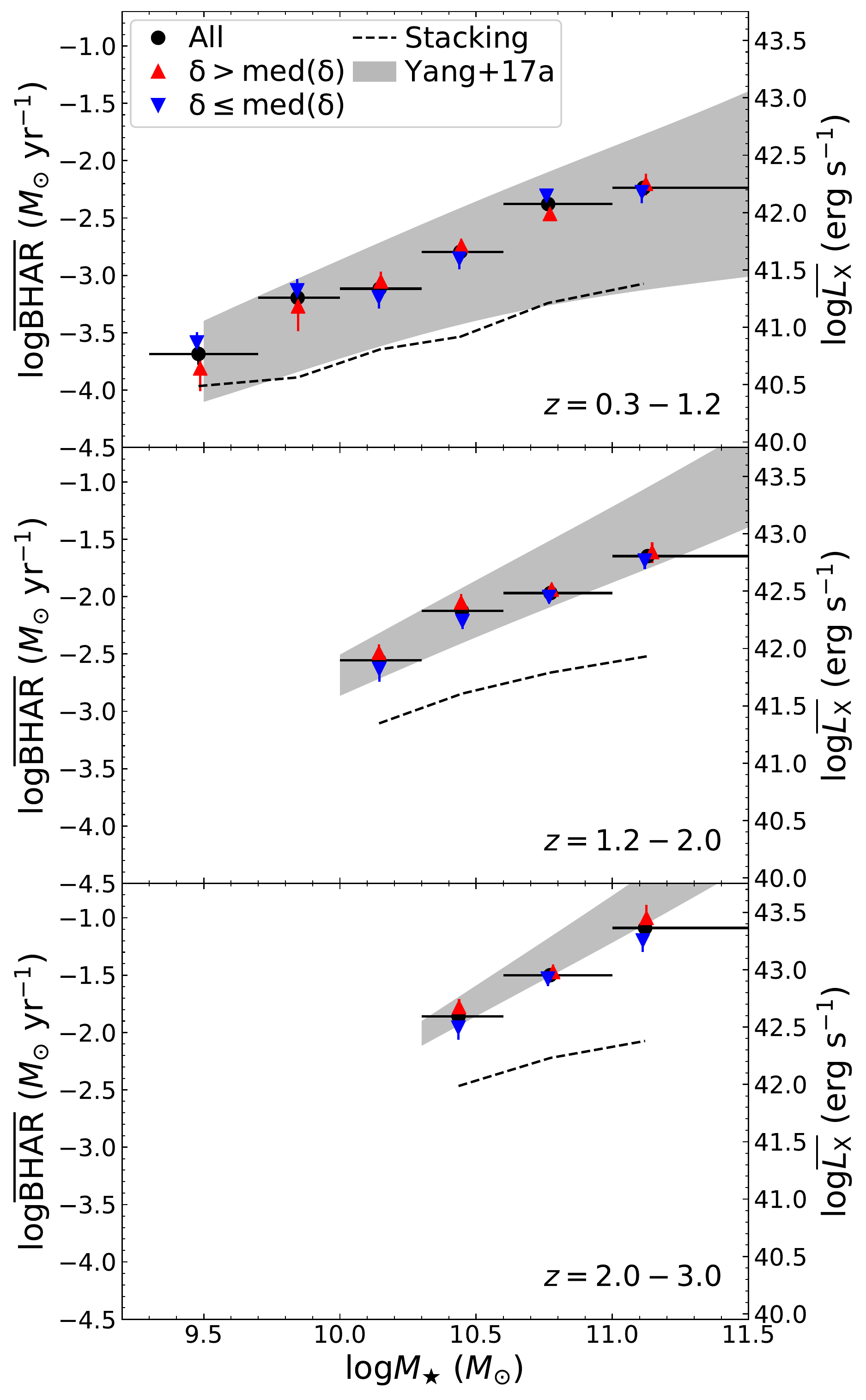}
\includegraphics[width=0.49\linewidth]{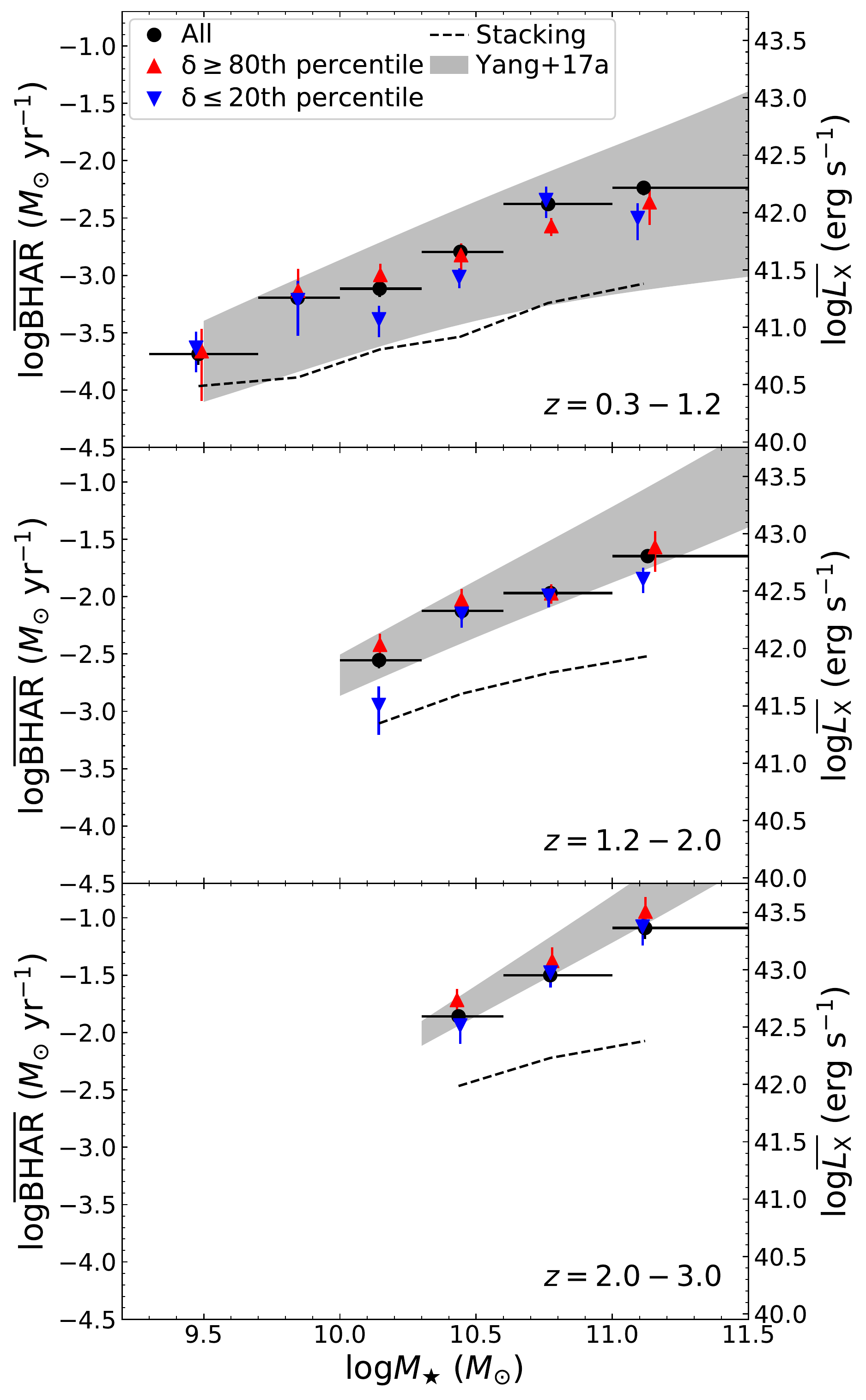}
\caption{Same format as Fig.~\ref{fig:bhar_vs_delta} but
for $\bharbar$ vs.\ $\mstar$. 
In both panels, the shaded regions show the $\bharbar$-$\mstar$
relations from \citet{yang18}.
The lower and upper boundaries of the \citet{yang18} relations 
correspond to the $\bharbar$ at the low and high limits of the 
redshift bin, respectively, except for the 
redshift bin of $z=0.3\text{--}1.2$.
For $z=0.3\text{--}1.2$, the lower 
boundary of the shaded region represents $\bharbar$ 
at $z=0.4$, which is the lowest redshift probed in 
\citet{yang18}.
The $\bharbar$ for high-overdensity and low-overdensity
subsamples are similar in general.
}
\label{fig:bhar_vs_M_od}
\end{figure*}

We calculate $\bharbar$ for all the $\mstar$ samples
and overdensity subsamples. 
The results are shown in Fig.~\ref{fig:bhar_vs_M_od} (left).
The high-overdensity and low-overdensity subsamples do not 
appear to have significantly different $\bharbar$, indicating 
that SMBH growth does not have a strong dependence on 
local environment at a given $\mstar$. 
{Any small apparent ${\bharbar}$ 
differences between the 
high-overdensity and low-overdensity subsamples are likely 
just due to statistical fluctuations, as some blue points
in Fig.~\ref{fig:bhar_vs_M_od} (left) are 
slightly above the corresponding red points while other
blue points are below.
The subsamples' median ${\bharbar}$ 
${1\sigma}$ uncertainty is 0.09~dex. 
Therefore, if the high-overdensity and low-overdensity 
subsamples had ${\bharbar}$ differing
by ${\gtrsim 0.09}$~dex systematically, 
the blue and red points would be significantly separated in 
Fig.~\ref{fig:bhar_vs_M_od} (left).}
In all redshift bins, $\bharbar$ rises toward the high 
$\mstar$ regime. 
\citet{yang18} have modeled the $\bharbar$-$\mstar$ 
relations at different redshifts in detail, and our 
data points are consistent with their results 
(see Fig.~\ref{fig:bhar_vs_M_od}).
The $\bharbar$ contributed from stacking is generally 
$\approx 0.5$~dex lower than the total $\bharbar$.
Therefore, most \xray\ emission is from the \xray\ 
detected sources. 

The two-subsample split (Fig.~\ref{fig:bhar_vs_M_od} left) 
guarantees that both subsamples 
have half the number of sources in each $\mstar$ bin, 
and thus the $\bharbar$ uncertainties are relatively small 
{(median uncertainty ${= 0.09}$~dex)} 
for the subsamples.
We also probe more extreme overdensity regimes by 
comparing $\bharbar$ of subsamples with the highest 20\% 
of overdensities and the lowest 20\% of overdensities
(see Fig.~\ref{fig:bhar_vs_M_od} right).
The high-overdensity and low-overdensity subsamples
also have similar $\bharbar$ in general, although 
the $\bharbar$ uncertainties become larger
{(median uncertainty ${=}$~0.13~dex)} 
compared to those in Fig.~\ref{fig:bhar_vs_M_od}~(left) due 
to reduced subsample sizes.
{In Fig.~\ref{fig:bhar_vs_M_od}, there are 2 
pairs of high-overdensity and low-overdensity points 
that are separated above a ${2\sigma}$ confidence level.
These deviations are likely due to statistical fluctuations.
There are a total of 26 pairs of points in 
Fig.~\ref{fig:bhar_vs_M_od} (i.e., 26 trials), and we expect to find 
${\lesssim 4}$ such deviations (99\% confidence 
level, calculated with a binomial distribution).
Also, our detailed quantitative analyses in \S\ref{sec:pca} 
do not find statistically significant 
${\bharbar}$-overdensity relation
when ${\mstar}$ is controlled.}


\subsubsection{Partial Correlation Analyses}
\label{sec:pca}
In \S\ref{sec:split_det}, we qualitatively showed that
the high-overdensity subsamples have similar $\bharbar$
as the corresponding low-overdensity subsamples when 
controlling for $\mstar$. 
This result indicates that SMBH growth, at a given 
$\mstar$, does not significantly depend on local environment.
We further quantitatively verify this point via 
partial-correlation (PCOR) analysis 
\citep[e.g.][]{johnson02}.

\begin{figure*}
\includegraphics[width=\linewidth]{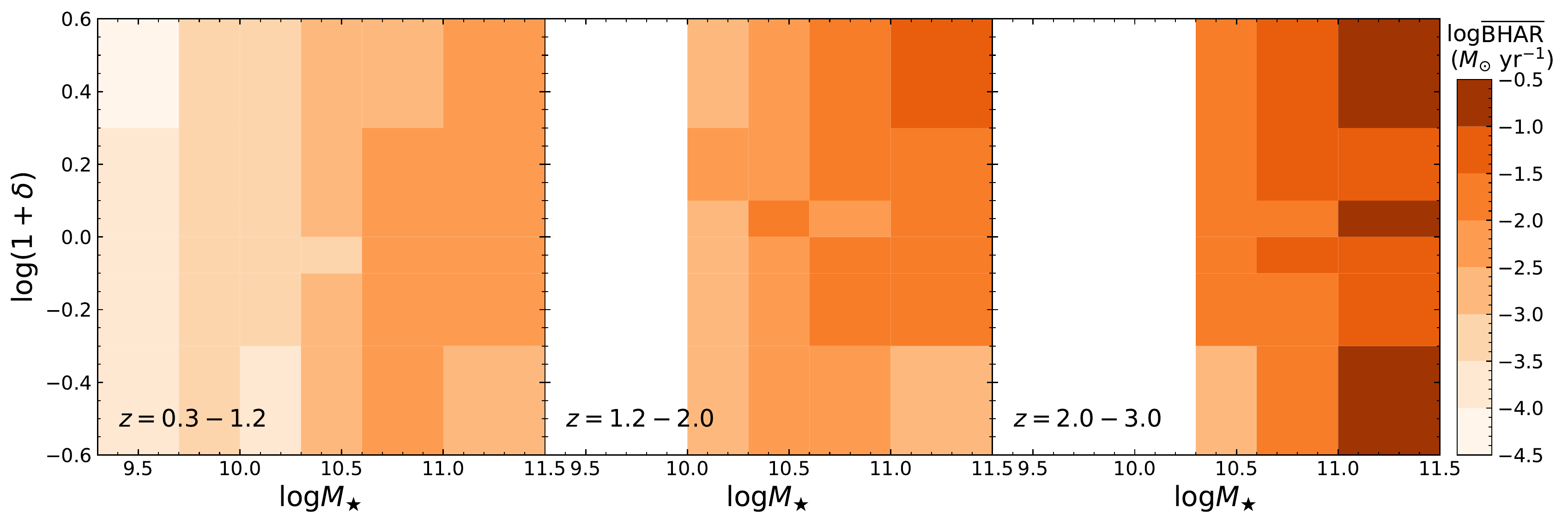}
\caption{$\bharbar$ as a function of overdensity and $\mstar$
for different redshift ranges. 
Darker color indicates higher $\bharbar$ as labeled. 
White color indicates $\bharbar$ is not available, because 
of large uncertainties on $\bharbar$ or $\mstar$ lying below 
the completeness limits.
The $\bharbar$ in each bin has an uncertainty of 
$\lesssim 0.3$~dex.
}
\label{fig:bhar_on_det_M_grids}
\end{figure*}

Following \cite{yang17}, we utilize {\sc pcor.r} in the {\sc r}
statistical package to perform PCOR analyses 
{\hbox{\citep{kim15}}}.
We bin sources in the overdensity-$\mstar$ plane and 
derive $\bharbar$ for each bin. 
The bin boundaries of overdensity and $\mstar$ are the same as 
in \S\ref{sec:split_det}. 
Fig.~\ref{fig:bhar_on_det_M_grids} shows the resulting 
$\bharbar$ on overdensity-$\mstar$ grids.
Similar to \cite{yang17}, we input the $\log\bharbar$, 
$\log(1+\delta)$, and $\log\mstar$ values into {\sc pcor.r}, where the 
$\log(1+\delta)$ and $\log\mstar$ are the medians in each bin. 
We study the $\bharbar$-overdensity ($\bharbar$-$\mstar$) correlation 
while controlling for the effects of $\mstar$ (overdensity) via all 
three statistics available in {\sc pcor.r} (Pearson, Spearman, 
and Kendall). 
{The Pearson statistic assumes log-linear relations, while
the Spearman and Kendall non-parametric statistics are rank-based 
and do not have such assumptions.}

\begin{figure}
\includegraphics[width=\linewidth]{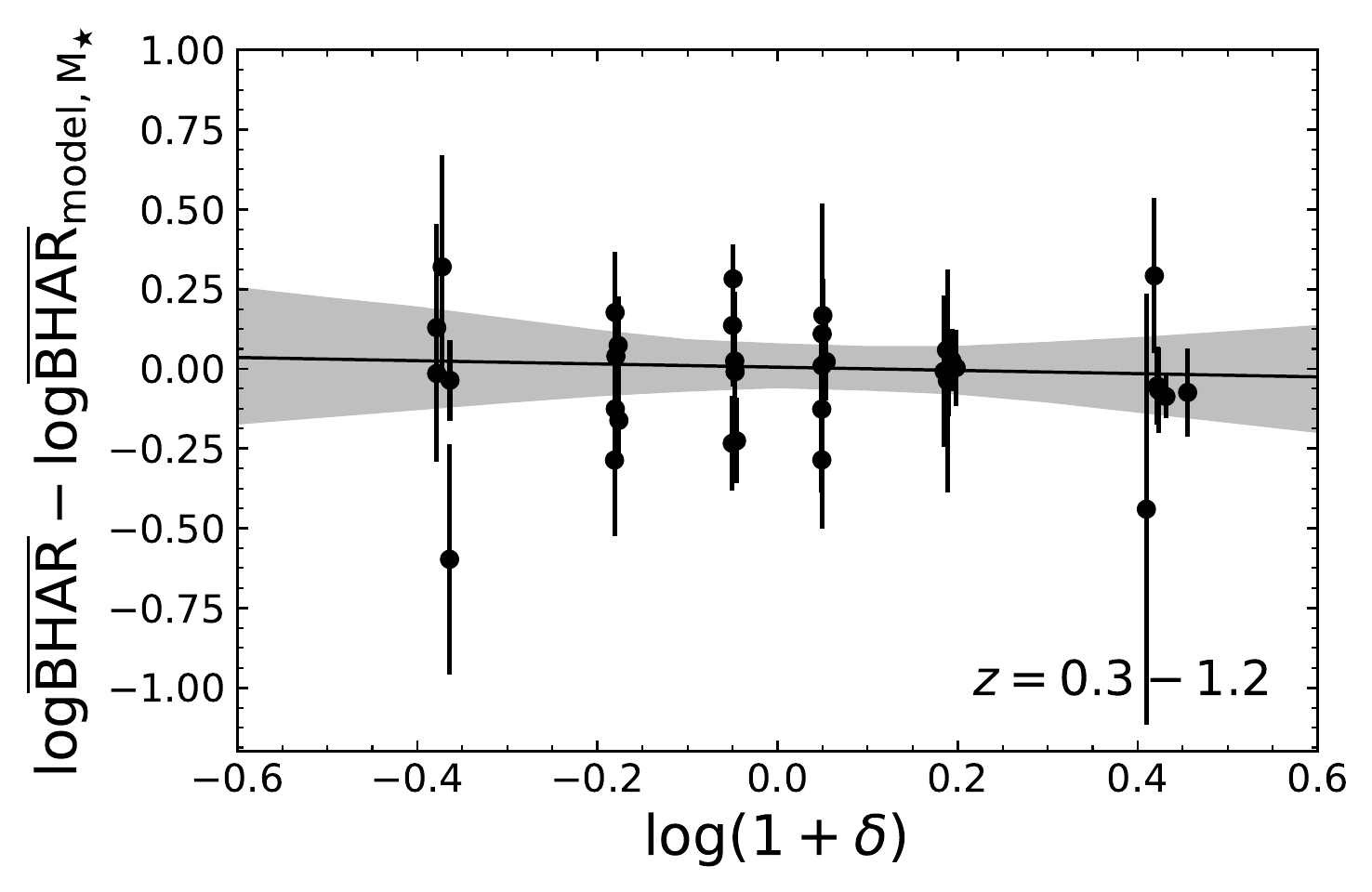}\\
\includegraphics[width=\linewidth]{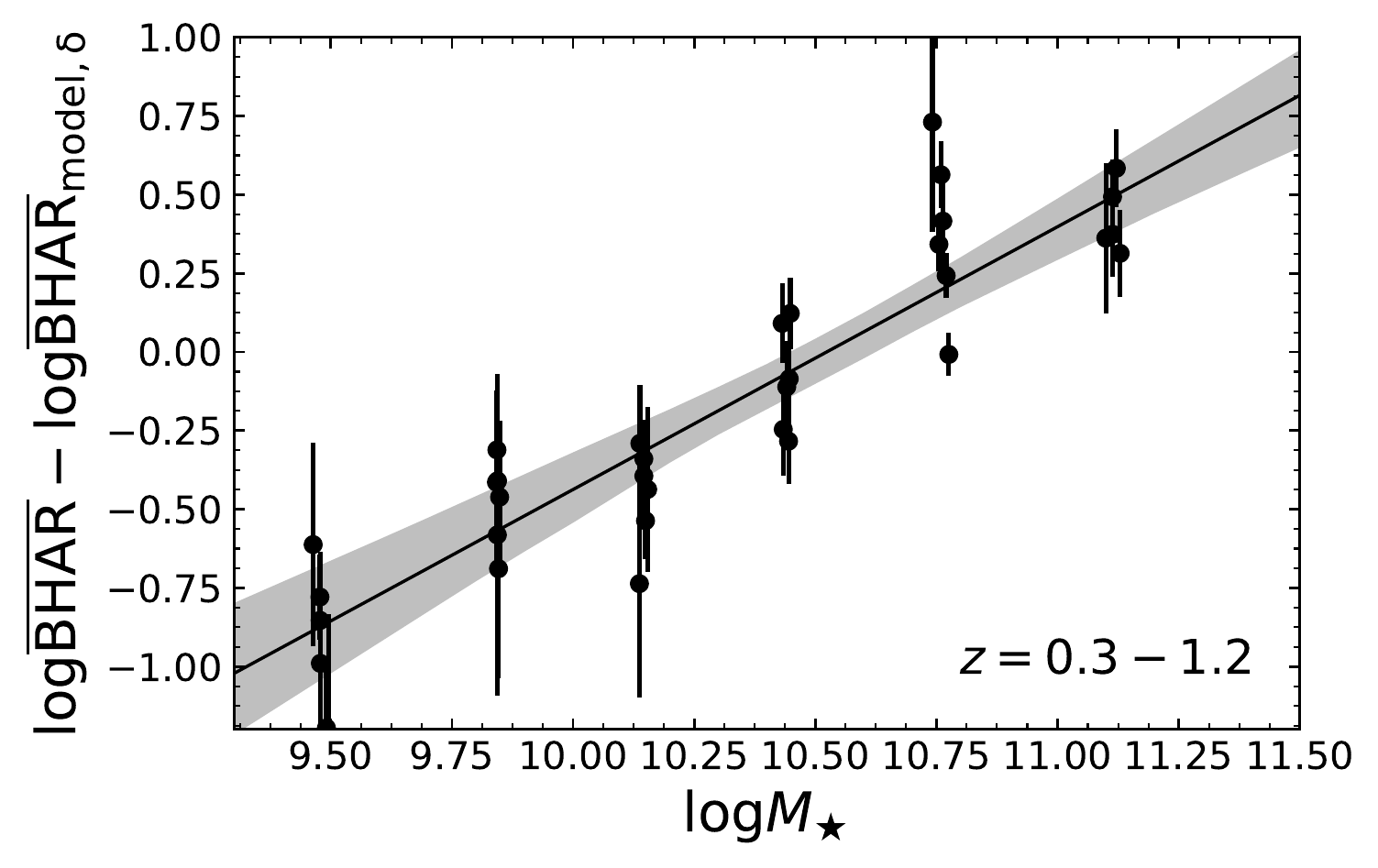}
\caption{$\bharbar$ residuals (of the $\bharbar$-$\mstar$ 
fit) vs.\ overdensity (top) and $\bharbar$ residuals 
(of the $\bharbar$-overdensity fit) vs.\ $\mstar$ (bottom)
for $z=0.3\text{--}1.2$.
The black lines represent the best fit of the data points;
the shaded regions indicate the 3$\sigma$ uncertainties. 
In the top panel, the residual $\bharbar$ is relatively 
flat as a function of overdensity; 
in the bottom panel, the residual $\bharbar$ rises steeply
toward high $\mstar$.
Results are found similar at $z=1.2\text{--}2.0$ and 
$z=2.0\text{--}3.0$.
}
\label{fig:residual_bhar}
\end{figure}

The results are listed in Tab.~\ref{tab:pcor}.
The $\bharbar$-$\mstar$ correlation is statistically significant 
for all statistical techniques and across all redshift ranges. 
In contrast, the $\bharbar$-overdensity correlation is not 
significant ($<3\sigma$) under any statistic  
in any redshift range. 
To visualize the PCOR results, we perform a least-$\chi^2$ 
log-linear fit of the $\bharbar$-$\mstar$ ($\bharbar$-overdensity) 
relation. 
We then fit the residual $\bharbar$ as a function of 
overdensity ($\mstar$).
{This procedure is similar to the Pearson statistic in 
PCOR analyses.}
The uncertainties of the best fit are estimated based on Markov 
chain Monte Carlo (MCMC) sampling with {\sc emcee} 
\citep{foreman13}. 
Fig.~\ref{fig:residual_bhar} displays the fitting results
for $z=0.3\text{--}1.2$.
The best fit of the residual $\bharbar$-overdensity relation is 
consistent with a flat model at a $3\sigma$ confidence 
level, while the residual $\bharbar$-$\mstar$ relation is 
steep. 
The conclusion also holds for the other two redshift ranges.
{In Fig.~\ref{fig:residual_bhar}-bottom, the data points
at ${\log\mstar\approx 10.75}$ tend to be above the fit. 
This perhaps indicates a log-linear model (assumed in the Pearson 
statistic) is not enough to describe fully the 
${\bharbar}$-${\mstar}$ relation, 
consistent with previous works (e.g. \hbox{\citealt{georgakakis17}};
\hbox{\citealt{aird18}}; \hbox{\citealt{yang18}}).
The Spearman and Kendall statistics do not assume a log-linear 
model, and they lead to qualitatively similar results as the
Pearson statistic (see Tab.~\ref{tab:pcor}). 
Also, we note that since there are a total of such 12 trials in 
Fig.~\ref{fig:residual_bhar} (considering both the top and bottom
panels), it is not surprising that 
one such deviation happens by chance (see \S\ref{sec:split_det}).
}

Therefore, we conclude that the $\bharbar$ significantly depends 
on $\mstar$ but not overdensity. 
The conclusion is qualitatively supported by 
Figs.~\ref{fig:all_vs_z}, \ref{fig:det_vs_M}, and 
\ref{fig:bhar_on_det_M_grids}. 
In Figs.~\ref{fig:all_vs_z} and \ref{fig:det_vs_M}, the \xray\
detected sources preferentially appear in the high-$\mstar$
regime but do not show much dependence on overdensity. 
In Fig.~\ref{fig:det_vs_M}, \xray\ sources have median overdensity 
values similar to those of all galaxies. 
In Fig.~\ref{fig:bhar_on_det_M_grids}, the $\bharbar$ gradient is 
strong in the $x$ ($\mstar$) direction but weak in the $y$ 
(overdensity) direction.

\begin{figure}
\includegraphics[width=\linewidth]{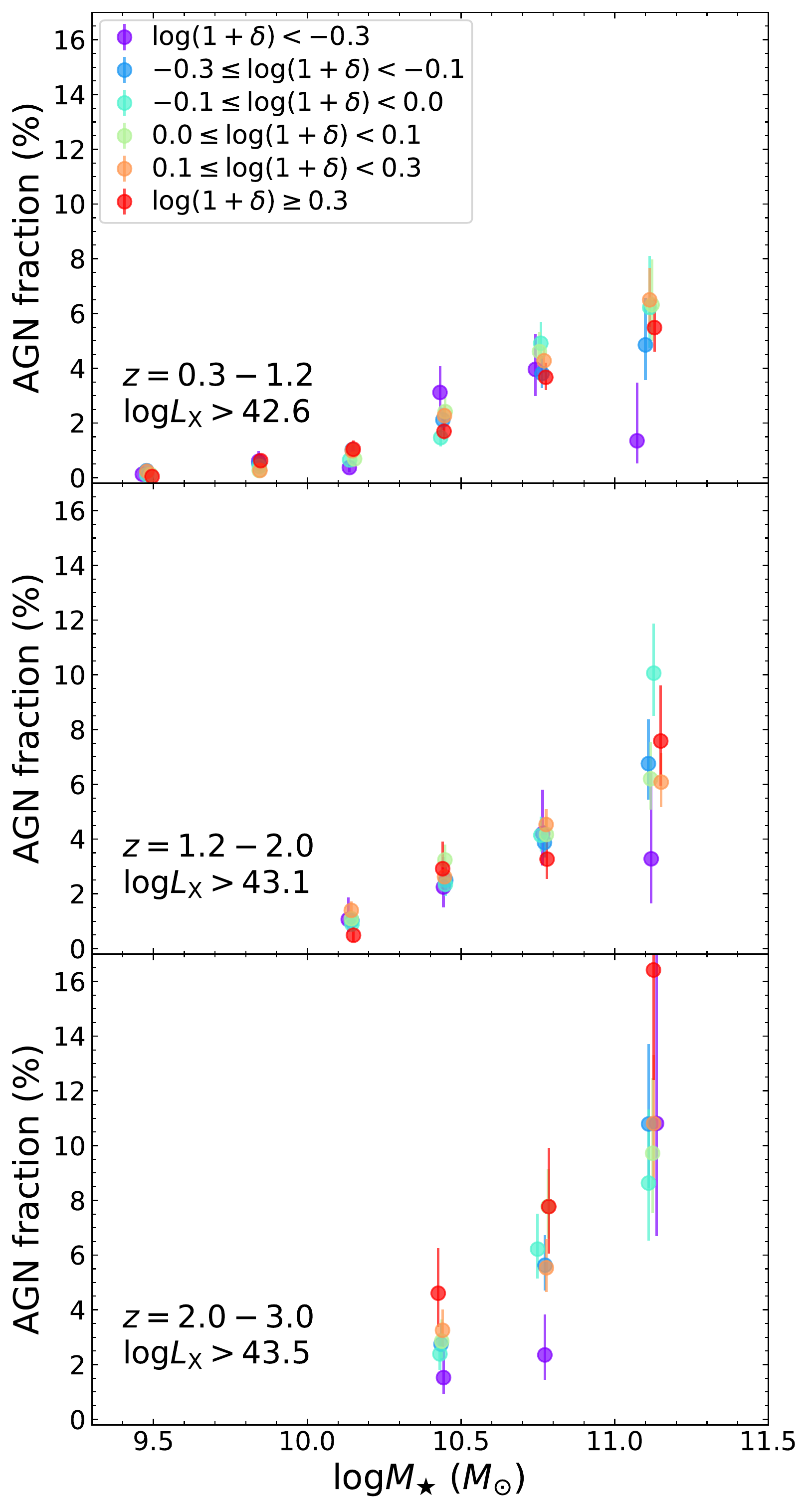}
\caption{AGN fraction as a function $\mstar$ for overdensity bins. 
The data are from Tab.~\ref{tab:agn_frac_det}.
At a given $\mstar$, different overdensity bins have similar 
AGN fractions. 
Some of the data points are overlapping due to this similarity.
}
\label{fig:AGNfrac_on_det_M_grids}
\end{figure}

Our analyses above are based on the $\bharbar$ technique to 
assess SMBH growth.  
Another common technique in the literature is to consider AGN fractions 
above a given $\lx$ threshold \citep[e.g.][]{silverman09b, xue10}. 
{Compared to our ${\bharbar}$ approach, the
AGN-fraction approach is less informative and less physical, 
because it needs a 
pre-defined ${\lx}$ threshold which depends on \xray\ 
survey sensitivity, and it does not consider \xray\ emission from
undetected sources.}
Also, unlike the $\bharbar$ approach, the AGN-fraction method 
weights low-$\lx$ and high-$\lx$ AGNs equally, as long as they 
are above the $\lx$ threshold, {thereby sacrificing information}. 
However, the AGN-fraction method still serves as a common alternative 
way to assess SMBH accretion (e.g. \hbox{\citealt{lehmer13}}; 
\hbox{\citealt{martini13}}). 
For a consistency check, we also present the AGN fractions in 
different $\mstar$ bins for different local environments in 
Tab.~\ref{tab:agn_frac_det} and Fig.~\ref{fig:AGNfrac_on_det_M_grids}.
The AGN fractions are calculated as the fractions of \xray\ detected 
sources above $\log\lx=42.6$ ($z=0.3\text{--}1.2$), $\log\lx=43.1$ 
($z=1.2\text{--}2.0$), and $\log\lx=43.5$ ($z=2.0\text{--}3.0$), 
respectively. 
For each redshift range, the threshold is chosen as the $\log\lx$ 
completeness limit at the redshift upper boundary (see 
Fig.~\ref{fig:all_vs_z}).
Due to the differences in $\lx$ thresholds, the AGN fractions 
(Tab.~\ref{tab:agn_frac_det}) at different redshift ranges are 
not directly comparable. 
The AGN-fraction errors are derived as $1\sigma$ binomial 
uncertainties using the {\sc astropy} module 
``{\sc binom\_conf\_interval}'' \citep[e.g.][]{cameron11}.
From Tab.~\ref{tab:agn_frac} and Fig.~\ref{fig:AGNfrac_on_det_M_grids}, 
at given $\mstar$ and redshift, the 
AGN fractions are generally similar in different overdensity bins, 
consistent with our PCOR analyses.
This consistency indicates that our conclusions are not affected 
by the \xray\ stacking procedures (especially the masking of \xray\ 
extended sources; see \S\ref{sec:xray_undet}), since the AGN fractions 
are based on \xray\ detected sources only.
Also, the AGN fraction rises toward massive galaxies, as expected 
from the strong $\bharbar$-$\mstar$ relation 
(Fig.~\ref{fig:bhar_vs_M_od}).
{In Fig.~\ref{fig:AGNfrac_on_det_M_grids} (both top and middle panels),
the lowest overdensity bin appears to have lower AGN fraction than
other overdensity bins at ${\log\mstar \approx 11.1}$. 
However, the differences are not significant at a 2${\sigma}$ 
confidence level. 
Also, considering there are many (78) points in 
Fig.~\ref{fig:AGNfrac_on_det_M_grids}, it is natural that a few
deviations happen considering the ``number of trials'' effect
detailed in \S\ref{sec:split_det}.
Therefore, we consider the apparent differences are likely due to 
statistical fluctuations.}

\begin{table}
\begin{center}
\caption{$p$-values (Significances) of 
	Partial Correlation Analyses (see \ref{sec:pca}).}
\label{tab:pcor}
\begin{tabular}{cccc}\hline\hline
\multicolumn{4}{c}{$z=0.3\text{--}1.2$ (48,581 galaxies)} \\ \hline
Relation & Pearson & Spearman & Kendall \\
 $\bharbar$-overdensity & 0.9 (0.1$\sigma$) & 0.2 (1.2$\sigma$) & 0.6 (0.6$\sigma$) \\
 $\bharbar$-$\mstar$ & 10$^{-48}$ (14.7$\sigma$) & 10$^{-45}$ (14.1$\sigma$) & 10$^{-10}$ (6.3$\sigma$) \\
\hline\hline
\multicolumn{4}{c}{$z=1.2\text{--}2.0$ (18,944 galaxies)} \\ \hline
Relation & Pearson & Spearman & Kendall \\
 $\bharbar$-overdensity & 0.08 (1.7$\sigma$) & 0.4 (0.8$\sigma$) & 0.4 (0.8$\sigma$) \\
 $\bharbar$-$\mstar$ & 10$^{-33}$ (12.0$\sigma$) & 10$^{-21}$ (9.5$\sigma$) & 10$^{-7}$ (5.0$\sigma$) \\
\hline\hline
\multicolumn{4}{c}{$z=2.0\text{--}3.0$ (5,932 galaxies)} \\ \hline
Relation & Pearson & Spearman & Kendall \\
 $\bharbar$-overdensity & 0.08 (1.8$\sigma$) & 0.1 (1.6$\sigma$) & 0.4 (0.9$\sigma$) \\
 $\bharbar$-$\mstar$ & 10$^{-19}$ (8.9$\sigma$) & 10$^{-19}$ (9.0$\sigma$) & 10$^{-5}$ (4.3$\sigma$) \\
\hline
\end{tabular}
\end{center}
\begin{flushleft}
{\sc Note.} ---
Here, the numbers of galaxies are different from Tab.~\ref{tab:sample}, 
because only sources above the limiting $\mstar$ are used in the analyses
of the $\bharbar$-$\mstar$-environment relation (see \S\ref{sec:split_det}).
\end{flushleft}
\end{table}

\begin{table*}
\begin{center}
\caption{AGN fractions (\%) in different overdensity bins}
\label{tab:agn_frac_det}
\begin{tabular}{c|ccccccc}\hline\hline
\multicolumn{7}{c}{$z=0.3\text{--}1.2$ ($\log\lx>42.6$)} \\ \hline
$\log\mstar$ & 9.3--9.7 & 9.7--10.0 & 10.0--10.3 & 
	     10.3--10.6 & 10.6--11.0& 11.0--11.5& \\ \hline
$\log(1+\delta)<-0.3$ & $0.1^{+0.13}_{-0.07}$ & $0.6^{+0.38}_{-0.24}$ & $0.4^{+0.37}_{-0.19}$ & $3.1^{+0.97}_{-0.75}$ & $4.0^{+1.28}_{-0.98}$ & $1.4^{+2.13}_{-0.83}$ \\
$-0.3\leq\log(1+\delta)<-0.1$ & $0.3^{+0.10}_{-0.07}$ & $0.5^{+0.20}_{-0.14}$ & $1.0^{+0.29}_{-0.22}$ & $2.1^{+0.44}_{-0.37}$ & $3.8^{+0.63}_{-0.54}$ & $4.9^{+1.73}_{-1.29}$ \\
$-0.1\leq\log(1+\delta)<0.0$ & $0.1^{+0.09}_{-0.05}$ & $0.4^{+0.21}_{-0.14}$ & $0.7^{+0.27}_{-0.19}$ & $1.5^{+0.43}_{-0.33}$ & $4.9^{+0.77}_{-0.67}$ & $6.2^{+1.89}_{-1.47}$ \\
$0.0\leq\log(1+\delta)<0.1$ & $0.1^{+0.09}_{-0.05}$ & $0.3^{+0.17}_{-0.10}$ & $0.7^{+0.29}_{-0.20}$ & $2.4^{+0.53}_{-0.43}$ & $4.6^{+0.71}_{-0.62}$ & $6.3^{+1.65}_{-1.33}$ \\
$0.1\leq\log(1+\delta)<0.3$ & $0.2^{+0.09}_{-0.07}$ & $0.3^{+0.13}_{-0.09}$ & $1.0^{+0.25}_{-0.20}$ & $2.3^{+0.38}_{-0.33}$ & $4.3^{+0.50}_{-0.45}$ & $6.5^{+1.16}_{-1.00}$ \\
$\log(1+\delta)\geq0.3$ & $0.0^{+0.08}_{-0.03}$ & $0.6^{+0.25}_{-0.18}$ & $1.0^{+0.32}_{-0.24}$ & $1.7^{+0.39}_{-0.32}$ & $3.7^{+0.54}_{-0.47}$ & $5.5^{+1.04}_{-0.88}$ \\
All & $0.2^{+0.04}_{-0.03}$ & $0.4^{+0.07}_{-0.06}$ & $0.9^{+0.11}_{-0.10}$ & $2.1^{+0.18}_{-0.17}$ & $4.2^{+0.26}_{-0.24}$ & $5.7^{+0.57}_{-0.52}$ \\
\hline\hline
\multicolumn{7}{c}{$z=1.2\text{--}2.0$ ($\log\lx>43.1$)} \\ \hline
$\log\mstar$ & -- & -- & 10.0--10.3 & 
	     10.3--10.6 & 10.6--11.0& 11.0--11.5& \\ \hline
$\log(1+\delta)<-0.3$ & -- & -- & $1.1^{+0.81}_{-0.46}$ & $2.2^{+1.10}_{-0.74}$ & $4.2^{+1.60}_{-1.17}$ & $3.3^{+3.14}_{-1.63}$ \\
$-0.3\leq\log(1+\delta)<-0.1$ & -- & -- & $1.0^{+0.31}_{-0.24}$ & $2.5^{+0.48}_{-0.41}$ & $3.9^{+0.63}_{-0.54}$ & $6.8^{+1.61}_{-1.32}$ \\
$-0.1\leq\log(1+\delta)<0.0$ & -- & -- & $0.9^{+0.31}_{-0.23}$ & $2.3^{+0.52}_{-0.43}$ & $4.1^{+0.68}_{-0.59}$ & $10.1^{+1.81}_{-1.56}$ \\
$0.0\leq\log(1+\delta)<0.1$ & -- & -- & $1.1^{+0.33}_{-0.25}$ & $3.2^{+0.57}_{-0.48}$ & $4.2^{+0.65}_{-0.57}$ & $6.2^{+1.34}_{-1.12}$ \\
$0.1\leq\log(1+\delta)<0.3$ & -- & -- & $1.4^{+0.33}_{-0.27}$ & $2.6^{+0.45}_{-0.39}$ & $4.5^{+0.56}_{-0.50}$ & $6.1^{+1.06}_{-0.91}$ \\
$\log(1+\delta)\geq0.3$ & -- & -- & $0.5^{+0.48}_{-0.24}$ & $2.9^{+1.00}_{-0.75}$ & $3.3^{+0.94}_{-0.73}$ & $7.6^{+2.03}_{-1.63}$ \\
All & -- & -- & $1.1^{+0.14}_{-0.12}$ & $2.7^{+0.23}_{-0.21}$ & $4.1^{+0.28}_{-0.26}$ & $7.0^{+0.61}_{-0.57}$ \\
\hline\hline
\multicolumn{7}{c}{$z=2.0\text{--}3.0$ ($\log\lx>43.5$)} \\ \hline
$\log\mstar$ & -- & -- & -- & 
	     10.3--10.6 & 10.6--11.0& 11.0--11.5& \\ \hline 
$\log(1+\delta)<-0.3$ & -- & -- & -- & $1.5^{+0.96}_{-0.59}$ & $2.4^{+1.47}_{-0.91}$ & $10.8^{+6.17}_{-4.11}$ \\
$-0.3\leq\log(1+\delta)<-0.1$ & -- & -- & -- & $2.7^{+0.74}_{-0.58}$ & $5.6^{+1.10}_{-0.93}$ & $10.8^{+2.92}_{-2.36}$ \\
$-0.1\leq\log(1+\delta)<0.0$ & -- & -- & -- & $2.4^{+0.75}_{-0.57}$ & $6.2^{+1.29}_{-1.08}$ & $8.6^{+2.69}_{-2.10}$ \\
$0.0\leq\log(1+\delta)<0.1$ & -- & -- & -- & $2.9^{+0.79}_{-0.62}$ & $7.8^{+1.36}_{-1.17}$ & $9.7^{+2.75}_{-2.20}$ \\
$0.1\leq\log(1+\delta)<0.3$ & -- & -- & -- & $3.3^{+0.75}_{-0.62}$ & $5.5^{+1.05}_{-0.89}$ & $10.8^{+2.50}_{-2.08}$ \\
$\log(1+\delta)\geq0.3$ & -- & -- & -- & $4.6^{+1.64}_{-1.23}$ & $7.8^{+2.15}_{-1.72}$ & $16.4^{+5.01}_{-4.03}$ \\
All & -- & -- & -- & $2.8^{+0.33}_{-0.29}$ & $6.1^{+0.52}_{-0.48}$ & $10.7^{+1.21}_{-1.10}$ \\
\hline
\end{tabular}
\end{center}
\end{table*}

\subsection{$\bharbar$ vs.\ Cosmic-Web Environment}
\label{sec:bhar_vs_web}
\begin{figure*}
\includegraphics[width=0.49\linewidth]{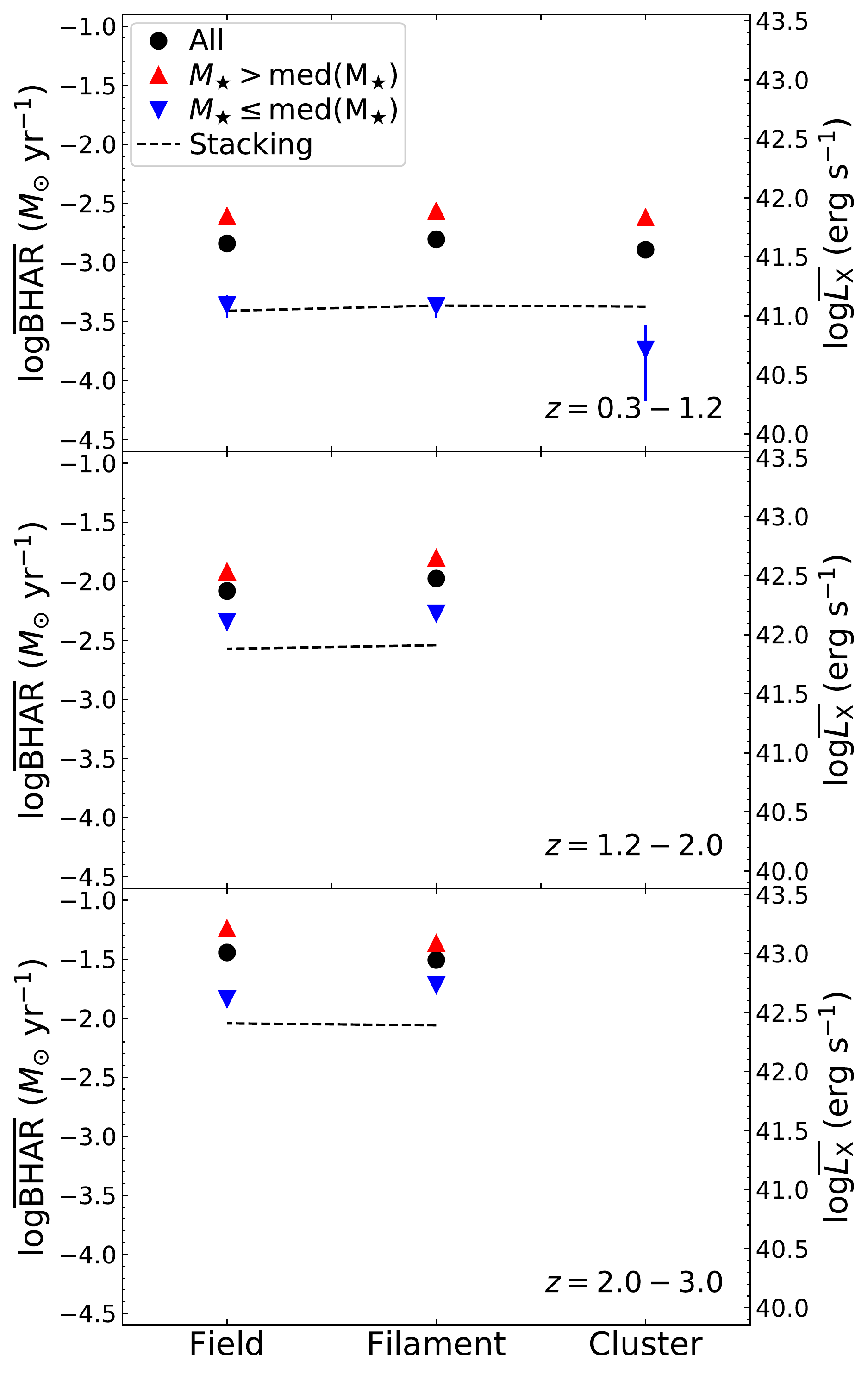}
\includegraphics[width=0.49\linewidth]{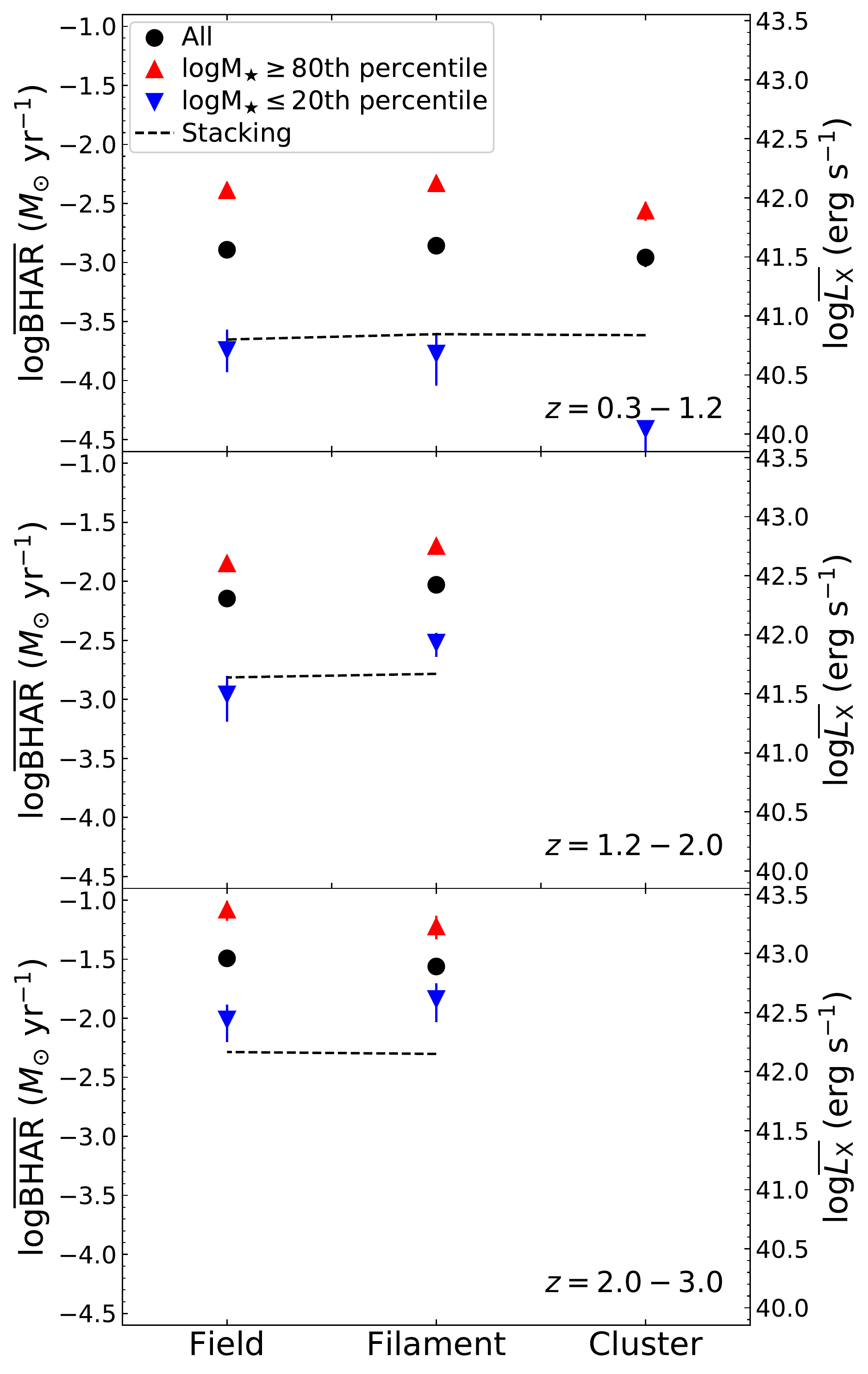}
\caption{Same format as Fig.~\ref{fig:bhar_vs_delta} but
for $\bharbar$ vs.\ cosmic-web environment. 
We do not assign cluster environment at redshifts above $z=1.2$ 
due to its generally weak signals (see \S\ref{sec:cosmic_web}). 
The high-$\mstar$ subsamples have $\bharbar$ significantly 
higher than their corresponding low-$\mstar$ subsamples.}
\label{fig:bhar_vs_web}
\end{figure*}

In \S\ref{sec:bhar_vs_det}, we show that the $\bharbar$ is 
not related to overdensity (on \hbox{sub-Mpc} scales) at 
given $\mstar$.
In this section, we investigate the dependence of $\bharbar$
on cosmic-web environment ($\sim 1\text{--}10$~Mpc scales).
Following \S\ref{sec:split_det}, we derive the $\bharbar$ 
for galaxies in field, filament, and cluster environments,
respectively. 
The cosmic web describes global environment on 
$\approx 1\text{--}10$~Mpc scales (\S\ref{sec:cosmic_web}).
The results are displayed in Fig.~\ref{fig:bhar_vs_web}.
The $\bharbar$ does not show a significant trend as a function
of cosmic-web environment.  
For each environment bin, we further divide the sources 
into high-$\mstar$ and low-$\mstar$ subsamples, respectively,
and calculate $\bharbar$ for each subsample. 
Similar to Fig.~\ref{fig:bhar_vs_delta}, 
the high-$\mstar$ subsamples have significantly higher 
$\bharbar$ than the corresponding low-$\mstar$ subsamples,
consistent with the dominant $\bharbar$-$\mstar$ relation 
(\S\ref{sec:bhar_vs_det}). 

\begin{figure}
\includegraphics[width=\linewidth]{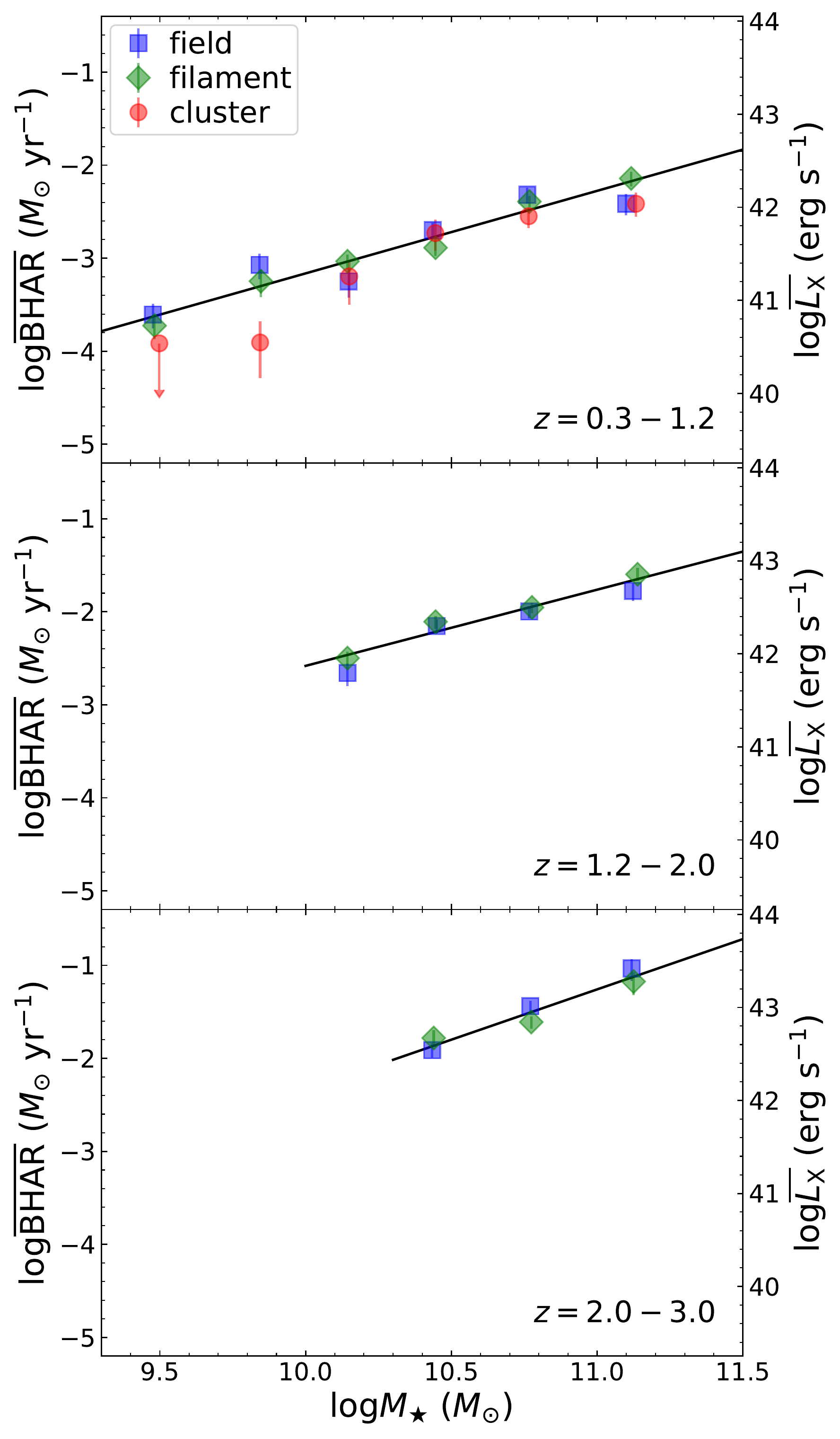}
\caption{$\bharbar$ as a function of $\mstar$ for different 
cosmic-web environments.
{${\lxbar}$ is marked on the right-hand 
side of each panel.} 
The black line represents the best-fit log-linear model 
using all the data points. 
At a given $\mstar$, the $\bharbar$ values are similar 
for different cosmic-web environments.
}
\label{fig:bhar_vs_M_web}
\end{figure}

We also bin our samples based on $\mstar$, and further 
divide each bin into subsamples for different parts 
of the cosmic web. 
We calculate $\bharbar$ for each subsample and show the 
results in Fig.~\ref{fig:bhar_vs_M_web}.
The $\bharbar$ values are not systematically different for 
different cosmic-web environments. 

Now we test the $\bharbar$ dependence on cosmic-web 
environment quantitatively. 
Unlike in \S\ref{sec:pca}, we do not perform a PCOR 
analysis, because the cosmic-web environment (field, 
filament, and cluster) is not a continuous quantity. 
Instead, we employ another statistical analysis based on 
the Akaike information criterion (AIC; \citealt{akaike74}).
The AIC is designed for model selection and defined as 
$\mathrm{AIC}=C+2k$, where $C$ is the fitting statistic 
($\chi^2$ for least squares fitting) and $k$ is the 
number of free parameters in the model. 
If one model has an AIC value much smaller than another 
model ($\Delta \mathrm{AIC}<\Delta \mathrm{AIC_{thresh}}$), 
then the former is considered superior to the latter
(see, e.g. \S2.6 of \citealt{burnham02}).
In our analyses, we choose $\mathrm{\Delta AIC_{thresh}=-7}$, 
corresponding to a 3$\sigma$ confidence level under the 
situation where the model parameter uncertainties are Gaussian 
(e.g. \citealt{murtaugh14}).

We apply the AIC technique to our data points in 
Fig.~\ref{fig:bhar_vs_M_web}.
For each redshift range, we perform a least-$\chi^2$ fit to all
the data points with a log-linear model, 
$\log\bharbar = A\times \log\mstar + B$, where $A$ and $B$ are 
free model parameters.\footnote{The single upper limit point in 
Fig.~\ref{fig:bhar_vs_M_web} is not used in the fitting.}
We calculate the AIC value (AIC$_1$) for this fitting.
The best-fit models are displayed in Fig.~\ref{fig:bhar_vs_M_web}.
We then create a set of three independent log-linear models, i.e., 
$\log\bharbar = A_{\rm field}\times \log\mstar + B_{\rm field}$,
$\log\bharbar = A_{\rm filament}\times \log\mstar + B_{\rm filament}$, and
$\log\bharbar = A_{\rm cluster}\times \log\mstar + B_{\rm cluster}$,
to fit the data.\footnote{For $z>1.2$, we only create the field 
and filament models, as we do not assign cluster environment 
(see \S\ref{sec:cosmic_web}).}
As the subscripts indicate, each model is used to fit the data 
points of the corresponding cosmic-web environment in 
Fig.~\ref{fig:bhar_vs_M_web}. 
We derive the AIC value (AIC$_2$) for this multi-model fitting. 
If $\Delta\mathrm{AIC}=\mathrm{AIC_2}-\mathrm{AIC_1}<-7$, then the 
$\bharbar$-$\mstar$ relations might be different for different 
cosmic-web environments. 
The resulting AIC values are listed in Tab.~\ref{tab:aic}.
For all three redshift ranges, the $\Delta\mathrm{AIC}$ values 
are above $-7$. 
Therefore, the differences among the $\bharbar$-$\mstar$ relations 
for different cosmic-web environments are not statistically 
significant.
{However, the non-detection of a 
${\bharbar}$-environment correlation might be, in 
principle, due to the limited sensitivity of our data. 
\hbox{\citet{martini09}} found that the AGN fraction in rich 
clusters is ${\approx 0.7}$~dex below that in the 
field at ${z\lesssim 1}$. 
A natural question is whether our data are sensitive enough to 
detect such ${\bharbar}$ differences, i.e., 
${\bharbar}$ drops by 0.7~dex from the field to cluster 
environments.
To answer this question, we perform a test. 
For our ${z=0.3\text{--}1.2}$ bin, we systematically 
shift our cluster (field) ${\bharbar}$ by ${-0.35}$~dex 
(${+0.35}$~dex) and re-calculate 
${\Delta\mathrm{AIC}}$. 
We find ${\Delta\mathrm{AIC}=-114}$, much lower than 
our threshold (${-7}$). 
Therefore, if our ${\bharbar}$ dropped by 0.7~dex from 
the field to cluster environments, we would definitely detect the 
environmental dependence of ${\bharbar}$. 
In fact, we find that our data are sensitive at a 
${\approx 3\sigma}$ level to a ${\approx 0.2}$~dex 
difference of ${\bharbar}$ from the field to cluster 
environments at ${z=0.3\text{--}1.2}$.
This difference between our work and \hbox{\citet{martini09}}
might be due to the lack of rich clusters in our sample (see 
\S\ref{sec:phy} for more discussion).}

\begin{figure}
\includegraphics[width=\linewidth]{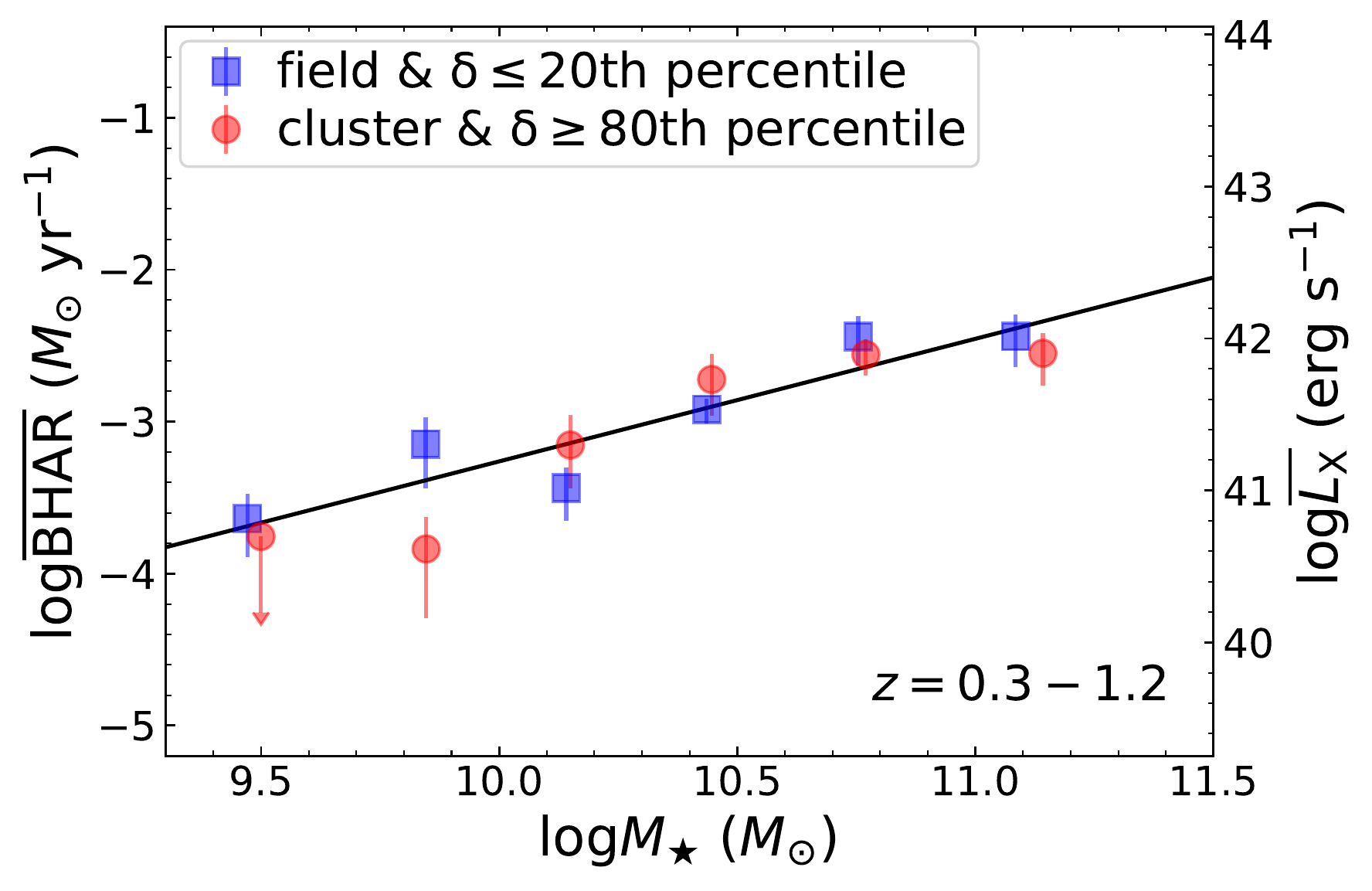}
\caption{{Same format as Fig.~\ref{fig:bhar_vs_M_web} (top) but
for cluster vs.\ field environments. 
Here, we limit the cluster (field) galaxies to those with the 
highest (lowest) 20\% overdensity in each ${\mstar}$ bin. 
At a given ${\mstar}$, the ${\bharbar}$ values 
are similar for cluster and field environments.}
}
\label{fig:bhar_vs_M_web_extreme}
\end{figure}

{In Fig.~\ref{fig:bhar_vs_M_web_extreme}, we also compare
${\bharbar}$ for cluster and field environments at 
${z=0.3\text{--}1.2}$.
Here, we limit the cluster (field) galaxies to those with the highest 
(lowest) 20\% overdensity in each ${\mstar}$ 
bin.\footnote{{Here, we do not limit cluster galaxies to 
those also identified by \citet{george11},  
because this would lead to too few sources (only ${<10}$ AGNs) 
for our analyses, as the cluster-member catalog in \citet{george11}
is not complete (see \S\ref{sec:cosmic_web}).}}
In this way, we probe the most-extreme environments. 
We perform AIC analyses and obtain ${\Delta\mathrm{AIC}=3.9}$,
above the threshold (${-7}$). 
Therefore, the ${\bharbar}$-${\mstar}$ 
relations for these two extreme environments are also not 
statistically different.
At higher redshift, we have also performed similar analyses for 
filament vs.\ field environments and reached the same 
conclusion.}

\begin{figure}
\includegraphics[width=\linewidth]{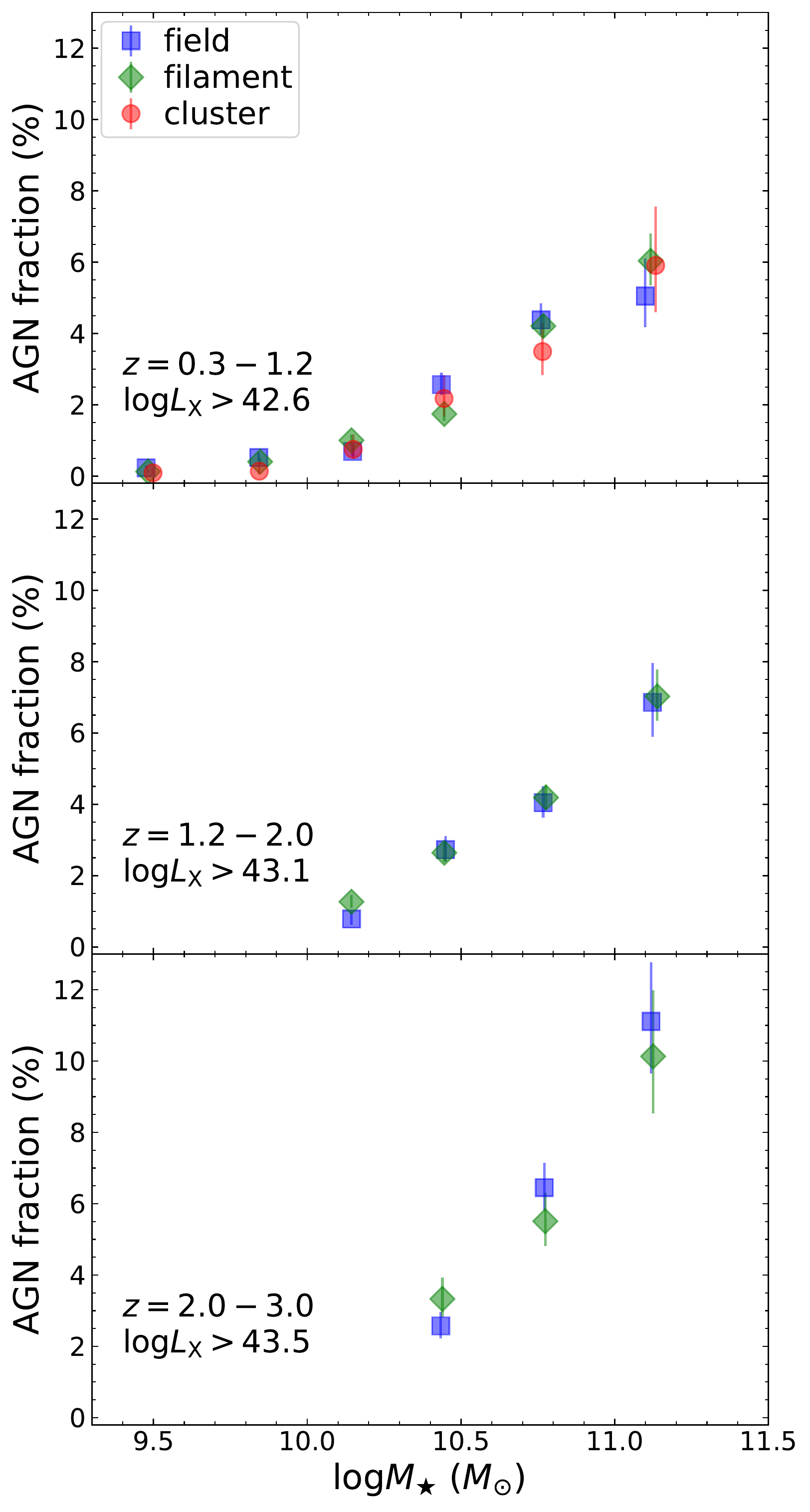}
\caption{AGN fraction as a function $\mstar$ for different 
cosmic-web environments. 
The data are from Tab.~\ref{tab:agn_frac}.
At a given $\mstar$, AGN fractions are similar for different
cosmic-web environments. 
}
\label{fig:AGNfrac_vs_M_web}
\end{figure}

As in \S\ref{sec:pca}, we also calculate AGN fractions 
for different cosmic-web environments for a consistency check. 
The results are presented in Tab.~\ref{tab:agn_frac} and 
Fig.~\ref{fig:AGNfrac_vs_M_web}.
The AGN fractions are generally similar for different cosmic-web 
environments when controlling for $\mstar$, consistent with our 
AIC analyses.
Our AGN fractions for $\log\mstar>10.3$ are 
\hbox{$\approx 2\%\text{--}6\%$} at \hbox{$z=0.3\text{--}1.2$}.
This range is consistent with the results of 
\citet[][see their Tab.~2]{silverman09b}, who found an AGN 
fraction of \hbox{$\approx 3\%$} for $\log\mstar>10.4$ at $z<1$, 
independent of environment.
{Our AGN fractions for ${\log\mstar=11\text{--}11.5}$ 
at ${z=0.3\text{--}1.2}$ are ${\approx 6\%}$, 
similar to that derived for SDSS galaxies of similar 
${\mstar}$ at ${z\approx 0.5}$ 
\hbox{\citep{haggard10}}.
}

\begin{table}
\begin{center}
\caption{Best-fit AIC values of the $\bharbar$-$\mstar$ relation 
(see \ref{sec:bhar_vs_web})}
\label{tab:aic}
\begin{tabular}{cccc}\hline\hline
Redshift & AIC$_1$ & AIC$_2$ & $\Delta \mathrm{AIC}$ \\ \hline 
0.3--1.2 & 32.48 & 33.35 & $0.9$ \\
1.2--2.0 & 12.34 & 14.38 & $2.0$ \\
2.0--3.0 & 9.82 & 9.75 & $-0.1$ \\
\hline
\end{tabular}
\end{center}
\end{table}

\begin{table*}
\begin{center}
\caption{AGN fractions (\%) for different cosmic-web environments}
\label{tab:agn_frac}
\begin{tabular}{c|ccccccc}\hline\hline
\multicolumn{7}{c}{$z=0.3\text{--}1.2$ ($\log\lx>42.6$)} \\ \hline
$\log\mstar$ & 9.3--9.7 & 9.7--10.0 & 10.0--10.3 & 
	     10.3--10.6 & 10.6--11.0& 11.0--11.5& \\ \hline
Field & $0.2^{+0.07}_{-0.05}$ & $0.5^{+0.14}_{-0.11}$ & $0.7^{+0.17}_{-0.14}$ & $2.6^{+0.34}_{-0.30}$ & $4.4^{+0.46}_{-0.42}$ & $5.0^{+1.05}_{-0.88}$ \\
Filament & $0.1^{+0.05}_{-0.03}$ & $0.4^{+0.10}_{-0.08}$ & $1.0^{+0.17}_{-0.14}$ & $1.7^{+0.23}_{-0.20}$ & $4.2^{+0.34}_{-0.32}$ & $6.0^{+0.77}_{-0.69}$ \\
Cluster & $0.1^{+0.15}_{-0.06}$ & $0.1^{+0.22}_{-0.09}$ & $0.8^{+0.42}_{-0.27}$ & $2.2^{+0.65}_{-0.51}$ & $3.5^{+0.81}_{-0.66}$ & $5.9^{+1.66}_{-1.31}$ \\
All & $0.2^{+0.04}_{-0.03}$ & $0.4^{+0.07}_{-0.06}$ & $0.9^{+0.11}_{-0.10}$ & $2.1^{+0.18}_{-0.17}$ & $4.2^{+0.26}_{-0.24}$ & $5.7^{+0.57}_{-0.52}$ \\
\hline\hline
\multicolumn{7}{c}{$z=1.2\text{--}2.0$ ($\log\lx>43.1$)} \\ \hline
$\log\mstar$ & -- & -- & 10.0--10.3 & 
	     10.3--10.6 & 10.6--11.0& 11.0--11.5& \\ \hline
Field & -- & -- & $0.8^{+0.20}_{-0.16}$ & $2.7^{+0.38}_{-0.33}$ & $4.0^{+0.46}_{-0.42}$ & $6.9^{+1.11}_{-0.96}$ \\
Filament & -- & -- & $1.3^{+0.20}_{-0.17}$ & $2.6^{+0.29}_{-0.26}$ & $4.2^{+0.36}_{-0.33}$ & $7.0^{+0.75}_{-0.68}$ \\
All & -- & -- & $1.1^{+0.14}_{-0.12}$ & $2.7^{+0.23}_{-0.21}$ & $4.1^{+0.28}_{-0.26}$ & $7.0^{+0.61}_{-0.57}$ \\
\hline\hline
\multicolumn{7}{c}{$z=2.0\text{--}3.0$ ($\log\lx>43.5$)} \\ \hline
$\log\mstar$ & -- & -- & -- & 
	     10.3--10.6 & 10.6--11.0& 11.0--11.5& \\ \hline 
Field & -- & -- & -- & $2.6^{+0.40}_{-0.35}$ & $6.4^{+0.69}_{-0.63}$ & $11.1^{+1.66}_{-1.47}$ \\
Filament & -- & -- & -- & $3.3^{+0.60}_{-0.51}$ & $5.5^{+0.80}_{-0.70}$ & $10.1^{+1.86}_{-1.60}$ \\
All & -- & -- & -- & $2.8^{+0.33}_{-0.29}$ & $6.1^{+0.52}_{-0.48}$ & $10.7^{+1.21}_{-1.10}$ \\
\hline
\end{tabular}
\end{center}
\end{table*}

\subsection{Tests in Narrower Redshift Bins}
\label{sec:narrow_bin}
\begin{figure}
\includegraphics[width=\linewidth]{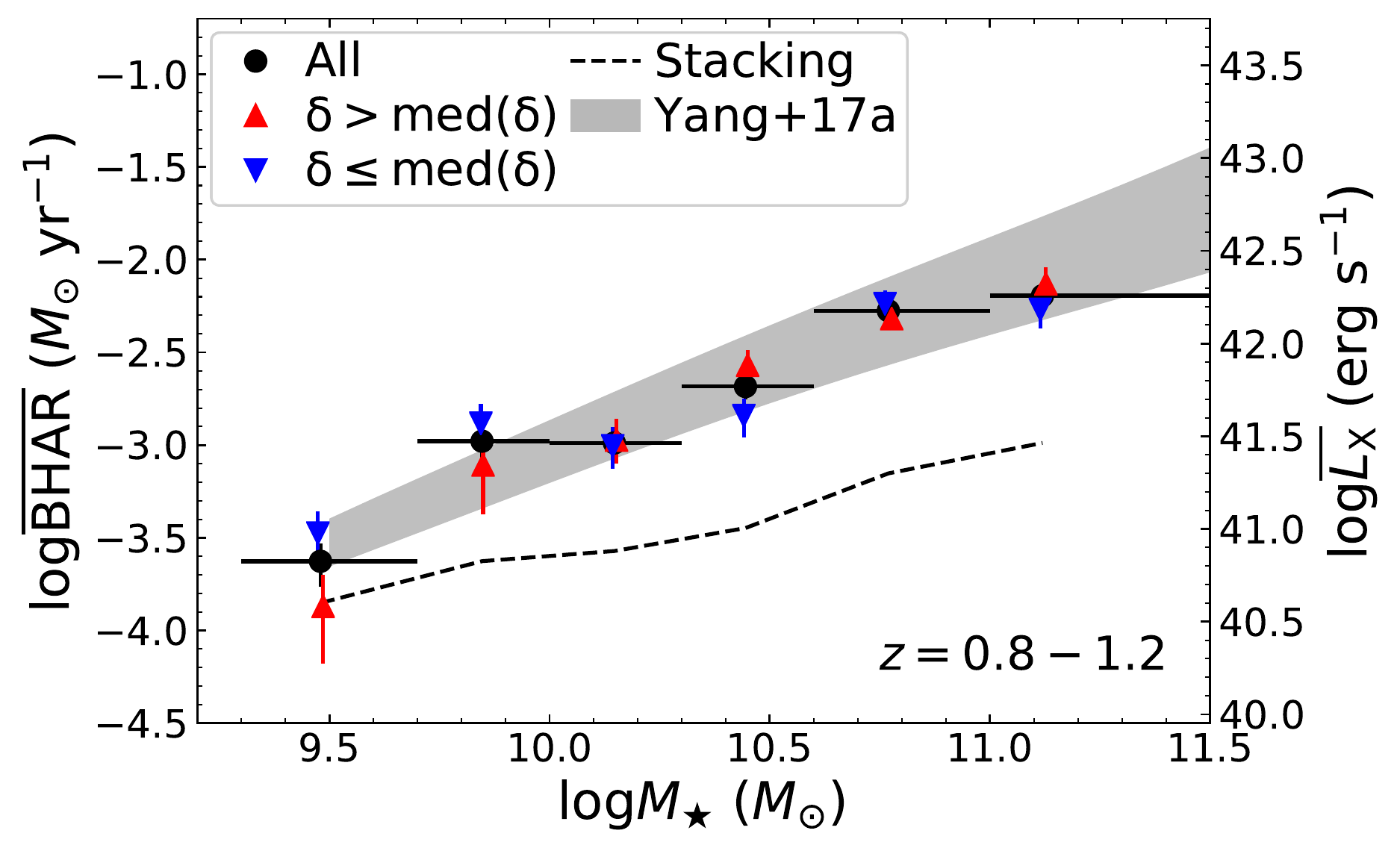}
\includegraphics[width=\linewidth]{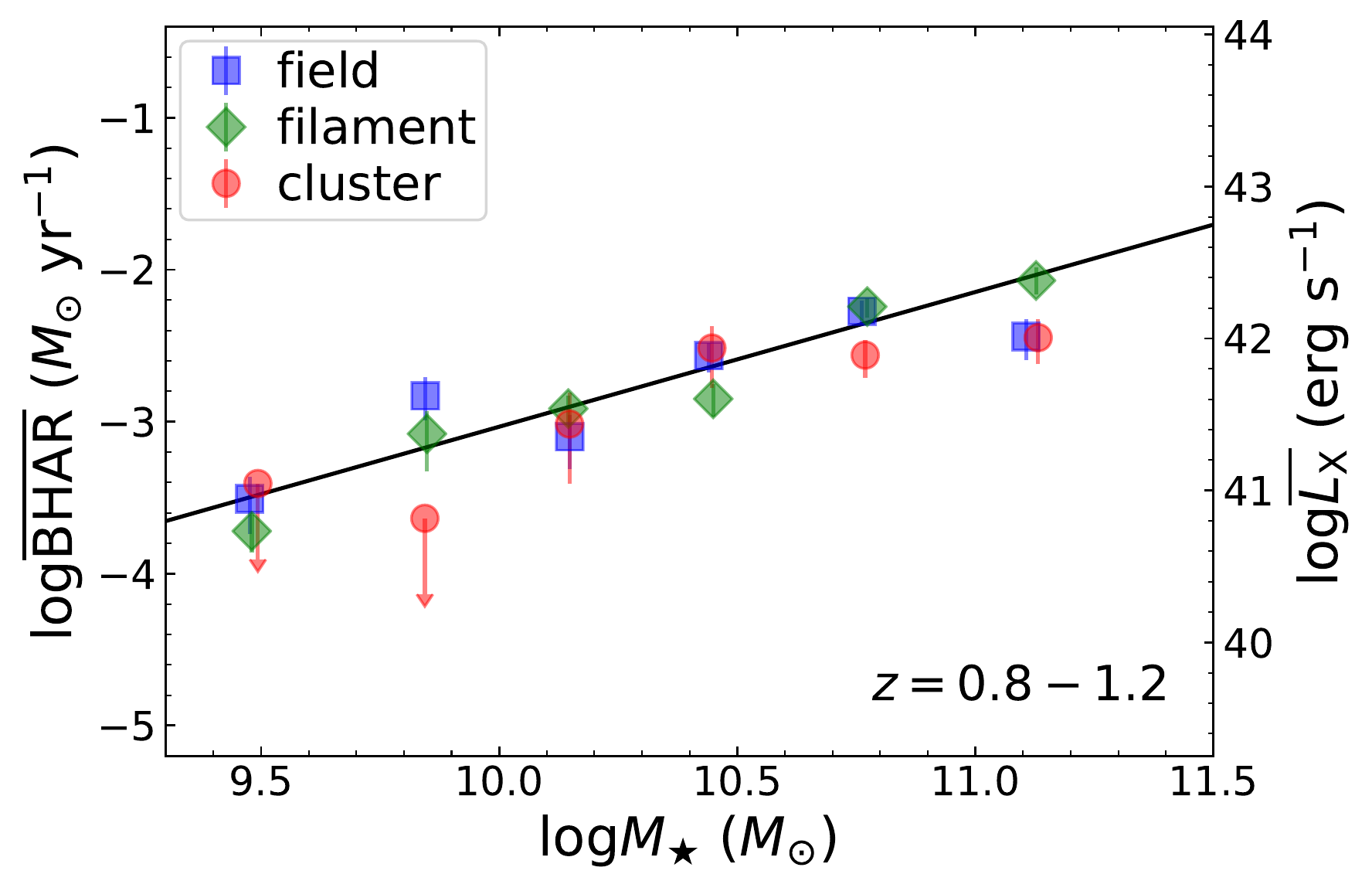}
\caption{{The top and bottom panels follow the same formats 
as Figs.~\ref{fig:bhar_vs_M_od} and Fig.~\ref{fig:bhar_vs_M_web} 
but for the narrower redshift bin of ${z=0.8{--}1.2}$. 
Still, we do not find significant $\bharbar$ dependence on environment. 
This conclusion also applies for other narrower redshift bins in 
\S\ref{sec:narrow_bin}.
}}
\label{fig:bhar_vs_M_narrow_zbin}
\end{figure}

{Our analyses above adopt relatively wide redshift bins, i.e., 
${z=0.3\text{--}1.2}$, 1.2\text{--}2.0, and 2.0\text{--}3.0,
to retain relatively large sample size in each bin.
Considering that both galaxy and AGN properties as well as cosmic environment 
evolve with redshift, the ${\bharbar}$-environment relation 
might also have redshift dependence.
To test for possible redshift dependence, we repeat our analyses in 
\S\ref{sec:bhar_vs_det} and \S\ref{sec:bhar_vs_web} 
using narrower redshift bins, i.e., ${z=0.3\text{--}0.8}$, 
0.8\text{--}1.2, 1.2\text{--}1.6, 1.6\text{--}2.0, 
2.0\text{--}2.5, and 2.5\text{--}3.0.
This procedure reduces the sample size in each bin and thus 
increases the uncertainties on ${\bharbar}$ in general.
In the new analyses with narrower redshift bins, we still do not find 
any significant ${\bharbar}$
dependence on either overdensity or cosmic-web environment,
consistent with the results in \S\ref{sec:bhar_vs_det} and 
\S\ref{sec:bhar_vs_web}.
Fig.~\ref{fig:bhar_vs_M_narrow_zbin} shows some example figures for
${z=0.8\text{--}1.2}$, and the figures for other narrower 
redshift bins are qualitatively similar. 
Therefore, our main conclusions are unlikely to be affected by our 
choice of relatively wide redshift bins.
}

\section{Discussion}\label{sec:discuss}
We discuss the physical implications of our results in \S\ref{sec:phy}.
We compare our results with previous observations of BHAR-environment
relations in \S\ref{sec:comp_prev_work}.

\subsection{Physical Implications}
\label{sec:phy}
Our results indicate that SMBH accretion is fundamentally related 
to $\mstar$. 
At a given $\mstar$, our $\bharbar$ does not show significant 
dependence on host-galaxy environment. 
Since galaxy environment is largely determined by dark matter, which 
generally dominates the gravitational field on $\gtrsim$~subMpc 
scales, our conclusions suggest SMBH growth is primarily related 
to baryons rather than dark matter \citep[e.g.][]{kormendy13, 
yang18}.
The broad physical picture is likely that dark-matter density 
fluctuations lead to the formation of halos, allowing baryons 
to condense into the halo centers and form galaxies. 
The galaxies feed their SMBHs with cold gas via baryonic physics, 
e.g. disk instabilities and galaxy bars (e.g. 
\hbox{\citealt{alexander12}} and references therein).
This scenario indicates that, in future studies of SMBH-galaxy
coevolution, it is critical to focus on relations between 
BHAR and host-galaxy intrinsic properties (e.g. $\mstar$, SFR, 
and morphology) rather than the environment. 
Small but deep surveys such as the \chandra\ Deep Fields 
(e.g. \hbox{\citealt{xue16}}; \hbox{\citealt{luo17}}) 
and CANDELS 
(\hbox{\citealt{grogin11}}; \hbox{\citealt{koekemoer11}}) 
are ideal for studying SMBH-galaxy coevolution. 

\begin{figure}
\includegraphics[width=\linewidth]{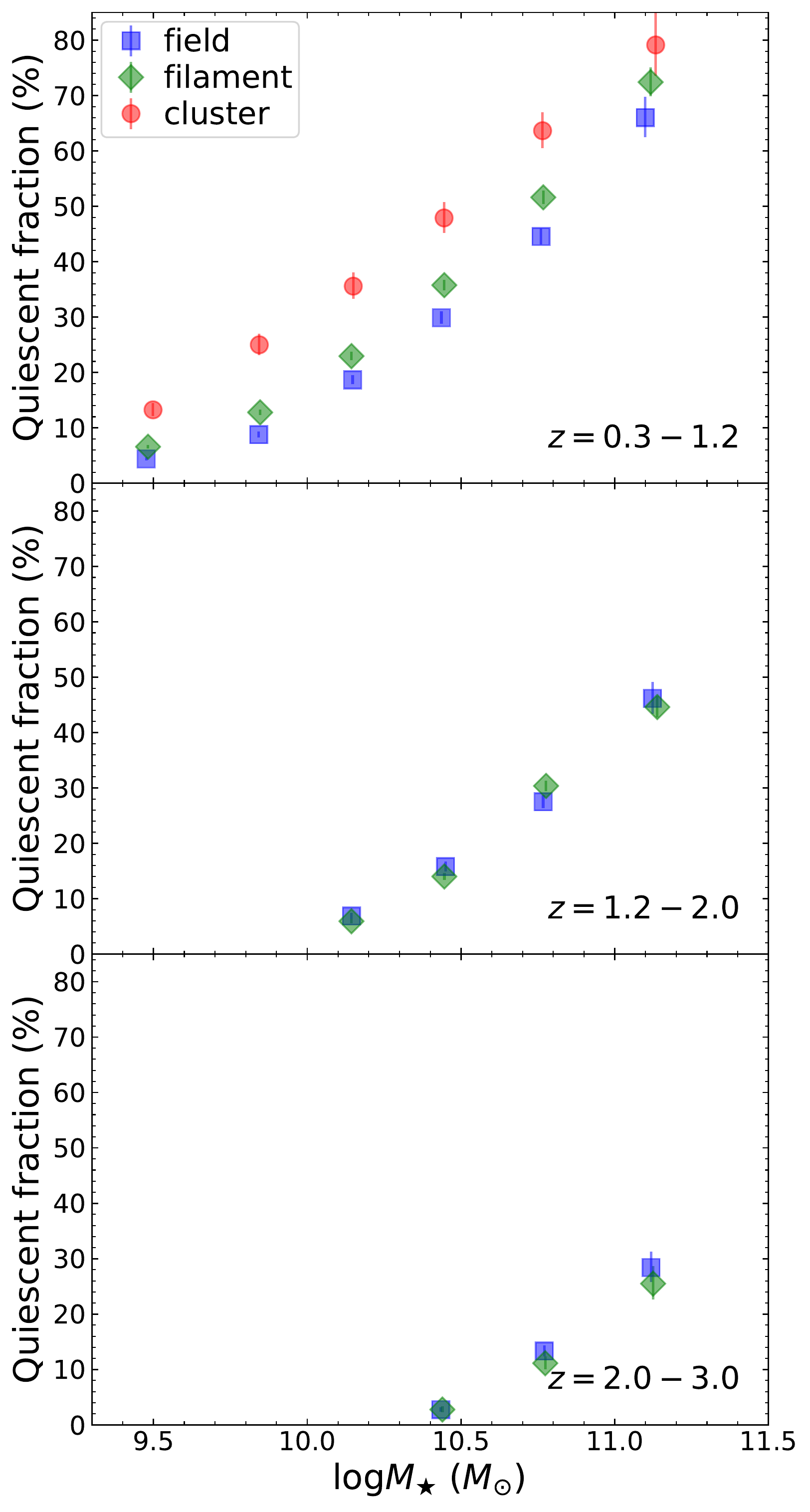}
\caption{Quiescent-galaxy fraction vs. stellar mass for 
different cosmic-web environments. 
The star-forming/quiescent classifications are based on 
a standard color-color scheme (see \S\ref{sec:Mstar}).
At $z=0.3\text{--1.2}$ and a given $\mstar$, the 
quiescent-galaxy fraction rises from the field to 
cluster environments (environmental quenching). 
At $z>1.2$, galaxies associated with the field and filament 
environments have similar quiescent-galaxy fractions. 
}
\label{fig:Qfrac_vs_M_web}
\end{figure}

It is well established that environment affects galaxy
evolution.
At a given $\mstar$ and at $z\lesssim 1$, the quiescent-galaxy 
fraction ({as defined in \S\ref{sec:Mstar}}) rises toward 
high-density regions, and this 
effect is often termed ``environmental quenching''
(e.g. \hbox{\citealt{peng10}}; \hbox{\citealt{scoville13}};
\hbox{\citealt{darvish15, darvish16, darvish17}}; 
\hbox{\citealt{laigle18}}).
Fig.~\ref{fig:Qfrac_vs_M_web} shows the quiescent-galaxy fraction
in our sample as a function of $\mstar$ for different cosmic-web 
environments (see \S\ref{sec:Mstar} for star-forming/quiescent 
classifications).
At $z=0.3\text{--}1.2$ and a given $\mstar$, the quiescent galaxy 
fraction significantly rises from the field to cluster environments;
at higher redshifts this environmental dependence seems to 
disappear.
Therefore, environmental quenching at $z=0.3\text{--}1.2$ is 
significantly observed in our sample.
For a given cosmic-web environment, the quiescent-galaxy fraction
also rises toward high $\mstar$ at all redshifts. 
This $\mstar$ dependence is expected from previous studies 
\citep[e.g.][]{brammer11, davidzon17}.

In contrast, our $\bharbar$ does not show a significant 
dependence on environment, if $\mstar$ is controlled 
(see \S\ref{sec:res}). 
These different behaviors of SMBH accretion and 
star formation suggest that SMBH and galaxy stellar-mass 
growth are not strongly coupled in general
(e.g. \hbox{\citealt{yang17, yang18}}), 
although we cannot rule out a weak secondary $\bharbar$-SFR relation 
(see \S\ref{sec:comp_prev_work}).  
The potential physical mechanisms responsible for 
environmental quenching such as tidal interaction and 
ram-pressure stripping (\S\ref{sec:intro}) might only have 
limited effects on SMBH accretion. 

{Since galaxy evolution has significant dependence on
environment at low redshift (${z \lesssim 1}$), the 
host-galaxy types of AGNs might also depend on environment.
In Tab.~\ref{tab:host_type}, we list the quiescent-galaxy
fractions for AGNs (as defined in \S\ref{sec:pca}) in different 
cosmic-web environments.  
At ${z=0.3\text{--}1.2}$, the quiescent-galaxy fraction 
of AGN hosts appears to rise from the field to clusters, 
similar to the trend for normal galaxies 
(see Fig.~\ref{fig:Qfrac_vs_M_web}).
At higher redshift, if anything, the trend seems to be the opposite 
going from the field to filament environments. 
However, these trends are not statistically significant at a 
3$\sigma$ confidence level due to our limited AGN sample size.
Future work with much larger AGN samples can determine these 
trends more accurately. 
}



\begin{table}
\begin{center}
\caption{Quiescent-galaxy fractions (\%) for AGN host galaxies}
\label{tab:host_type}
\begin{tabular}{ccccc}\hline\hline
Redshift & Field & Filament & Cluster & All \\ \hline
$0.3\text{--}1.2$ & $18.8^{+2.66}_{-2.40}$ & $26.5^{+2.38}_{-2.25}$ & $29.3^{+6.29}_{-5.59}$ & $24.0^{+1.70}_{-1.62}$ \\
$1.2\text{--}2.0$ & $13.9^{+2.62}_{-2.26}$ & $10.0^{+1.67}_{-1.45}$ & -- &$11.4^{+1.40}_{-1.26}$ \\
$2.0\text{--}3.0$ & $7.6^{+2.19}_{-1.73}$ & $6.6^{+2.63}_{-1.92}$ & -- &$7.2^{+1.63}_{-1.35}$ \\
\hline
\end{tabular}
\end{center}
\end{table}

\subsection{Previous Works on BHAR vs.\ Environment}
\label{sec:comp_prev_work}
{Based on sources at ${z\lesssim 1}$, 
observations have found 
that cluster (${\mhalo \lesssim 10^{14}\ M_\odot}$) and 
field environments have similar \xray\ AGN 
fractions among massive galaxies, consistent with 
our results (e.g. \hbox{\citealt{georgakakis08}}; 
\hbox{\citealt{silverman09b}}; \hbox{\citealt{koulouridis14}}).
However, these analyses are often restricted to low-redshift 
relatively bright galaxies.}
Thanks to the reliable \hbox{photo-$z$} measurements and our improved 
methodology for assessing SMBH accretion, our work is able to 
investigate $\bharbar$-environment relations for all galaxies 
above the $\mstar$ completeness limits up to $z=3$. 
Importantly, our study covers $z\approx 1.5\text{--}2.5$ where 
cosmic AGN activity peaks.

{Some studies find that, at ${z\lesssim 1}$, 
the \xray\ AGN fractions in rich clusters 
(${\mhalo \sim 10^{15}\ M_\odot}$) are generally 
lower than those in the field (e.g. \hbox{\citealt{martini09}}; 
\hbox{\citealt{ehlert14}}).
Due to the lack of excellent multiwavelength coverage for 
${\mstar}$ calculation, these studies often adopt a 
simple ${R}$-band magnitude cut to approximate an
${\mstar}$ cut of the galaxy population 
(e.g. \hbox{\citealt{cappellari16}}).
However, we note that consensus has not been widely reached on 
whether rich clusters have lower AGN fractions than the field.
For example, \hbox{\citet{haggard10}} found that rich clusters 
and the field have similar AGN fractions at 
${z=0.05\text{--}0.31}$, when the same magnitude 
and ${\lx}$ cuts are applied to the cluster and field 
populations. 
If AGN activity is indeed suppressed in rich clusters at a given
${\mstar}$, the physical reason might be different 
galaxy types in cluster and field environments. 
Rich clusters, especially in their central regions, are dominated 
by the quiescent galaxy population, which tends to have lower AGN 
fractions than the star-forming population at a given 
${\mstar}$ 
(e.g. \hbox{\citealt{wang17}}; \hbox{\citealt{aird18}};
\hbox{\citealt{yang18}}).
We cannot study such rich clusters in our work, because they are rare 
and generally absent in COSMOS, where the clusters typically have 
${\mhalo \lesssim 10^{14}\ M_\odot}$ 
(e.g. \hbox{\citealt{knobel09}}; \hbox{\citealt{george11}}).
However, only a small fraction of the galaxy population 
(${\lesssim 1\%}$) lives in rare rich clusters with 
${\mhalo \sim 10^{15}\ M_\odot}$ 
\hbox{\citep[e.g. Tab.~1 of ][]{bahcall99}}.
Our work covers a large comoving volume 
(${\approx 10^7}$~Mpc${^3}$ 
for each redshift bin; \S\ref{sec:sample}), and thus 
we probe the main range of cosmic environments in the overall 
Universe with 
$\mhalo \approx 10^{11} \text{--} 10^{14}$~$M_\odot$.
We do not find significant ${\bharbar}$-environment relations when 
controlling for ${\mstar}$, even at the extremes of the 
environments we sample (e.g. Figs.~\ref{fig:bhar_vs_M_od} and 
\ref{fig:bhar_vs_M_web_extreme}).
Therefore, we conclude that, for the overall galaxy population, 
${\bharbar}$ generally does not depend on cosmic environment once 
${\mstar}$ is controlled, although this conclusion might not hold
for galaxies living in rare rich clusters.}

At high redshift ($z\gtrsim 2$), observations suggest that some 
massive protoclusters might have elevated AGN activity 
(see \S\ref{sec:intro}; {but also see \citealt{macuga18}}).
For example, SSA~22 is a prominent protocluster at $z\approx 3.1$, 
and it is likely a progenitor of local rich clusters with 
$\mhalo \sim 10^{15}\ M_\odot$ \citep[e.g.][]{steidel98}. 
\citet{lehmer09} estimated that its \xray\ AGN fraction is 
$\approx 5\% \text{--} 10\%$ among Lyman Break Galaxies (LBGs) 
and Ly$\alpha$ Emitters (LAEs), considerably higher than 
the AGN fraction ($\approx 1\%\text{--}2\%$) among LBGs and 
LAEs in the field at similar redshifts.
However, the LBGs and LAEs in SSA~22 might have different 
$\mstar$ compared to the LBGs and LAEs in the field, and the
difference in $\mstar$ could drive the apparent differences 
in AGN fraction (see Tab.~\ref{tab:agn_frac}). 
Also, the SSA~22 AGN sample size is limited ($\approx 10$) and 
suffers from significant statistical uncertainty. 
SSA~22 has $\gtrsim 150$ galaxies in its central region of 
$\approx 10$~Mpc$^2$ \citep[e.g.][]{topping16}, translating to 
an overdensity value of $\log(1+\delta) \gtrsim 1.3$ 
(\S\ref{sec:density_field}).
We do not have such prominent structures in our sample at 
$z\gtrsim 2.5$ (Fig.~\ref{fig:all_vs_z}) and thus cannot probe 
their AGN activity using COSMOS. 
Note that the estimated overdensity value of SSA~22 is for
high redshifts ($z\gtrsim 2.5$). 
It is not directly comparable to our structures at low redshift,
since the normalization factors $\Sigma_{\rm median}$ are different 
at different redshifts (see Eq.~\ref{eq:over_density}).

{AGN-clustering studies also make use of 
source spatial distributions to probe SMBH-galaxy 
coevolution.
Early observations found that \xray\ AGNs and normal 
galaxies appear to have different clustering properties
(especially on ${\lesssim 1}$~Mpc~${h^{-1}}$
comoving scales),
suggesting different environmental effects for
central vs.\ satellite galaxies
(e.g. \hbox{\citealt{hickox09}}; \hbox{\citealt{miyaji11}}; 
\hbox{\citealt{allevato12}}).
However, galaxy properties (especially ${\mstar}$)
were not carefully controlled among these studies. 
Considering the strong ${\bharbar}$-${\mstar}$
correlation \hbox{\citep[e.g.][]{yang17, yang18}} and the 
well-established relation between galaxy clustering and 
${\mstar}$ \hbox{\citep[e.g.][]{coil06, coil17}}, it is 
critical to match ${\mstar}$ in AGN vs.\ galaxy 
clustering analyses. 
More recent observations show that the clustering 
properties of AGNs and galaxies are similar over a wide range 
of comoving scale 
(${\approx 0.1\text{--}30}$~Mpc~${h^{-1}}$)
when ${\mstar}$ is carefully matched, indicating
that ${\mstar}$ (rather than a central/satellite effect) 
mainly drives the observed clustering properties of AGNs
(e.g. \hbox{\citealt{georgakakis14}}; \hbox{\citealt{powell18}}).
Also, our analyses do not support different environmental effects 
for central vs.\ satellite galaxies, although we do not perform 
a central/satellite classification (\S\ref{sec:env}).
If, for example, environment only affects AGN activity in 
central galaxies, we would witness an increasingly strong
${\bharbar}$-environment relation in more massive 
galaxies which have a higher chance of being central 
(e.g. \hbox{\citealt{reddick13}}). 
However, we do not find any significant 
${\bharbar}$-environment correlation
over a wide range of stellar mass 
(${\log \mstar \approx 9.5\text{--}11.5}$; 
\S\ref{sec:res}).
Clustering analyses infer that AGNs typically have 
${\log\mhalo \approx 13}$ (e.g. 
\hbox{\citealt{allevato11, allevato14, allevato16}}; 
\hbox{\citealt{richardson13}}).
This is as expected, since low-mass halos 
(${\log\mhalo \lesssim 12}$) only host low-mass 
galaxies with weak ${\bharbar}$ and high-mass 
halos (${\log\mhalo \gtrsim 14}$) are rare.
}

\section{Summary and Future Work}\label{sec:summary}
We have studied the $\bharbar$ dependence on $\mstar$ and 
environment in redshift bins of $z=0.3\text{--}1.2$, 
$z=1.2\text{--}2.0$, and $z=2.0\text{--}3.0$, based on 
sources in the COSMOS field. 
Our main procedures and results are summarized below:
\begin{enumerate}

\item We have compiled a large galaxy sample in the 
COSMOS field ($\approx$~170,000 sources; \S\ref{sec:sample}) 
and estimated their $\mstar$ via SED fitting 
(\S\ref{sec:Mstar}). 
We have measured surface overdensity (\hbox{sub-Mpc} scales) 
and cosmic-web environment ($\approx 1\text{--}10$~Mpc scales) 
for our sources (\S\ref{sec:env}).

\item We have derived $\bharbar$ for different samples, 
considering both \xray\ detected and undetected sources
(\S\ref{sec:bhar}).
For \xray\ detected sources, we adopt, in order of priority, 
hard-band, full-band, and soft-band fluxes, in our 
calculations (\S\ref{sec:xray_det}). 
This choice is to minimize the effects of \xray\ obscuration. 
We include the \xray\ emission from \xray\ undetected sources
via stacking (\S\ref{sec:xray_undet}).

\item We do not find a statistically significant $\bharbar$ 
dependence on overdensity or cosmic-web environment 
($\approx 1\text{--}10$~Mpc) for $\mstar$ controlled samples
(\S\ref{sec:res}). 
Instead, $\bharbar$ is always strongly related to 
$\mstar$, regardless of environment. 
These results suggest that $\bharbar$ might be primarily  
related to the host galaxies rather than cosmic environment 
on scales of $\approx 0.1\text{--}10$~Mpc, which is determined 
by dark matter (\S\ref{sec:phy}). 
Thanks to the large comoving volume sampled ($\approx 10^7$~Mpc$^3$
for each redshift bin), we can probe the main range of cosmic 
environments in the overall Universe (\S\ref{sec:comp_prev_work}).
Therefore, we conclude that, for the overall galaxy population, 
$\bharbar$ generally does not depend on cosmic environment once 
$\mstar$ is controlled, although this conclusion might not hold
for the $\lesssim 1\%$ of galaxies living in rare rich clusters 
with $\mhalo \sim 10^{15}$~$M_\odot$.

\item In contrast to SMBH accretion, star formation activity 
significantly depends on environment at $z\lesssim 1$ 
(\S\ref{sec:phy}). 
For our sample, the quiescent-galaxy fraction rises from 
the field to cluster environment for $\mstar$-controlled 
samples at $z=0.3\text{--}1.2$, consistent with previous 
observations. 
The different behaviors of SMBH accretion and star formation 
suggest that SMBH and galaxy growth are not strongly coupled 
in general. 
Environment-related mechanisms such as tidal interaction and 
ram-pressure stripping that could shape galaxy evolution 
do not appear to strongly affect SMBH growth. 

\end{enumerate}

Future work can probe the $\bharbar$ dependence on environment
for larger physical scales ($\approx 10\text{--}100$~Mpc). 
Since COSMOS alone cannot sample the full range of cosmic 
environments on these scales (e.g. \hbox{\citealt{meneux09}}; 
\hbox{\citealt{skibba14}}),
these studies will need several COSMOS-like fields, e.g. 
\hbox{XMM-LSS} \citep{chen18}, \hbox{Wide-CDF-S}, and \hbox{ELAIS-S1},
or much larger fields such as Stripe~82 \citep{lamassa13} and 
\hbox{XMM-XXL} \citep{pierre16}.
{Such larger fields can also be used to probe SMBH growth 
in rare rich clusters/protoclusters (see \S\ref{sec:comp_prev_work}) 
while controlling for host-galaxy properties, especially 
${\mstar}$.}
In addition, future work may study $\bharbar$ in galaxy close 
pairs on $\approx 10\text{--}100$~kpc scales \citep[e.g.][]{mundy17}.

{The upcoming \textit{eROSITA} all-sky \xray\ survey will yield
a sample of ${\sim 10^{6}}$~AGNs at 
${z\lesssim 1}$ (e.g. \hbox{\citealt{merloni12}}), 
allowing studies of ${\bharbar}$-environment relations 
at low-to-moderate redshift with overwhelming source statistics.
In this work, we do not find significant environmental 
dependence of average BHAR.
It is still possible that the full distribution of BHAR depends 
on environment, although this would require ``finely tuned''  
BHAR distributions in different environments to maintain constant
${\bharbar}$. 
A full characterization of the BHAR distribution as a function 
of ${\mstar}$, environment, and redshift (e.g. 
\hbox{\citealt{georgakakis17}}; \hbox{\citealt{aird18}}; 
\hbox{\citealt{yang18}}) requires future \xray\ observatories 
like \textit{Athena} and \textit{Lynx}, which are necessary to 
sample the faint end of the BHAR distribution in COSMOS-like 
(or larger) fields (e.g. \hbox{\citealt{georgakakis18}}).}

{With the advance of environment-measurement 
methodology, new environmental metrics 
other than overdensity (and the consequent field/filament/cluster 
classification) may be developed. 
Future work can study the ${\bharbar}$ dependence on 
these new environmental metrics (e.g. mass density 
instead of number density as used in this work).
Future spectroscopic observations with Extremely Large 
Telescopes (ELTs) should improve the spec-${z}$ 
completeness for COSMOS and other fields by a large factor, 
allowing environmental measurements with superior accuracy 
(e.g. reducing the projection distance from 
${\approx 100}$~Mpc to ${\approx 10}$~Mpc;
Appendix~\ref{app:ill}).  
Based on such new spec-${z}$ data, studies can revisit 
the ${\bharbar}$-environment-${\mstar}$ 
connection, even for central/satellite galaxies separately.  
}


\section*{Acknowledgements}
We thank the referee for helpful feedback that improved this work.
We thank Francesca Civano, Clotilde Laigle, and Mara Salvato 
for providing relevant data.
We thank Robin Ciardullo, Antonis Georgakakis, Ryan Hickox,
Donghui Jeong, Paul Martini, 
Qingling Ni, John Silverman, Michael Strauss, 
and  Rosemary Wyse for helpful discussions.  
G.Y., W.N.B., C.T.C., and F.V.\ acknowledge support 
from \chandra\ \xray\ Center grants \hbox{GO4-15130A} 
and \hbox{AR8-19016X}, NASA grant NNX17AF07G, and the 
NASA Astrophysics Data Analysis Program (ADAP). 
B.D.\ acknowledges financial support from NASA
through the ADAP, grant number NNX12AE20G, and the 
National Science Foundation, grant number 1716907.
D.M.A.\ acknowledges the Science and Technology 
Facilities Council (STFC) through grant ST/P000541/1.
F.E.B.\ acknowledges support from CONICYT-Chile 
(Basal-CATA PFB-06/2007, FONDO ALMA 31160033) and 
the Ministry of Economy, Development, and Tourism's 
Millennium Science Initiative through grant IC120009, 
awarded to The Millennium Institute of Astrophysics, MAS.

This project uses {\sc astropy} (a Python package; see 
\hbox{\citealt{astropy13, astropy}}) and the SVO Filter Profile 
Service (http://svo2.cab.intacsic.es/theory/fps/).




\bibliographystyle{mnras}
\bibliography{all.bib} 



\appendix

\section{Explanation of Environment Measurements}
\label{app:ill}
\begin{figure*}
\includegraphics[width=\linewidth]{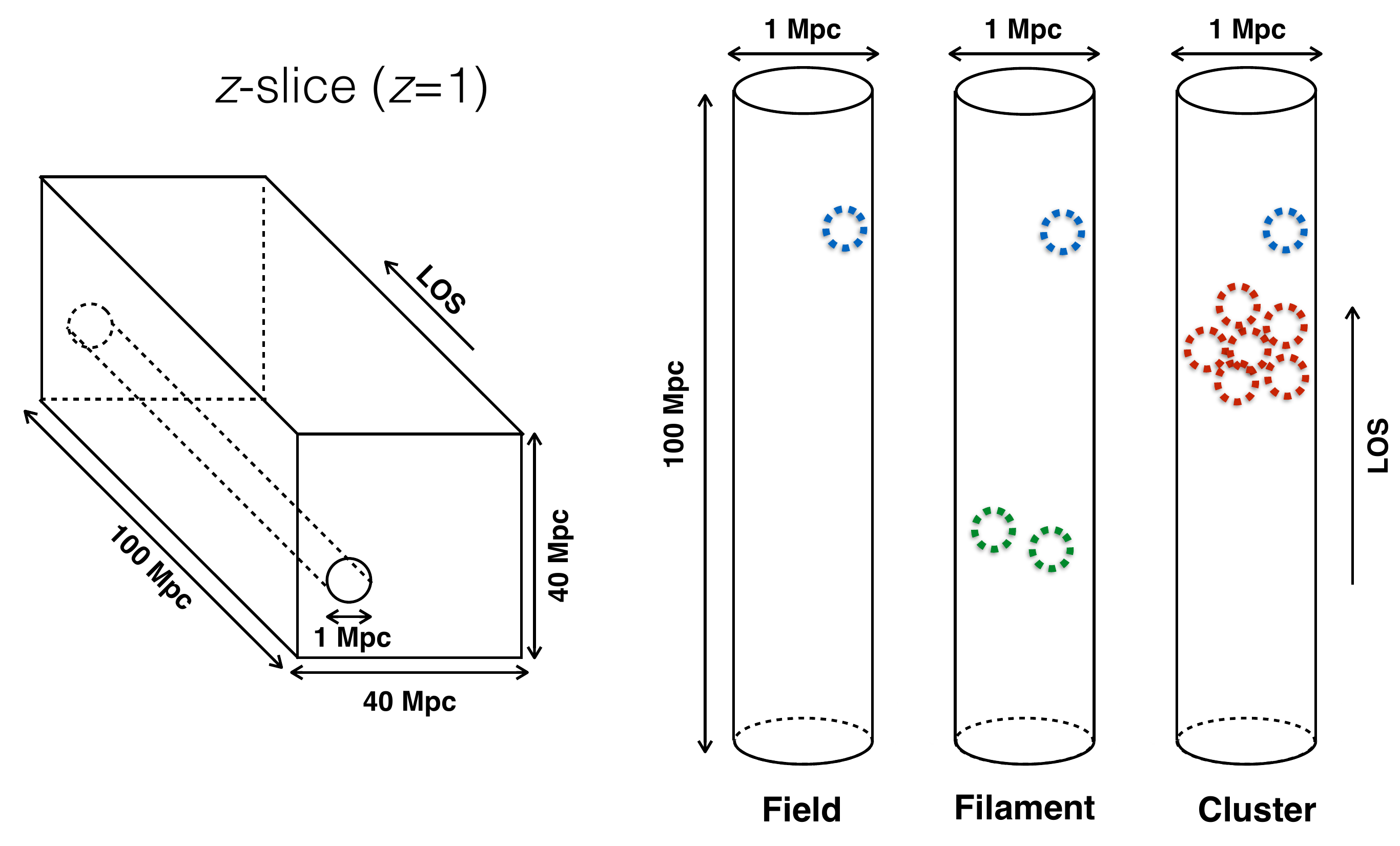}
\caption{Schematic plot for overdensity measurements 
(not drawn to scale).
The left panel shows the redshift slice at $z=1$. 
The $z$-slice contains galaxies projected along the LOS over 
$\approx 100$~Mpc. 
Our surface density is measured in cylinders with radii 
$\approx 0.5$~Mpc as marked.
The right panel illustrates three typical cylinders for the
field, filament, and cluster environments, respectively. 
The dashed circles denote galaxies inside the cylinders.
Blue, green, and red colors indicate galaxies associated 
with 3D field, filament, and cluster environments, 
respectively.
The numbers of galaxies plotted reflect the typical galaxy 
numbers in our measurements at $z\approx 1$.
}
\label{fig:scheme}
\end{figure*}

Our environment estimation in \S\ref{sec:env} is 
two-dimensional in nature (2D; projected over 
$\approx 80\text{--}200$~Mpc along the line-of-sight (LOS); 
see Fig.~\ref{fig:scheme}).
Admittedly, the 2D environment measurements have 
limitations and cannot fully recover the entire 3D 
environment.
However, through intensive tests on simulated data, 
studies have found that the 2D environment estimates 
can reliably trace the intrinsic 3D environments. 
For example, \citet{scoville13} found that the 
projected 2D densities are monotonically related to 
the true 3D volume densities with a power-law slope 
of $\approx 0.67$.
\citet{laigle18} found that the 2D measured filaments 
robustly match their 3D counterparts. 
These strong 2D-3D correlations result from the 
fact that, in the projection, the chance for different 
structures to overlap is low.
The low overlapping probability is caused by the facts 
that most ($\gtrsim 80\%$) of the 3D space is the 
field environment in $\Lambda$CDM simulations
(e.g. \hbox{\citealt{aragon_calvo10}}; 
\hbox{\citealt{cautun14}}) 
and that low-mass halos might not host galaxies and 
are thus unobservable (e.g. \hbox{\citealt{desjacques16}} 
and references therein).
Assuming Poisson fluctuations, we estimate the chance for 
two (or more) overlapping filaments along a LOS 
is $\lesssim 3\%$, 
based on the fact that the filament environment covers 
$\lesssim 30\%$ of the total area (see 
Figs.~\ref{fig:det_web_z1} and \ref{fig:det_web_z2}). 
Although a rigorous quantitative demonstration on 
simulated data is beyond the scope of this work, 
we qualitatively explain our 2D environment measurements 
in a straightforward way below. 

Taking our $z$-slice at $z=1$ as an example 
(Fig.~\ref{fig:det_web_z1}), we show the scheme of our 
environment measurements in 
Fig.~\ref{fig:scheme}.\footnote{Technically, the measurements
are more complicated (see \S\ref{sec:env}),
the schematic plot here is just for demonstration purposes.}
Our surface-density field is measured within a 2D circle with 
a radius of $\approx 0.5$~Mpc, projected from a 3D cylinder of 
length $\approx 100$~Mpc (Fig.~\ref{fig:scheme} left). 
Fig.~\ref{fig:scheme} (right) shows typical field, 
filament, and cluster environments. 
The numbers of galaxies plotted reflect the typical galaxy 
numbers in our measurements for different environments at 
$z\approx 1$.
For the field environment, our surface density is averaged 
over the whole cylinder with a volume of 
$\pi \times 0.5^2 \times 100 $~Mpc$^3$. 
This relatively large volume is necessary to include
$\gtrsim 1$ galaxies.
For the filament environment, the density enhancement is mainly 
due to the 3D dense region with scale similar to the filament
``thickness'' ($\lesssim 1$~Mpc scales; see 
Figs.~\ref{fig:det_web_z1} and \ref{fig:det_web_z2}).
The situation for the cluster environment is similar to that 
for filament environment. 

Admittedly, environment mis-classification might happen in 
some cases. 
For example, a filament, when it aligns with the LOS, might be 
mis-classified as a cluster. 
However, this situation should be rare because filaments  
are often not straight and have curved shapes (see 
Figs.~\ref{fig:det_web_z1} and \ref{fig:det_web_z2}).
Also, galaxies in the cluster environment generally have 
significantly lower SFR than those in the filament environment
at $z\lesssim 1$
(e.g. \hbox{\citealt{darvish17}}; Fig.~\ref{fig:Qfrac_vs_M_web}). 
This physical phenomenon would not be observed if our 
classified cluster population is heavily polluted by 
an intrinsic filament population. 

Due the existence of various projection effects, 
any quantitative correlation with 2D environment should 
not be literally interpreted as a quantitative correlation 
with the intrinsic 3D environment. 
For example, a quantity ``$A$'' is found to be positively 
correlated with 2D overdensity with a power-law index 
of $\alpha$. 
We can only conclude qualitatively that $A$ is positively 
related to 3D overdensity, but not quantitatively that 
the relation between $A$ and 3D overdensity is also a power 
law with an index of of $\alpha$.

%


\bsp	
\label{lastpage}
\end{document}